\documentclass[prd,showpacs,showkeys,floatfix,onecolumn,amsmath,amssymb,floatfix]{revtex4}
\usepackage{graphicx,color,dcolumn,booktabs,bm}
\usepackage{longtable,lscape}
\usepackage{amssymb}
\usepackage{indentfirst}
\usepackage{feynmf}   %{feynmp}
\usepackage{epstopdf}   %{feynmp}
\usepackage{slashed}  %for Feynman symbols
\usepackage{cases}
\usepackage{color}
\usepackage{multirow}
\usepackage[colorlinks,citecolor=blue,anchorcolor=red,menucolor=red,linkcolor=red,filecolor=red,runcolor=red,urlcolor=blue,frenchlinks=red]{hyperref}
\usepackage{dcolumn}
\usepackage{bm}
\usepackage{enumerate}

\begin{document}

\title{D-wave charmed and bottomed baryons from QCD sum rules}

\author{Hua-Xing Chen$^1$}
\author{Qiang Mao$^{1,2}$}
\email{maoqiang@ahszu.edu.cn}
%\author{Wei Chen$^3$}
%\email{wec053@mail.usask.ca}
\author{Atsushi Hosaka$^{3,4}$}
\email{hosaka@rcnp.osaka-u.ac.jp}
\author{Xiang Liu$^{5,6}$}
\email{xiangliu@lzu.edu.cn}
\author{Shi-Lin Zhu$^{7,8,9}$}
\email{zhusl@pku.edu.cn}
\affiliation{
$^1$School of Physics and Beijing Key Laboratory of Advanced Nuclear Materials and Physics, Beihang University, Beijing 100191, China \\
$^2$Department of Electrical and Electronic Engineering, Suzhou University, Suzhou 234000, China\\
$^3$Research Center for Nuclear Physics, Osaka University, Ibaraki 567--0047, Japan \\
$^4$J-PARC Branch, KEK Theory Center, Institute of Particle and Nuclear Studies, KEK, Tokai, Ibaraki 319-1106, Japan \\
$^5$School of Physical Science and Technology, Lanzhou University, Lanzhou 730000, China\\
$^6$Research Center for Hadron and CSR Physics, Lanzhou University and Institute of Modern Physics of CAS, Lanzhou 730000, China\\
$^7$School of Physics and State Key Laboratory of Nuclear Physics and Technology, Peking University, Beijing 100871, China \\
$^8$Collaborative Innovation Center of Quantum Matter, Beijing 100871, China \\
$^9$Center of High Energy Physics, Peking University, Beijing 100871, China}

\begin{abstract}
We study the $D$-wave charmed baryons of $SU(3)$ flavor $\mathbf{\bar3}_F$ using the method of QCD sum rules in the framework of heavy quark effective theory.
We find that the $\Lambda_c(2880)$, $\Xi_c(3055)$ and $\Xi_c(3080)$ can be well described by the $D$-wave $SU(3)$ $\mathbf{\bar 3}_F$ charmed baryon multiplets of $J^P=3/2^+$ and $5/2^+$, which contain two $\lambda$-mode orbital excitations, i.e., the $\Lambda_c(2880)$ has $J^P=5/2^+$, and the $\Xi_c(3055)$ and $\Xi_c(3080)$ have $J^P = 3/2^+$ and $5/2^+$, respectively.
Our results also suggest that the $\Lambda_c(2880)$ has a partner state, the $\Lambda_c(3/2^+)$ of $J^P=3/2^+$. Its mass is around $2.81 ^{+0.33}_{-0.18}$ GeV, and the mass difference between it and the $\Lambda_c(2880)$ is $28^{+45}_{-24}$ MeV. We also evaluate the masses of their bottom partners.
\end{abstract}

\maketitle

\section{Introduction}

In the past years important experimental progresses have been made in the field of charmed baryons.
All the $1S$ charmed baryons have been well established~\cite{pdg}. Moreover, the $1P$ states
$\Lambda_c(2595)$, $\Lambda_c(2625)$, $\Xi_c(2790)$ and $\Xi_c(2815)$ have
also been well observed and complete two $SU(3)$ flavor $\mathbf{\bar3}_F$ multiplets of $J^P=1/2^-$ and $3/2^-$~\cite{Albrecht:1993pt,Frabetti:1993hg,Edwards:1994ar,Alexander:1999ud}.
Besides them, there still exist many higher states, i.e.,
the $\Lambda_c(2765)$ ($J^P=?^?$)~\cite{Artuso:2000xy},
the $\Lambda_c(2880)$ ($J^P=5/2^+$)~\cite{Artuso:2000xy},
the $\Lambda_c(2940)$ ($J^P=?^?$)~\cite{Aubert:2006sp,Abe:2006rz},
the $\Sigma_c(2800)$ ($J^P=?^?$)~\cite{Mizuk:2004yu},
the $\Xi_c(2930)$ ($J^P=?^?$)~\cite{Aubert:2007eb},
the $\Xi_c(2980)$ ($J^P=?^?$)~\cite{Chistov:2006zj,Yelton:2016fqw},
the $\Xi_c(3055)$ ($J^P=?^?$)~\cite{Aubert:2007dt,Kato:2016hca},
the $\Xi_c(3080)$ ($J^P=?^?$)~\cite{Chistov:2006zj},
and the $\Xi_c(3123)$ ($J^P=?^?$)~\cite{Aubert:2007dt}.
Some of them may belong to the $1P$ $SU(3)$ flavor $\mathbf{6}_F$ multiplets, while some of them are good $D$-wave charmed baryon candidates.
Especially, in this paper we shall concentrate on the $\Lambda_c(2880)$, $\Xi_c(3055)$ and $\Xi_c(3080)$, which
were proposed (or detailly discussed) in Refs.~\cite{Ebert:2011kk,Chen:2014nyo,Cheng:2015rra} to be $1D$ charmed baryons
of the quantum numbers $J^P=5/2^+$, $3/2^+$ and $5/2^+$, respectively.
More assignments can be found in Refs.~\cite{Cheng:2006dk,Garcilazo:2007eh,Gerasyuta:2007un,Zhong:2007gp,Selem:2006nd}, and we refer to reviews~\cite{Chen:2016spr,Cheng:2015rra} for their recent progress.

The charmed baryons have been investigated using many phenomenological methods/models in the past two decades, including
various quark models~\cite{Capstick:1986bm,Ebert:2007nw,Ortega:2012cx,Shah:2016nxi,Thakkar:2016dna},
the combined expansion in $1/m_Q$ and $1/N_c$~\cite{Jenkins:1996de},
the hyperfine interaction~\cite{Copley:1979wj,Karliner:2008sv},
the Feynman-Hellmann theorem~\cite{Roncaglia:1995az},
the variational approach~\cite{Roberts:2007ni},
the unitarized dynamical model~\cite{GarciaRecio:2012db},
the extended local hidden gauge approach~\cite{Liang:2014eba},
the unitarized chiral perturbation theory~\cite{Lu:2014ina},
and the Lattice QCD~\cite{Bowler:1996ws,Burch:2008qx,Brown:2014ena}, etc.
Their pionic decays and related pion induced reactions have also been studied in Refs.~\cite{Zhong:2007gp,Nagahiro:2016nsx,Kim:2014qha}.
See reviews in Refs.~\cite{Korner:1994nh,Bianco:2003vb,Klempt:2009pi,Crede:2013sze}.

We have also systematically studied the charmed baryons, i.e.,
the $S$-wave bottom baryons~\cite{Liu:2007fg},
the $P$-wave charmed baryons~\cite{Chen:2015kpa},
and the $P$-wave bottom baryons~\cite{Mao:2015gya},
using the method of QCD sum rules~\cite{Shifman:1978bx,Reinders:1984sr} in the framework of heavy quark effective theory (HQET)~\cite{Grinstein:1990mj,Eichten:1989zv,Falk:1990yz}.
This scheme has been successfully applied to study heavy mesons and baryons containing a single heavy
quark~\cite{Bagan:1991sg,Neubert:1991sp,Neubert:1993mb,Broadhurst:1991fc,Ball:1993xv,Huang:1994zj,Dai:1996yw,Dai:1993kt,Dai:1996qx,Colangelo:1998ga,Dai:2003yg,Zhou:2014ytp,Zhou:2015ywa,Shuryak:1981fza,Grozin:1992td,Bagan:1993ii,Dai:1995bc,Dai:1996xv,Groote:1996em,Zhu:2000py,Lee:2000tb,Huang:2000tn,Wang:2003zp},
while other studies using the method of QCD sum rules but not in HQET can be found in Refs.~\cite{Bagan:1992tp,Bagan:1991sc,Duraes:2007te,Wang:2007sqa,Chen:2016qju,Chen:2015moa}.

In this paper we study the $D$-wave charmed baryons of $SU(3)$ flavor $\mathbf{\bar3}_F$ $(\Lambda_c, \Xi_c)$ using the method of QCD sum rules within HQET.
This paper is organized as follows. First we
systematically construct the interpolating currents for the $D$-wave charmed baryons in Sec.~\ref{sec:current}. Then
we select some of them to perform the QCD sum rule analysis at both the leading order in Sec.~\ref{sec:leading}
and the order ${\mathcal O}(1/m_Q)$ in Sec.~\ref{sec:nexttoleading}.
During the calculations, we shall take the ${\mathcal O}(1/m_c)$ corrections ($m_c$ is the heavy quark mass) into account, and extract the chromomagnetic splitting.
In Sec.~\ref{sec:numerical} we perform numerical analyses and discuss the obtained results. A short summary is given in Sec.~\ref{sec:summary}.

\section{interpolating fields for the $P$-wave charmed baryon}
\label{sec:current}

The charmed baryons of $P$- and $D$-waves have been systemically classified in Ref.~\cite{Chen:2007xf}, where their strong decays were systematically investigated using the $^3P_0$ model.
The $P$-wave charmed baryon interpolating fields have been systematically constructed in Refs.~\cite{Chen:2015kpa,Mao:2015gya} using the same notations, i.e., $l_\rho$ denotes the orbital angular momentum between the two light quarks and $l_\lambda$ denotes the orbital angular momentum between the charm quark and the two-light-quark system.

In this paper we follow the same approach of Refs.~\cite{Chen:2015kpa,Mao:2015gya},
and construct the $D$-wave ($L = 2$) charmed baryon interpolating fields.
We use the notation $J^{\alpha_1\cdots\alpha_{j-1/2}}_{j,P,F,j_l,s_l,\rho\rho/\lambda\lambda/\rho\lambda}$
to denote the $D$-wave charmed baryon interpolating field having the total angular momentum $j$ and parity $P$,
and belonging to the spin doublet $[F,j_l,s_l,\rho\rho/\lambda\lambda/\rho\lambda]$.
Here $F$ denotes the $SU(3)$ flavor representation, either $\mathbf{\bar 3}_F$ or $\mathbf{6}_F$;
$j_l$ and $s_l$ denote the total angular momentum and spin angular momentum of the light components;
$[\rho\rho]$ denotes $l_\rho = 2$ and $l_\lambda = 0$, $[\lambda\lambda]$ denotes $l_\rho = 0$ and $l_\lambda = 2$, and $[\rho\lambda]$ denotes $l_\rho = 1$ and $l_\lambda = 1$.
We have the relations $L = l_\lambda \otimes l_\rho$, $j_l = L \otimes s_l$ and $j = j_l \otimes s_Q$, where $s_Q = 1/2$ is the spin of the heavy quark.

\begin{figure*}[htb]
\begin{center}
\scalebox{0.6}{\includegraphics{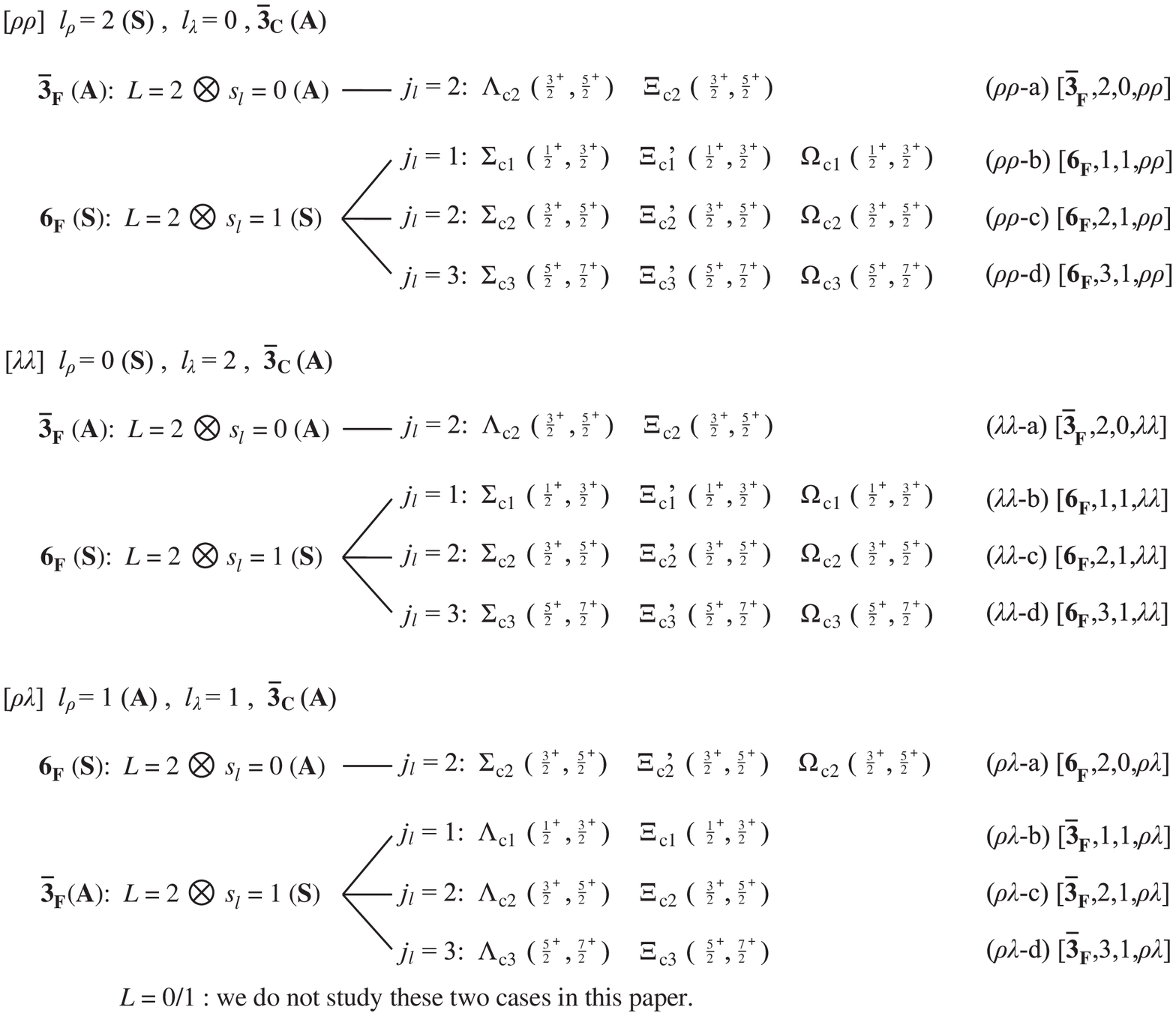}}
\end{center}
\caption{The notations for $D$-wave charmed baryons: $\mathbf{6}_F$ ($\mathbf{S}$) and $\mathbf{\bar 3}_F$ ($\mathbf{A}$) denote the $SU(3)$ flavor representations; $\mathbf{\bar 3}_C$ ($\mathbf{A}$) denotes the $SU(3)$ color representation; $s_l$ is the spin angular momentum of the two light quarks; $j_l = L \otimes s_l = l_\lambda \otimes l_\rho \otimes s_l$ is the total angular momentum of the two light quarks.
\label{fig:pwave}}
\end{figure*}

We summarize all the possible configurations of the $D$-wave ($L = 2$) charmed baryons in Fig.~\ref{fig:pwave}, where
$\mathbf{A}$ and $\mathbf{S}$ denote the structure to be antisymmetric and symmetric, respectively.
We note that the type $[\rho\lambda]$ ($l_\rho = 1$ and $l_\lambda = 1$) can actually have total orbital angular momenta $L = 0$, $1$ and $2$, but
in this paper we only concentrate on the $L=2$ for the $D$-wave case.

Generally, the interpolating field for charmed baryons can be written as a combination of a diquark field and a heavy quark field£º
\begin{eqnarray}
J(x) \sim \epsilon_{abc} \left( q^{aT}(x) \mathbb{C} \Gamma_1 q^b (x) \right) \Gamma_2 h_v^c (x) \, ,
\label{eq:baryonfield}
\end{eqnarray}
where $a$, $b$ and $c$ are color indices; $\epsilon_{abc}$ is the totally antisymmetric tensor; the superscript $T$ represents the transpose of the Dirac indices; the matrices $\Gamma_{1,2}$ are Dirac matrices which describe the Lorentz structure; $\mathbb{C}$ is the charge-conjugation operator; $q(x)$ denotes the light quark field at location $x$, and it can be either $u(x)$ or $d(x)$ or $s(x)$; $h_v(x)$ denotes the heavy quark field, and we have used the Fierz transformation to move it to the rightmost place.
Besides these notations, $\gamma^t_\mu = \gamma_\mu - v\!\!\!\slash v_\mu$, $D_\mu = \partial_\mu - i g A_\mu$, $D_\mu^t = D_\mu - (D \cdot v) v_\mu$, $v$ is the velocity of the heavy quark, and $g_t^{\alpha_1\alpha_2}=g^{\alpha_1\alpha_2} - v^{\alpha_1} v^{\alpha_2}$ is the transverse metric tensor.

To describe the orbital angular momenta, we directly apply two derivatives containing two symmetric Lorentz indices on the light diquark field
(see Refs.~\cite{Chen:2015kpa,Mao:2015gya,Zhu:2000py,Lee:2000tb,Huang:2000tn,Wang:2003zp} for more details) to construct the $D$-wave diquark fields of the configuration $[\rho\rho/\lambda\lambda/\rho\lambda]$:
\begin{eqnarray*}
~[\rho\rho]~[^1D_2] &:& \, l_\rho=2~(\mathbf{S}) \, , \, l_\lambda=0 \, , \, L=2 \, , \, s_l=0~(\mathbf{A}) \, , \, j_l=2
\\ \nonumber \epsilon_{abc} &\times& \Big ( [\mathcal{D}_{\mu_1} \mathcal{D}_{\mu_2} q^{aT}(x)] \mathbb{C} \gamma_5 q^b(x) - 2 [\mathcal{D}_{\mu_1} q^{aT}(x)] \mathbb{C} \gamma_5 [\mathcal{D}_{\mu_2} q^b(x)] + q^{aT}(x) \mathbb{C} \gamma_5 [\mathcal{D}_{\mu_1} \mathcal{D}_{\mu_2} q^b(x)] \Big ) + \mu_1 \leftrightarrow \mu_2 \, ,
\\
~[\rho\rho]~[^3D_{1/2/3}] &:& \, l_\rho=2~(\mathbf{S}) \, , \, l_\lambda=0 \, , \, L=2 \, , \, s_l=1~(\mathbf{A}) \, , \, j_l=1/2/3
\\ \nonumber \epsilon_{abc} &\times& \Big ( [\mathcal{D}_{\mu_1} \mathcal{D}_{\mu_2} q^{aT}(x)] \mathbb{C} \gamma_\mu q^b(x) - 2 [\mathcal{D}_{\mu_1} q^{aT}(x)] \mathbb{C} \gamma_\mu [\mathcal{D}_{\mu_2} q^b(x)] + q^{aT}(x) \mathbb{C} \gamma_\mu [\mathcal{D}_{\mu_1} \mathcal{D}_{\mu_2} q^b(x)] \Big ) + \mu_1 \leftrightarrow \mu_2 \, ,
\\ ~[\lambda\lambda]~[^1S_0] &:& \, l_\rho=0~(\mathbf{S}) \, , \, l_\lambda=2 \, , \, L=2 \, , \, s_l=0~(\mathbf{A}) \, , \, j_l=2
\\ \nonumber \epsilon_{abc} &\times& \Big ( [\mathcal{D}_{\mu_1} \mathcal{D}_{\mu_2} q^{aT}(x)] \mathbb{C} \gamma_5 q^b(x) + 2 [\mathcal{D}_{\mu_1} q^{aT}(x)] \mathbb{C} \gamma_5 [\mathcal{D}_{\mu_2} q^b(x)] + q^{aT}(x) \mathbb{C} \gamma_5 [\mathcal{D}_{\mu_1} \mathcal{D}_{\mu_2} q^b(x)] \Big ) + \mu_1 \leftrightarrow \mu_2 \, ,
\\ ~[\lambda\lambda]~[^3S_1] &:& \, l_\rho=0~(\mathbf{S}) \, , \, l_\lambda=2 \, , \, L=2 \, , \, s_l=1~(\mathbf{A}) \, , \, j_l=1/2/3
\\ \nonumber \epsilon_{abc} &\times& \Big ( [\mathcal{D}_{\mu_1} \mathcal{D}_{\mu_2} q^{aT}(x)] \mathbb{C} \gamma_\mu q^b(x) + 2 [\mathcal{D}_{\mu_1} q^{aT}(x)] \mathbb{C} \gamma_\mu [\mathcal{D}_{\mu_2} q^b(x)] + q^{aT}(x) \mathbb{C} \gamma_\mu [\mathcal{D}_{\mu_1} \mathcal{D}_{\mu_2} q^b(x)] \Big ) + \mu_1 \leftrightarrow \mu_2 \, ,
\\ ~[\rho\lambda]~[^1P_1] &:& \, l_\rho=1~(\mathbf{A}) \, , \, l_\lambda=1 \, , \, L=2 \, , \, s_l=0~(\mathbf{A}) \, , \, j_l=2
\\ \nonumber \epsilon_{abc} &\times& \Big ( [\mathcal{D}_{\mu_1} \mathcal{D}_{\mu_2} q^{aT}(x)] \mathbb{C} \gamma_5 q^b(x) - q^{aT}(x) \mathbb{C} \gamma_5 [\mathcal{D}_{\mu_1} \mathcal{D}_{\mu_2} q^b(x)] \Big ) + \mu_1 \leftrightarrow \mu_2 \, ,
\\ ~[\rho\lambda]~[^3P_{0/1/2}] &:& \, l_\rho=1~(\mathbf{A}) \, , \, l_\lambda=1 \, , \, L=2 \, , \, s_l=1~(\mathbf{A}) \, , \, j_l=1/2/3
\\ \nonumber \epsilon_{abc} &\times& \Big ( [\mathcal{D}_{\mu_1} \mathcal{D}_{\mu_2} q^{aT}(x)] \mathbb{C} \gamma_\mu q^b(x) - q^{aT}(x) \mathbb{C} \gamma_\mu [\mathcal{D}_{\mu_1} \mathcal{D}_{\mu_2} q^b(x)] \Big ) + \mu_1 \leftrightarrow \mu_2 \, ,
\\ ~[\rho\lambda]~[\cdots] &:& \, l_\rho=1~(\mathbf{A}) \, , \, l_\lambda=1 \, , \, L=0/1 \, , \, s_l=0/1~(\mathbf{A/S}) \, , \, j_l=0/1/2
\\ \nonumber && \mbox{we do not study these cases in this paper.}
\end{eqnarray*}
In these expressions, we have used $[^{2s_l+1} \big( l_\rho \big) _{l_\rho \otimes s_l} ]$ to denote the spin, orbital and total angular momenta of the diquark, where $l_\lambda$ (the orbital angular momentum between the charm quark and the diquark) is not taken into account. Especially, $[^3D_{1/2/3}]$ means $l_\rho \otimes s_l$ can be 1, 2 and 3, while $[^3P_{0/1/2}]$ means $l_\rho \otimes s_l$ can be 0, 1 and 2.

Based on these $D$-wave diquark fields, we can construct the $D$-wave ($L = 2$) charmed baryons of the configuration $[\rho\rho/\lambda\lambda/\rho\lambda]$:
\begin{itemize}

\item $[\rho\rho]$ ($l_\rho = 2$ ($\mathbf{S}$) and $l_\lambda = 0$):

\begin{enumerate}[({$\rho\rho$}-a)]

\item $[\mathbf{\bar 3}_F, 2, 0, \rho\rho]$ with $s_l=0$ ($\mathbf{A}$) and $j_l = 2$. Now the diquark has color $\mathbf{\bar 3}_C$ ($\mathbf{A}$) and flavor $\mathbf{\bar 3}_F$ ($\mathbf{A}$), and we obtain a spin doublet $(j^P = 3/2^+ , 5/2^+)$:
\begin{eqnarray}
&& J^{\alpha}_{3/2,+,\mathbf{\bar 3}_F,2,0,\rho\rho}(x)
\label{eq:current1}
\\ \nonumber &=& \epsilon_{abc} \Big ( [\mathcal{D}^t_{\mu_1} \mathcal{D}^t_{\mu_2} q^{aT}(x)] \mathbb{C} \gamma_5 q^b(x) - 2 [\mathcal{D}^t_{\mu_1} q^{aT}(x)] \mathbb{C} \gamma_5 [\mathcal{D}^t_{\mu_2} q^b(x)] + q^{aT}(x) \mathbb{C} \gamma_5 [\mathcal{D}^t_{\mu_1} \mathcal{D}^t_{\mu_2} q^b(x)] \Big )
\\ \nonumber && ~~~~~~~~~~~~
\times \Big ( {1\over2} g_t^{\mu_1 \alpha} g_t^{\mu_2 \mu_4} + {1\over2} g_t^{\mu_2 \alpha} g_t^{\mu_1 \mu_4} - {1\over3} g_t^{\mu_1 \mu_2} g_t^{\mu_4 \alpha} \Big ) \times \gamma^t_{\mu_4} \gamma_5 h_v^c(x) \, ,
\\ && J^{\alpha_1\alpha_2}_{5/2,+,\mathbf{\bar 3}_F,2,0,\rho\rho}(x)
\label{eq:current2}
\\ \nonumber &=& \epsilon_{abc} \Big ( [\mathcal{D}^t_{\mu_1} \mathcal{D}^t_{\mu_2} q^{aT}(x)] \mathbb{C} \gamma_5 q^b(x) - 2 [\mathcal{D}^t_{\mu_1} q^{aT}(x)] \mathbb{C} \gamma_5 [\mathcal{D}^t_{\mu_2} q^b(x)] + q^{aT}(x) \mathbb{C} \gamma_5 [\mathcal{D}^t_{\mu_1} \mathcal{D}^t_{\mu_2} q^b(x)] \Big )
\\ \nonumber && ~~~~~~~~~~~~
\times \Gamma_t^{\alpha_1\alpha_2,\mu_1\mu_2} \times  h_v^c(x) \, ,
\end{eqnarray}
where $\Gamma_t^{\alpha_1\alpha_2,\mu_1\mu_2}$ is the projection operator projecting into pure spin 2, whose explicit form is given in Appendix~\ref{sec:project}.

\item $[\mathbf{6}_F, 1, 1, \rho\rho]$ with $s_l = 1$ ($\mathbf{S}$) and $j_l = 1$. Now the diquark has color $\mathbf{\bar 3}_C$ ($\mathbf{A}$) and flavor $\mathbf{6}_F$ ($\mathbf{S}$), and we obtain a spin doublet $(1/2^+, 3/2^+)$:
\begin{eqnarray}
&& J_{1/2,+,\mathbf{6}_F,1,1,\rho\rho}(x)
\label{eq:current3}
\\ \nonumber &=& \epsilon_{abc} \Big ( [\mathcal{D}^t_{\mu_1} \mathcal{D}^t_{\mu_2} q^{aT}(x)] \mathbb{C} \gamma^t_{\mu_3} q^b(x) - 2 [\mathcal{D}^t_{\mu_1} q^{aT}(x)] \mathbb{C} \gamma^t_{\mu_3} [\mathcal{D}^t_{\mu_2} q^b(x)] + q^{aT}(x) \mathbb{C} \gamma^t_{\mu_3} [\mathcal{D}^t_{\mu_1} \mathcal{D}^t_{\mu_2} q^b(x)] \Big )
\\ \nonumber && ~~~~~~~~~~~~
\times \Big ( g_t^{\mu_1 \mu_3} g_t^{\mu_2 \mu_4} + g_t^{\mu_2 \mu_3} g_t^{\mu_1 \mu_4} \Big ) \times \gamma^t_{\mu_4} \gamma_5 h_v^c(x) \, ,
\\ && J^{\alpha}_{3/2,+,\mathbf{6}_F,1,1,\rho\rho}(x)
\label{eq:current4}
\\ \nonumber &=& \epsilon_{abc} \Big ( [\mathcal{D}^t_{\mu_1} \mathcal{D}^t_{\mu_2} q^{aT}(x)] \mathbb{C} \gamma^t_{\mu_3} q^b(x) - 2 [\mathcal{D}^t_{\mu_1} q^{aT}(x)] \mathbb{C} \gamma^t_{\mu_3} [\mathcal{D}^t_{\mu_2} q^b(x)] + q^{aT}(x) \mathbb{C} \gamma^t_{\mu_3} [\mathcal{D}^t_{\mu_1} \mathcal{D}^t_{\mu_2} q^b(x)] \Big )
\\ \nonumber && ~~~~~~~~~~~~
\times \Big ( {1\over2} g_t^{\mu_1 \mu_3} g_t^{\mu_2 \alpha} + {1\over2} g_t^{\mu_2 \mu_3} g_t^{\mu_1 \alpha} - {1\over3} g_t^{\mu_1 \mu_2} g_t^{\mu_3 \alpha} \Big ) \times h_v^c(x) \, .
\end{eqnarray}

\item $[\mathbf{6}_F, 2, 1, \rho\rho]$ with $s_l = 1$ ($\mathbf{S}$) and $j_l = 2$. Now the diquark has color $\mathbf{\bar 3}_C$ ($\mathbf{A}$) and flavor $\mathbf{6}_F$ ($\mathbf{S}$), and we obtain a spin doublet $(3/2^+, 5/2^+)$. We failed to construct these currents because we do not know how to explicitly combine angular momenta $J = 2$ and $J = 1$ to be $J = 2$, i.e., how to use two symmetric indices $\{ \mu_1 \mu_2 + \mu_2 \mu_1 \}$ and another index $\mu_3$ to obtain two symmetric indices $\{ \alpha_1 \alpha_2 + \alpha_2 \alpha_1 \}$. To estimate the masses of these states, we shall use the currents of ($\rho\rho$-b) and ($\rho\rho$-d) as explained in Sec.~\ref{sec:summary}.

\item $[\mathbf{6}_F, 3, 1, \rho\rho]$ with $s_l = 1$ ($\mathbf{S}$) and $j_l = 3$. Now the diquark has color $\mathbf{\bar 3}_C$ ($\mathbf{A}$) and flavor $\mathbf{6}_F$ ($\mathbf{S}$), and we obtain a spin doublet $(5/2^+, 7/2^+)$:
\begin{eqnarray}
&& J^{\alpha_1\alpha_2}_{5/2,+,\mathbf{6}_F,3,1,\rho\rho}(x)
\label{eq:current5}
\\ \nonumber &=& \epsilon_{abc} \Big ( [\mathcal{D}^t_{\mu_1} \mathcal{D}^t_{\mu_2} q^{aT}(x)] \mathbb{C} \gamma^t_{\mu_3} q^b(x) - 2 [\mathcal{D}^t_{\mu_1} q^{aT}(x)] \mathbb{C} \gamma^t_{\mu_3} [\mathcal{D}^t_{\mu_2} q^b(x)] + q^{aT}(x) \mathbb{C} \gamma^t_{\mu_3} [\mathcal{D}^t_{\mu_1} \mathcal{D}^t_{\mu_2} q^b(x)] \Big )
\\ \nonumber && ~~~~~~~~~~~~
\times {\Gamma_t^{\alpha_1\alpha_2,}}_{\nu_1\nu_2} \times \Big ( g_t^{\mu_1 \nu_1} g_t^{\mu_2 \nu_2} g_t^{\mu_3 \mu_4} + g_t^{\mu_3 \nu_1} g_t^{\mu_2 \nu_2} g_t^{\mu_1 \mu_4} + g_t^{\mu_3 \nu_1} g_t^{\mu_1 \nu_2} g_t^{\mu_2 \mu_4} \Big ) \times \gamma^t_{\mu_4} \gamma_5 h_v^c(x) \, ,
\\ && J^{\alpha_1\alpha_2\alpha_3}_{7/2,+,\mathbf{6}_F,3,1,\rho\rho}(x)
\label{eq:current6}
\\ \nonumber &=& \epsilon_{abc} \Big ( [\mathcal{D}^t_{\mu_1} \mathcal{D}^t_{\mu_2} q^{aT}(x)] \mathbb{C} \gamma^t_{\mu_3} q^b(x) - 2 [\mathcal{D}^t_{\mu_1} q^{aT}(x)] \mathbb{C} \gamma^t_{\mu_3} [\mathcal{D}^t_{\mu_2} q^b(x)] + q^{aT}(x) \mathbb{C} \gamma^t_{\mu_3} [\mathcal{D}^t_{\mu_1} \mathcal{D}^t_{\mu_2} q^b(x)] \Big )
\\ \nonumber && ~~~~~~~~~~~~
\times \Gamma_t^{\alpha_1\alpha_2\alpha_3,\mu_1\mu_2\mu_3} \times h_v^c(x) \, ,
\end{eqnarray}
where $\Gamma_t^{\alpha_1\alpha_2\alpha_3,\mu_1\mu_2\mu_3}$ is the projection operator projecting into pure spin 3.

\end{enumerate}

\item $[\lambda\lambda]$ ($l_\rho = 0$ ($\mathbf{S}$) and $l_\lambda = 2$):

\begin{enumerate}[({$\lambda\lambda$}-a)]

\item $[\mathbf{\bar 3}_F, 2, 0, \lambda\lambda]$ with $s_l=0$ ($\mathbf{A}$) and $j_l = 2$. Now the diquark has color $\mathbf{\bar 3}_C$ ($\mathbf{A}$) and flavor $\mathbf{\bar 3}_F$ ($\mathbf{A}$), and we obtain a spin doublet $(j^P = 3/2^+ , 5/2^+)$:
\begin{eqnarray}
&& J^{\alpha}_{3/2,+,\mathbf{\bar 3}_F,2,0,\lambda\lambda}(x)
\label{eq:current7}
\\ \nonumber &=& \epsilon_{abc} \Big ( [\mathcal{D}^t_{\mu_1} \mathcal{D}^t_{\mu_2} q^{aT}(x)] \mathbb{C} \gamma_5 q^b(x) + 2 [\mathcal{D}^t_{\mu_1} q^{aT}(x)] \mathbb{C} \gamma_5 [\mathcal{D}^t_{\mu_2} q^b(x)] + q^{aT}(x) \mathbb{C} \gamma_5 [\mathcal{D}^t_{\mu_1} \mathcal{D}^t_{\mu_2} q^b(x)] \Big )
\\ \nonumber && ~~~~~~~~~~~~
\times \Big ( {1\over2} g_t^{\mu_1 \alpha} g_t^{\mu_2 \mu_4} + {1\over2} g_t^{\mu_2 \alpha} g_t^{\mu_1 \mu_4} - {1\over3} g_t^{\mu_1 \mu_2} g_t^{\mu_4 \alpha} \Big ) \times \gamma^t_{\mu_4} \gamma_5 h_v^c(x) \, ,
\\ && J^{\alpha_1\alpha_2}_{5/2,+,\mathbf{\bar 3}_F,2,0,\lambda\lambda}(x)
\label{eq:current8}
\\ \nonumber &=& \epsilon_{abc} \Big ( [\mathcal{D}^t_{\mu_1} \mathcal{D}^t_{\mu_2} q^{aT}(x)] \mathbb{C} \gamma_5 q^b(x) + 2 [\mathcal{D}^t_{\mu_1} q^{aT}(x)] \mathbb{C} \gamma_5 [\mathcal{D}^t_{\mu_2} q^b(x)] + q^{aT}(x) \mathbb{C} \gamma_5 [\mathcal{D}^t_{\mu_1} \mathcal{D}^t_{\mu_2} q^b(x)] \Big )
\\ \nonumber && ~~~~~~~~~~~~
\times \Gamma_t^{\alpha_1\alpha_2,\mu_1\mu_2} \times  h_v^c(x) \, .
\end{eqnarray}

\item $[\mathbf{6}_F, 1, 1, \lambda\lambda]$ with $s_l = 1$ ($\mathbf{S}$) and $j_l = 1$. Now the diquark has color $\mathbf{\bar 3}_C$ ($\mathbf{A}$) and flavor $\mathbf{6}_F$ ($\mathbf{S}$), and we obtain a spin doublet $(1/2^+, 3/2^+)$:
\begin{eqnarray}
&& J_{1/2,+,\mathbf{6}_F,1,1,\lambda\lambda}(x)
\label{eq:current9}
\\ \nonumber &=& \epsilon_{abc} \Big ( [\mathcal{D}^t_{\mu_1} \mathcal{D}^t_{\mu_2} q^{aT}(x)] \mathbb{C} \gamma^t_{\mu_3} q^b(x) + 2 [\mathcal{D}^t_{\mu_1} q^{aT}(x)] \mathbb{C} \gamma^t_{\mu_3} [\mathcal{D}^t_{\mu_2} q^b(x)] + q^{aT}(x) \mathbb{C} \gamma^t_{\mu_3} [\mathcal{D}^t_{\mu_1} \mathcal{D}^t_{\mu_2} q^b(x)] \Big )
\\ \nonumber && ~~~~~~~~~~~~
\times \Big ( g_t^{\mu_1 \mu_3} g_t^{\mu_2 \mu_4} + g_t^{\mu_2 \mu_3} g_t^{\mu_1 \mu_4} \Big ) \times \gamma^t_{\mu_4} \gamma_5 h_v^c(x) \, ,
\\ && J^{\alpha}_{3/2,+,\mathbf{6}_F,1,1,\lambda\lambda}(x)
\label{eq:current10}
\\ \nonumber &=& \epsilon_{abc} \Big ( [\mathcal{D}^t_{\mu_1} \mathcal{D}^t_{\mu_2} q^{aT}(x)] \mathbb{C} \gamma^t_{\mu_3} q^b(x) + 2 [\mathcal{D}^t_{\mu_1} q^{aT}(x)] \mathbb{C} \gamma^t_{\mu_3} [\mathcal{D}^t_{\mu_2} q^b(x)] + q^{aT}(x) \mathbb{C} \gamma^t_{\mu_3} [\mathcal{D}^t_{\mu_1} \mathcal{D}^t_{\mu_2} q^b(x)] \Big )
\\ \nonumber && ~~~~~~~~~~~~
\times \Big ( {1\over2} g_t^{\mu_1 \mu_3} g_t^{\mu_2 \alpha} + {1\over2} g_t^{\mu_2 \mu_3} g_t^{\mu_1 \alpha} - {1\over3} g_t^{\mu_1 \mu_2} g_t^{\mu_3 \alpha} \Big ) \times h_v^c(x) \, .
\end{eqnarray}

\item $[\mathbf{6}_F, 2, 1, \lambda\lambda]$ with $s_l = 1$ ($\mathbf{S}$) and $j_l = 2$. Now the diquark has color $\mathbf{\bar 3}_C$ ($\mathbf{A}$) and flavor $\mathbf{6}_F$ ($\mathbf{S}$), and we obtain a spin doublet $(3/2^+, 5/2^+)$. We failed to construct these currents.

\item $[\mathbf{6}_F, 3, 1, \lambda\lambda]$ with $s_l = 1$ ($\mathbf{S}$) and $j_l = 3$. Now the diquark has color $\mathbf{\bar 3}_C$ ($\mathbf{A}$) and flavor $\mathbf{6}_F$ ($\mathbf{S}$), and we obtain a spin doublet $(5/2^+, 7/2^+)$:
\begin{eqnarray}
&& J^{\alpha_1\alpha_2}_{5/2,+,\mathbf{6}_F,3,1,\lambda\lambda}(x)
\label{eq:current11}
\\ \nonumber &=& \epsilon_{abc} \Big ( [\mathcal{D}^t_{\mu_1} \mathcal{D}^t_{\mu_2} q^{aT}(x)] \mathbb{C} \gamma^t_{\mu_3} q^b(x) + 2 [\mathcal{D}^t_{\mu_1} q^{aT}(x)] \mathbb{C} \gamma^t_{\mu_3} [\mathcal{D}^t_{\mu_2} q^b(x)] + q^{aT}(x) \mathbb{C} \gamma^t_{\mu_3} [\mathcal{D}^t_{\mu_1} \mathcal{D}^t_{\mu_2} q^b(x)] \Big )
\\ \nonumber && ~~~~~~~~~~~~
\times {\Gamma_t^{\alpha_1\alpha_2,}}_{\nu_1\nu_2} \times \Big ( g_t^{\mu_1 \nu_1} g_t^{\mu_2 \nu_2} g_t^{\mu_3 \mu_4} + g_t^{\mu_3 \nu_1} g_t^{\mu_2 \nu_2} g_t^{\mu_1 \mu_4} + g_t^{\mu_3 \nu_1} g_t^{\mu_1 \nu_2} g_t^{\mu_2 \mu_4} \Big ) \times \gamma^t_{\mu_4} \gamma_5 h_v^c(x) \, ,
\\ && J^{\alpha_1\alpha_2\alpha_3}_{7/2,+,\mathbf{6}_F,3,1,\lambda\lambda}(x)
\label{eq:current12}
\\ \nonumber &=& \epsilon_{abc} \Big ( [\mathcal{D}^t_{\mu_1} \mathcal{D}^t_{\mu_2} q^{aT}(x)] \mathbb{C} \gamma^t_{\mu_3} q^b(x) + 2 [\mathcal{D}^t_{\mu_1} q^{aT}(x)] \mathbb{C} \gamma^t_{\mu_3} [\mathcal{D}^t_{\mu_2} q^b(x)] + q^{aT}(x) \mathbb{C} \gamma^t_{\mu_3} [\mathcal{D}^t_{\mu_1} \mathcal{D}^t_{\mu_2} q^b(x)] \Big )
\\ \nonumber && ~~~~~~~~~~~~
\times \Gamma_t^{\alpha_1\alpha_2\alpha_3,\mu_1\mu_2\mu_3} \times h_v^c(x) \, .
\end{eqnarray}

\end{enumerate}

\item $[\rho\lambda]$ ($l_\rho = 1$ ($\mathbf{A}$) and $l_\lambda = 1$):

\begin{enumerate}[({$\rho\lambda$}-a)]

\item $[\mathbf{6}_F, 2, 0, \rho\lambda]$ with $s_l=0$ ($\mathbf{A}$) and $j_l = 2$. Now the diquark has color $\mathbf{\bar 3}_C$ ($\mathbf{A}$) and flavor $\mathbf{6}_F$ ($\mathbf{S}$), and we obtain a spin doublet $(j^P = 3/2^+ , 5/2^+)$:
\begin{eqnarray}
J^{\alpha}_{3/2,+,\mathbf{6}_F,2,0,\rho\lambda}(x)
&=& \epsilon_{abc} \Big ( [\mathcal{D}^t_{\mu_1} \mathcal{D}^t_{\mu_2} q^{aT}(x)] \mathbb{C} \gamma_5 q^b(x) - q^{aT}(x) \mathbb{C} \gamma_5 [\mathcal{D}^t_{\mu_1} \mathcal{D}^t_{\mu_2} q^b(x)] \Big )
\label{eq:current13}
\\ \nonumber && ~~~~~~~~~~~~
\times \Big ( {1\over2} g_t^{\mu_1 \alpha} g_t^{\mu_2 \mu_4} + {1\over2} g_t^{\mu_2 \alpha} g_t^{\mu_1 \mu_4} - {1\over3} g_t^{\mu_1 \mu_2} g_t^{\mu_4 \alpha} \Big ) \times \gamma^t_{\mu_4} \gamma_5 h_v^c(x) \, ,
\\ J^{\alpha_1\alpha_2}_{5/2,+,\mathbf{6}_F,2,0,\rho\lambda}(x)
&=& \epsilon_{abc} \Big ( [\mathcal{D}^t_{\mu_1} \mathcal{D}^t_{\mu_2} q^{aT}(x)] \mathbb{C} \gamma_5 q^b(x) - q^{aT}(x) \mathbb{C} \gamma_5 [\mathcal{D}^t_{\mu_1} \mathcal{D}^t_{\mu_2} q^b(x)] \Big )
\label{eq:current14}
\\ \nonumber && ~~~~~~~~~~~~
\times \Gamma_t^{\alpha_1\alpha_2,\mu_1\mu_2} \times  h_v^c(x) \, .
\end{eqnarray}

\item $[\mathbf{\bar 3}_F, 1, 1, \rho\lambda]$ with $s_l = 1$ ($\mathbf{S}$) and $j_l = 1$. Now the diquark has color $\mathbf{\bar 3}_C$ ($\mathbf{A}$) and flavor $\mathbf{\bar 3}_F$ ($\mathbf{A}$), and we obtain a spin doublet $(1/2^+, 3/2^+)$:
\begin{eqnarray}
J_{1/2,+,\mathbf{\bar 3}_F,1,1,\rho\lambda}(x)
&=& \epsilon_{abc} \Big ( [\mathcal{D}^t_{\mu_1} \mathcal{D}^t_{\mu_2} q^{aT}(x)] \mathbb{C} \gamma^t_{\mu_3} q^b(x) - q^{aT}(x) \mathbb{C} \gamma^t_{\mu_3} [\mathcal{D}^t_{\mu_1} \mathcal{D}^t_{\mu_2} q^b(x)] \Big )
\label{eq:current15}
\\ \nonumber && ~~~~~~~~~~~~
\times \Big ( g_t^{\mu_1 \mu_3} g_t^{\mu_2 \mu_4} + g_t^{\mu_2 \mu_3} g_t^{\mu_1 \mu_4} \Big ) \times \gamma^t_{\mu_4} \gamma_5 h_v^c(x) \, ,
\\ J^{\alpha}_{3/2,+,\mathbf{\bar 3}_F,1,1,\rho\lambda}(x)
&=& \epsilon_{abc} \Big ( [\mathcal{D}^t_{\mu_1} \mathcal{D}^t_{\mu_2} q^{aT}(x)] \mathbb{C} \gamma^t_{\mu_3} q^b(x) - q^{aT}(x) \mathbb{C} \gamma^t_{\mu_3} [\mathcal{D}^t_{\mu_1} \mathcal{D}^t_{\mu_2} q^b(x)] \Big )
\label{eq:current16}
\\ \nonumber && ~~~~~~~~~~~~
\times \Big ( {1\over2} g_t^{\mu_1 \mu_3} g_t^{\mu_2 \alpha} + {1\over2} g_t^{\mu_2 \mu_3} g_t^{\mu_1 \alpha} - {1\over3} g_t^{\mu_1 \mu_2} g_t^{\mu_3 \alpha} \Big ) \times h_v^c(x) \, .
\end{eqnarray}

\item $[\mathbf{\bar 3}_F, 2, 1, \rho\lambda]$ with $s_l = 1$ ($\mathbf{S}$) and $j_l = 2$. Now the diquark has color $\mathbf{\bar 3}_C$ ($\mathbf{A}$) and flavor $\mathbf{\bar 3}_F$ ($\mathbf{A}$), and we obtain a spin doublet $(3/2^+, 5/2^+)$. We failed to construct these currents.

\item $[\mathbf{\bar 3}_F, 3, 1, \rho\lambda]$ with $s_l = 1$ ($\mathbf{S}$) and $j_l = 3$. Now the diquark has color $\mathbf{\bar 3}_C$ ($\mathbf{A}$) and flavor $\mathbf{\bar 3}_F$ ($\mathbf{A}$), and we obtain a spin doublet $(5/2^+, 7/2^+)$:
\begin{eqnarray}
&& J^{\alpha_1\alpha_2}_{5/2,+,\mathbf{\bar 3}_F,3,1,\rho\lambda}(x)
\label{eq:current17}
\\ \nonumber &=& \epsilon_{abc} \Big ( [\mathcal{D}^t_{\mu_1} \mathcal{D}^t_{\mu_2} q^{aT}(x)] \mathbb{C} \gamma^t_{\mu_3} q^b(x) - q^{aT}(x) \mathbb{C} \gamma^t_{\mu_3} [\mathcal{D}^t_{\mu_1} \mathcal{D}^t_{\mu_2} q^b(x)] \Big )
\\ \nonumber && ~~~~~~~~~~~~
\times {\Gamma_t^{\alpha_1\alpha_2,}}_{\nu_1\nu_2} \times \Big ( g_t^{\mu_1 \nu_1} g_t^{\mu_2 \nu_2} g_t^{\mu_3 \mu_4} + g_t^{\mu_3 \nu_1} g_t^{\mu_2 \nu_2} g_t^{\mu_1 \mu_4} + g_t^{\mu_3 \nu_1} g_t^{\mu_1 \nu_2} g_t^{\mu_2 \mu_4} \Big ) \times \gamma^t_{\mu_4} \gamma_5 h_v^c(x) \, ,
\\ && J^{\alpha_1\alpha_2\alpha_3}_{7/2,+,\mathbf{\bar 3}_F,3,1,\rho\lambda}(x)
\label{eq:current18}
\\ \nonumber &=& \epsilon_{abc} \Big ( [\mathcal{D}^t_{\mu_1} \mathcal{D}^t_{\mu_2} q^{aT}(x)] \mathbb{C} \gamma^t_{\mu_3} q^b(x) - q^{aT}(x) \mathbb{C} \gamma^t_{\mu_3} [\mathcal{D}^t_{\mu_1} \mathcal{D}^t_{\mu_2} q^b(x)] \Big )
\\ \nonumber && ~~~~~~~~~~~~
\times \Gamma_t^{\alpha_1\alpha_2\alpha_3,\mu_1\mu_2\mu_3} \times h_v^c(x) \, .
\end{eqnarray}

\end{enumerate}

\end{itemize}
We note that all these interpolating fields have been projected to $j={1\over2}/{3\over2}/{5\over2}/{7\over2}$.
Identical sum rules can be obtained using either $J^{\alpha_1\cdots\alpha_{|j_l-1/2|}}_{|j_l-1/2|,P,F,j_l,s_l,\rho\rho/\lambda\lambda/\rho\lambda}$ or $J^{\alpha_1\cdots\alpha_{j_l+1/2}}_{j_l+1/2,P,F,j_l,s_l,\rho\rho/\lambda\lambda/\rho\lambda}$ in the same doublet,
both at the leading order and at the $O(1/m_Q)$ order~\cite{Dai:1993kt,Dai:1996yw,Dai:1996qx,Dai:2003yg}.
Hence, we only need to use one of them to perform QCD sum rule analyses.

There are altogether five baryon multiplets of $SU(3)$ flavor $\mathbf{\bar3}_F$, i.e., $[\mathbf{\bar 3}_F,2,0,\rho\rho]$, $[\mathbf{\bar 3}_F,2,0,\lambda\lambda]$, $[\mathbf{\bar 3}_F,1,1,\rho\lambda]$, $[\mathbf{\bar 3}_F,2,1,\rho\lambda]$ and $[\mathbf{\bar 3}_F,3,1,\rho\lambda]$.
In the next section we shall use $J^{\alpha}_{3/2,+,\mathbf{\bar 3}_F,2,0,\rho\rho}$, $J^{\alpha}_{3/2,+,\mathbf{\bar 3}_F,2,0,\lambda\lambda}$,
$J_{1/2,+,\mathbf{\bar 3}_F,1,1,\rho\lambda}$ and $J^{\alpha_1\alpha_2\alpha_3}_{7/2,+,\mathbf{\bar 3}_F,3,1,\rho\lambda}$ to perform QCD sum rule analyses.
We shall further replace $\mathbf{6}_F$ by $\Sigma_c$, $\Xi_c^\prime$, and $\Omega_c$, and $\mathbf{\bar3}_F$ by $\Lambda_c$ and $\Xi_c$ to explicitly denote the quark contents inside, such as $J^{\alpha}_{3/2,+,\Lambda_c,2,0,\lambda\lambda}$ and $J^{\alpha}_{3/2,+,\Xi_c,2,0,\lambda\lambda}$ belonging to $[\Lambda_c,2,0,\lambda\lambda]$ and $[\Xi_c,2,0,\lambda\lambda]$, respectively:
\begin{eqnarray}
&& J^{\alpha}_{3/2,+,\Lambda_c,2,0,\lambda\lambda}(x)
\label{eq:current19}
\\ \nonumber &=& \epsilon_{abc} \Big ( [\mathcal{D}^t_{\mu_1} \mathcal{D}^t_{\mu_2} u^{aT}(x)] \mathbb{C} \gamma_5 d^b(x) + 2 [\mathcal{D}^t_{\mu_1} u^{aT}(x)] \mathbb{C} \gamma_5 [\mathcal{D}^t_{\mu_2} d^b(x)] + u^{aT}(x) \mathbb{C} \gamma_5 [\mathcal{D}^t_{\mu_1} \mathcal{D}^t_{\mu_2} d^b(x)] \Big )
\\ \nonumber && ~~~~~~~~~~~~
\times \Big ( {1\over2} g_t^{\mu_1 \alpha} g_t^{\mu_2 \mu_4} + {1\over2} g_t^{\mu_2 \alpha} g_t^{\mu_1 \mu_4} - {1\over3} g_t^{\mu_1 \mu_2} g_t^{\mu_4 \alpha} \Big ) \times \gamma^t_{\mu_4} \gamma_5 h_v^c(x) \, ,
\\ && J^{\alpha}_{3/2,+,\Xi_c,2,0,\lambda\lambda}(x)
\label{eq:current20}
\\ \nonumber &=& \epsilon_{abc} \Big ( [\mathcal{D}^t_{\mu_1} \mathcal{D}^t_{\mu_2} u^{aT}(x)] \mathbb{C} \gamma_5 s^b(x) + 2 [\mathcal{D}^t_{\mu_1} u^{aT}(x)] \mathbb{C} \gamma_5 [\mathcal{D}^t_{\mu_2} s^b(x)] + u^{aT}(x) \mathbb{C} \gamma_5 [\mathcal{D}^t_{\mu_1} \mathcal{D}^t_{\mu_2} s^b(x)] \Big )
\\ \nonumber && ~~~~~~~~~~~~
\times \Big ( {1\over2} g_t^{\mu_1 \alpha} g_t^{\mu_2 \mu_4} + {1\over2} g_t^{\mu_2 \alpha} g_t^{\mu_1 \mu_4} - {1\over3} g_t^{\mu_1 \mu_2} g_t^{\mu_4 \alpha} \Big ) \times \gamma^t_{\mu_4} \gamma_5 h_v^c(x) \, .
\end{eqnarray}

\section{Sum Rules at the Leading Order}
\label{sec:leading}

In the previous section we have partly classified the $D$-wave charmed baryon interpolating fields, and in this and next sections we use them to further perform QCD sum rule analyses. When classifying these fields, we have taken into account their inner structures by fixing their inner quantum numbers $j_l$, $s_l$, $l_\rho$, and $l_\lambda$.
Although the physical state is probably a mixed state containing components with various inner quantum numbers, at the beginning we can always assume the state $|j,P,F,j_l,s_l,\rho\rho/\lambda\lambda/\rho\lambda\rangle$ exists, which has the quantum numbers $j$, $P$, $F$ and the inner quantum numbers $j_l$, $s_l$, and $[\rho\rho/\lambda\lambda/\rho\lambda]$ in the $m_Q \rightarrow \infty$ limit. It belongs to the spin doublet of the spin $j = j_l \otimes s_Q = j_l \pm 1/2$ with $[F,j_l,s_l,\rho\rho/\lambda\lambda/\rho\lambda]$, and coupled by the interpolating field $J^{\alpha_1\cdots\alpha_{j-1/2}}_{j,P,F,j_l,s_l,\rho\rho/\lambda\lambda/\rho\lambda}$ through
\begin{eqnarray}
\langle 0| J^{\alpha_1\cdots\alpha_{j-1/2}}_{j,P,F,j_l,s_l,\rho\rho/\lambda\lambda/\rho\lambda} |j,P,F,j_l,s_l,\rho\rho/\lambda\lambda/\rho\lambda \rangle
= f_{F,j_l,s_l,\rho\rho/\lambda\lambda/\rho\lambda} u^{\alpha_1\cdots\alpha_{j-1/2}} \, ,
\end{eqnarray}
where $f_{F,j_l,s_l,\rho\rho/\lambda\lambda/\rho\lambda}$ is the decay constant, and $u^{\alpha_1\cdots\alpha_j}$ is the relevant spinor. For examples, $u(x)$ and $u^\alpha(x)$ are the Dirac and Rarita-Schwinger spinors, respectively.
Then the two-point correlation function can be written as
\begin{eqnarray}
\Pi^{\alpha_1\cdots\alpha_{j-1/2},\beta_1\cdots\beta_{j-1/2}}_{F,j_l,s_l,\rho\rho/\lambda\lambda/\rho\lambda} (\omega)
&=& i \int d^4 x e^{i k x} \langle 0 |
T[J^{\alpha_1\cdots\alpha_{j-1/2}}_{j,P,F,j_l,s_l,\rho\rho/\lambda\lambda/\rho\lambda}(x)
\bar J^{\beta_1\cdots\beta_{j-1/2}}_{j,P,F,j_l,s_l,\rho\rho/\lambda\lambda/\rho\lambda}(0)] | 0 \rangle
\\ \nonumber &=& \mathbb{S} [ g_t^{\alpha_1 \beta_1} \cdots g_t^{\alpha_{j-1/2} \beta_{j-1/2}} ] {1 + v\!\!\!\slash \over 2} \Pi_{F,j_l,s_l,\rho\rho/\lambda\lambda/\rho\lambda} (\omega) + \cdots \, ,
\label{eq:pi}
\end{eqnarray}
where $\omega$ is twice the external off-shell energy, $\omega = 2 v \cdot k$,
and $\mathbb{S} [\cdots]$ is used to denote symmetrization and subtracting the trace
terms in the sets $(\alpha_1 \cdots \alpha_{j-1/2})$ and $(\beta_1 \cdots
\beta_{j-1/2})$.
The leading term $\Pi_{F,j_l,s_l,\rho\rho/\lambda\lambda/\rho\lambda} (\omega)$ has been totally symmetrized and only contains the highest spin $j$ component, while $\cdots$ contains other spin components.
We note that we have omitted the quantum numbers $j$ and $P$ simply because the two currents in the same doublet give identical sum rules at the leading order in the heavy quark limit.

At the hadron level the correlation function (\ref{eq:pi}) can be simply written as
\begin{eqnarray}
\Pi_{F,j_l,s_l,\rho\rho/\lambda\lambda/\rho\lambda}(\omega) = {2 f_{F,j_l,s_l,\rho\rho/\lambda\lambda/\rho\lambda}^{2} \over 2 \overline{\Lambda}_{F,j_l,s_l,\rho\rho/\lambda\lambda/\rho\lambda}
- \omega} + \mbox{higher states} \, , \label{eq:pole}
\end{eqnarray}
where $\overline{\Lambda}_{F,j_l,s_l,\rho\rho/\lambda\lambda/\rho\lambda}$ is the difference between the mass of the lowest-lying heavy baryon state and the heavy quark mass:
\begin{eqnarray}
\overline{\Lambda}_{F,j_l,s_l,\rho\rho/\lambda\lambda/\rho\lambda} \equiv \lim_{m_Q \rightarrow \infty} (m_{j,P,F,j_l,s_l,\rho\rho/\lambda\lambda/\rho\lambda} - m_Q) \, .
\end{eqnarray}

At the quark and gluon level the correlation function (\ref{eq:pi}) can be evaluated using the method of operator production expansion (OPE)~\cite{Dai:1993kt,Dai:1996yw,Dai:1996qx,Dai:2003yg}. Using $J^{\alpha}_{3/2,+,\Lambda_c,2,0,\lambda\lambda}$ and $J^{\alpha}_{3/2,+,\Xi_c,2,0,\lambda\lambda}$ as examples, we insert Eqs.~(\ref{eq:current19}) and (\ref{eq:current20}) into Eq.~(\ref{eq:pi}), perform the Borel transformation, and then obtain
\begin{eqnarray}
\Pi_{\Lambda_c,2,0,\lambda\lambda}(\omega_c, T) &=& f_{\Lambda_c,2,0,\lambda\lambda}^2 e^{-2 \bar \Lambda_{\Lambda_c,2,0,\lambda\lambda} / T}
\label{eq:ope1}
= \int_{0}^{\omega_c} [\frac{5}{145152\pi^4}\omega^9 - \frac{\langle g_s^2 GG \rangle}{1728 \pi^4}\omega^5 ]e^{-\omega/T}d\omega \, ,
\\ \Pi_{\Xi_c,2,0,\lambda\lambda}(\omega_c, T) &=& f_{\Xi_c,2,0,\lambda\lambda}^2 e^{-2 \bar \Lambda_{\Xi_c,2,0,\lambda\lambda} / T}
\label{eq:ope2}
\\ \nonumber &=&
\int_{2m_s}^{\omega_c} [ \frac{5}{145152\pi^4}\omega^9 - \frac{m_s^2}{672\pi^4}\omega^7 - \frac{m_s\langle \bar q q \rangle}{72 \pi^2}\omega^5 + \frac{m_s\langle \bar s s \rangle}{48 \pi^2}\omega^5
\\ \nonumber && ~~~~ - \frac{\langle g_s^2 GG \rangle}{1728 \pi^4}\omega^5 + \frac{5m_s^2\langle g_s^2 GG \rangle}{576 \pi^4}\omega^3  - \frac{5m_s\langle g_s^2 GG \rangle\langle \bar s s \rangle}{216 \pi^2}\omega ]e^{-\omega/T}d\omega \, .
\end{eqnarray}
Sum rules for other currents are shown in Appendix~\ref{sec:others}.
We note that in our calculations we have used the software {\it Mathematica} with a package called $FeynCalc$~\cite{Mertig:1990an}.
The condensates and other parameters contained in these sum rules take the following
values~\cite{Dai:1993kt,Dai:1996yw,Dai:1996qx,Dai:2003yg,pdg,Yang:1993bp,Hwang:1994vp,Narison:2002pw,Gimenez:2005nt,Jamin:2002ev,Ioffe:2002be,Ovchinnikov:1988gk,colangelo}:
%
%%%%%%%%%%%%%%%%%%%%%%%%%%%%%%%%%%%%%%%%%%%%%%%%%%%%%%%%%%%%%%%%%%%%%%%%%%%%%%
\begin{eqnarray}
\nonumber && \langle \bar qq \rangle = \langle \bar uu \rangle = \langle \bar dd \rangle = - (0.24 \mbox{ GeV})^3 \, ,
\\ \nonumber && \langle \bar ss \rangle = (0.8\pm 0.1)\times \langle\bar qq \rangle \, ,
\\ \nonumber && \langle {\alpha_s\over\pi} GG\rangle = 0.005 \pm 0.004 \mbox{ GeV}^4\, ,
\\ \label{condensates} && m_s = 0.125 \mbox{ GeV} \, ,
\\
\nonumber && \langle g_s \bar q \sigma G q \rangle = M_0^2 \times \langle \bar qq \rangle\, ,
\\
\nonumber && \langle g_s \bar s \sigma G s \rangle = M_0^2 \times \langle \bar ss \rangle\, ,
\\
\nonumber && M_0^2= 0.8 \mbox{ GeV}^2\, .
\end{eqnarray}
%%%%%%%%%%%%%%%%%%%%%%%%%%%%%%%%%%%%%%%%%%%%%%%%%%%%%%%%%%%%%%%%%%%%%%%%%%%%%%
Finally, we differentiate Log[Eq.~(\ref{eq:ope1})] and Log[Eq.~(\ref{eq:ope2})] with respect to $[-2/T]$ to obtain $\overline{\Lambda}_{F,j_l,s_l,\rho\rho/\lambda\lambda/\rho\lambda}$:
%%%%%%%%%%%%%%%%%%%%%%%%%%%%%%%%%%%%%%%%%%%%%%%%%%%%%%%%%%%%%%%%%%%%%%%%%%%%%%
\begin{equation}
\overline{\Lambda}_{F,j_l,s_l,\rho\rho/\lambda\lambda/\rho\lambda}(\omega_c, T) = \frac{\frac{\partial}{\partial(-2/T)}\Pi_{F,j_l,s_l,\rho\rho/\lambda\lambda/\rho\lambda}(\omega_c, T)}{\Pi_{F,j_l,s_l,\rho\rho/\lambda\lambda/\rho\lambda}(\omega_c, T)} \, ,
\label{eq:mass}
\end{equation}
%%%%%%%%%%%%%%%%%%%%%%%%%%%%%%%%%%%%%%%%%%%%%%%%%%%%%%%%%%%%%%%%%%%%%%%%%%%%%%
which can be further used to obtain $f_{F,j_l,s_l,\rho\rho/\lambda\lambda/\rho\lambda}$:
\begin{equation}
f_{F,j_l,s_l,\rho\rho/\lambda\lambda/\rho\lambda}(\omega_c, T) = \sqrt{\Pi_{F,j_l,s_l,\rho\rho/\lambda\lambda/\rho\lambda}(\omega_c, T) \times e^{ 2 \overline{\Lambda}_{F,j_l,s_l,\rho\rho/\lambda\lambda/\rho\lambda}(\omega_c, T) / T}/2} \, .
\label{eq:coupling}
\end{equation}

There are two free parameters in Eq.~(\ref{eq:mass}), the Borel mass $T$ and the threshold value $\omega_c$. We have three criteria to constrain them. The first criterion is to require the high-order corrections to be less than 10\%:
%
%%%%%%%%%%%%%%%%%%%%%%%%%%%%%%%%%%%%%%%%%%%%%%%%%%%%%%%%%%%%%%%%%%%%%%%%%%%%%%
\begin{equation}
\label{eq_convergence}
\mbox{Convergence (CVG)} \equiv |\frac{ \Pi^{\rm high-order}_{F,j_l,s_l,\rho\rho/\lambda\lambda/\rho\lambda}(\infty, T) }{ \Pi_{F,j_l,s_l,\rho\rho/\lambda\lambda/\rho\lambda}(\infty, T) }| \leq 10\% \, ,
\end{equation}
%%%%%%%%%%%%%%%%%%%%%%%%%%%%%%%%%%%%%%%%%%%%%%%%%%%%%%%%%%%%%%%%%%%%%%%%%%%%%%
%
where $\Pi^{\rm high\mbox{-}order}_{F,j_l,s_l,\rho\rho/\lambda\lambda/\rho\lambda}(\omega_c, T)$ is used to denote the high-order corrections, for example,
%
%%%%%%%%%%%%%%%%%%%%%%%%%%%%%%%%%%%%%%%%%%%%%%%%%%%%%%%%%%%%%%%%%%%%%%%%%%%%%%
\begin{equation}
\Pi^{\rm high\mbox{-}order}_{\Xi_c,2,0,\lambda\lambda}(\omega_c, T) =
\int_{2m_s}^{\omega_c} [ - \frac{m_s\langle \bar q q \rangle}{72 \pi^2}\omega^5 + \frac{m_s\langle \bar s s \rangle}{48 \pi^2}\omega^5
- \frac{\langle g_s^2 GG \rangle}{1728 \pi^4}\omega^5 + \frac{5m_s^2\langle g_s^2 GG \rangle}{576 \pi^4}\omega^3  - \frac{5m_s\langle g_s^2 GG \rangle\langle \bar s s \rangle}{216 \pi^2}\omega ]e^{-\omega/T}d\omega \, .
\end{equation}
%%%%%%%%%%%%%%%%%%%%%%%%%%%%%%%%%%%%%%%%%%%%%%%%%%%%%%%%%%%%%%%%%%%%%%%%%%%%%%
%
The second criterion is to require the pole contribution (PC) to be larger than 10\%:
%
%%%%%%%%%%%%%%%%%%%%%%%%%%%%%%%%%%%%%%%%%%%%%%%%%%%%%%%%%%%%%%%%%%%%%%%%%%%%%%
\begin{equation}
\label{eq_pole}
\mbox{PC} \equiv \frac{ \Pi_{F,j_l,s_l,\rho\rho/\lambda\lambda/\rho\lambda}(\omega_c, T) }{ \Pi_{F,j_l,s_l,\rho\rho/\lambda\lambda/\rho\lambda}( \infty , T) } \geq 10\% \, .
\end{equation}
%%%%%%%%%%%%%%%%%%%%%%%%%%%%%%%%%%%%%%%%%%%%%%%%%%%%%%%%%%%%%%%%%%%%%%%%%%%%%%
%
Altogether we obtain an interval $T_{min}<T<T_{max}$ for a fixed threshold value $\omega_c$.

The small pole contribution used in Eq.~(\ref{eq_pole}) is mathematically due to the large powers of $s$ in the spectral function, which makes the suppression of the Borel transformation on the continuum not so effective. For example, see Ref.~\cite{Chen:2014vha} where the pole contribution of the $d^*(2380)$ is only about 0.0002 due to the large power of $s$ in its spectral function. However, actually we do not need a pole which is significant in the whole energy space, but just need it to be dominant inside our working region. Such a pole can be found as if the mass prediction does not depend on the other free parameter, the threshold value $\omega_c$. Hence, the third criterion is to require the dependence of $m_{j,P,F,j_l,s_l,\rho\rho/\lambda\lambda/\rho\lambda}$ (mass of the heavy baryon state) on the threshold value $\omega_c$ to be weak, which will be discussed in detail in Sec.~\ref{sec:numerical}. At the same time we shall also check the dependence of $m_{j,P,F,j_l,s_l,\rho\rho/\lambda\lambda/\rho\lambda}$ on the Borel mass $T$.

\begin{figure}[hbt]
\begin{center}
\scalebox{0.6}{\includegraphics{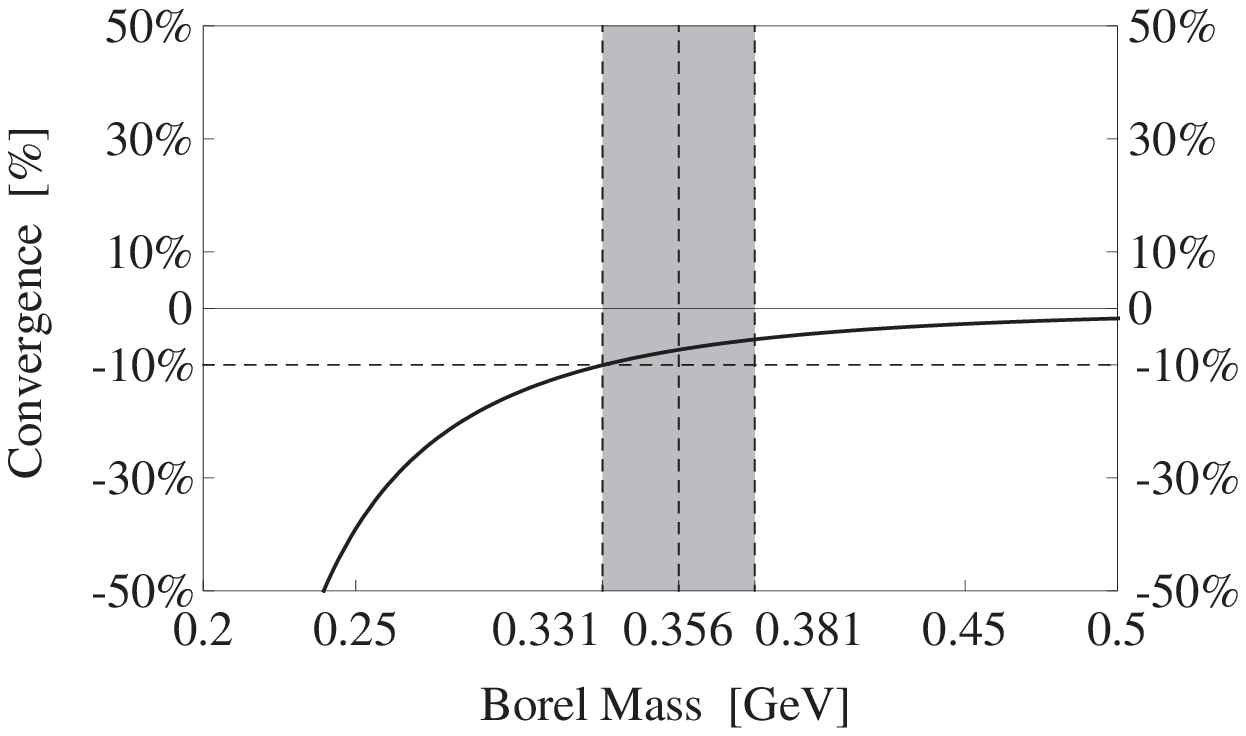}}
\scalebox{0.573}{\includegraphics{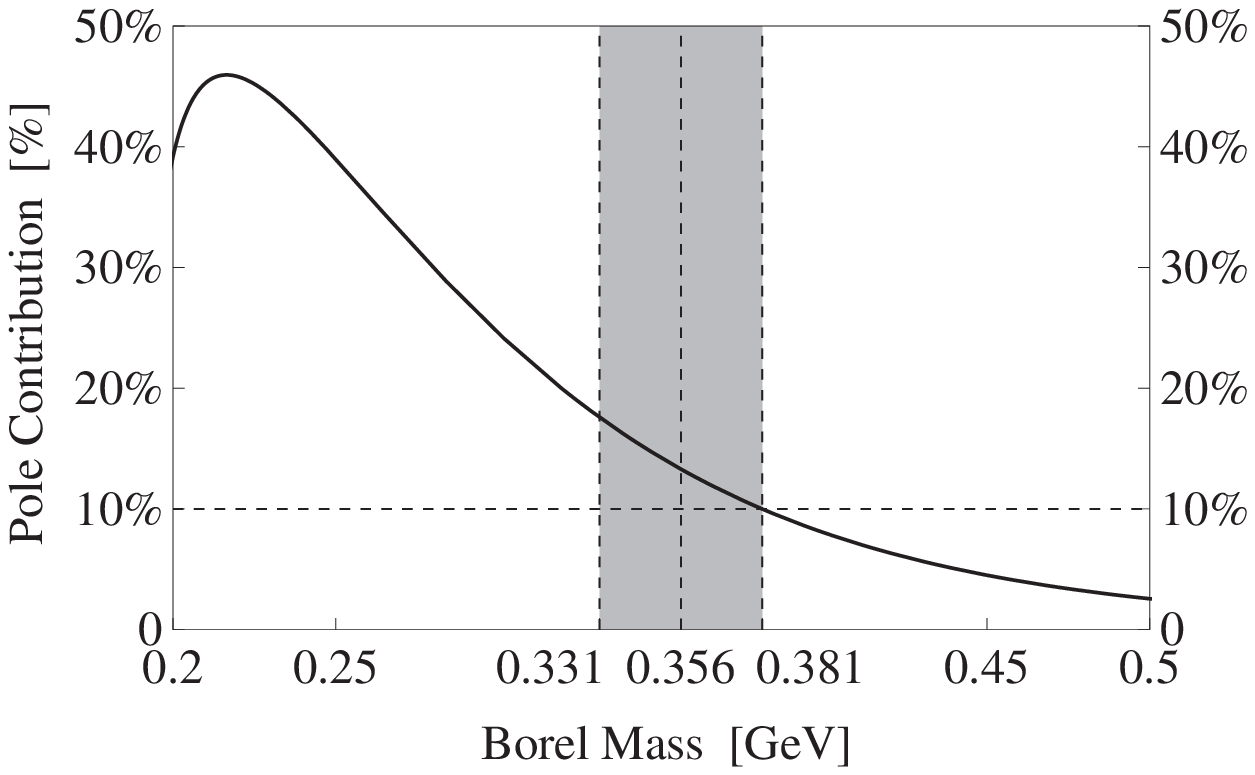}}
\caption{In the left panel we show the variation of CVG, defined in Eq.~(\ref{eq_convergence}), as a function of the Borel mass $T$.
In the right panel we show the variation of PC, defined in Eq.~(\ref{eq_pole}), as a function of the Borel mass $T$, where the threshold value is chosen to be $\omega_c$ = 2.5 GeV.
The current $J^{\alpha}_{3/2,+,\Lambda_c,2,0,\lambda\lambda}$ is used here.}
\label{fig:pole}
\end{center}
\end{figure}

\begin{figure}[hbt]
\begin{center}
\scalebox{0.585}{\includegraphics{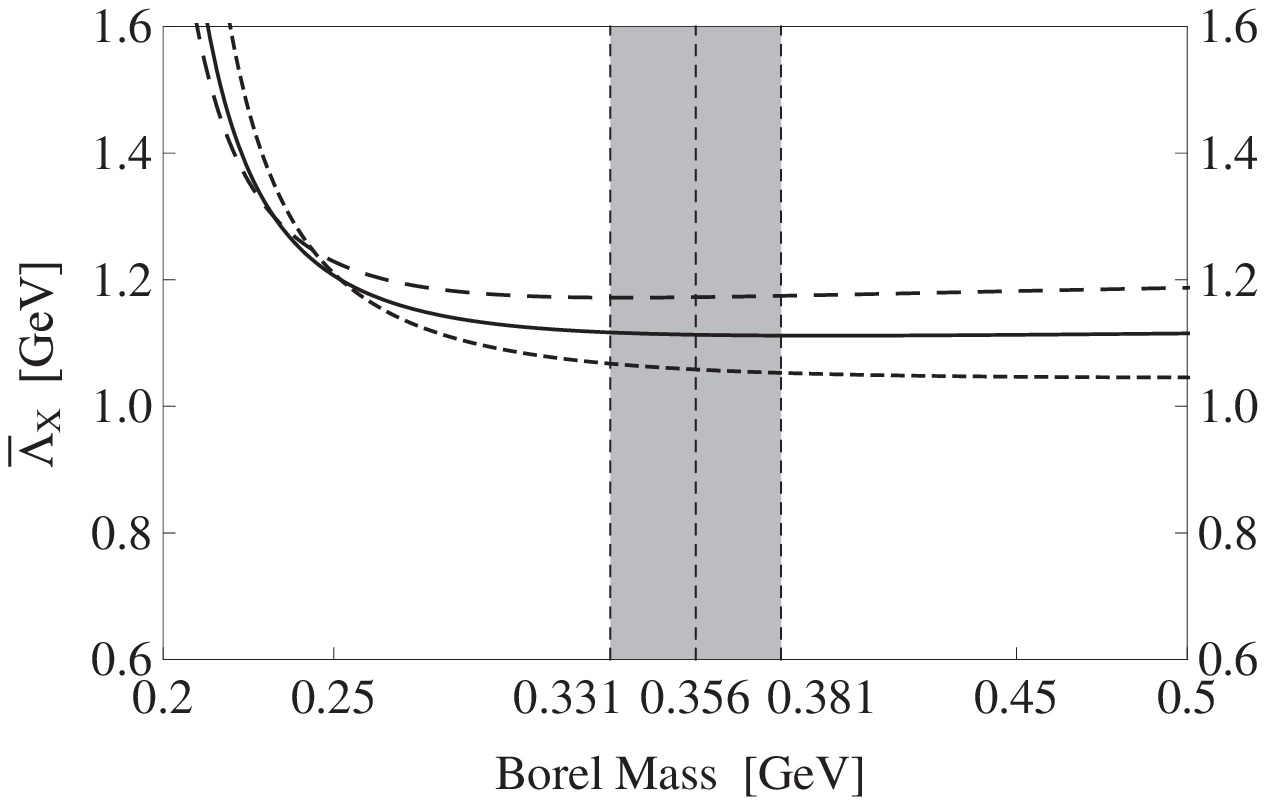}}
\scalebox{0.61}{\includegraphics{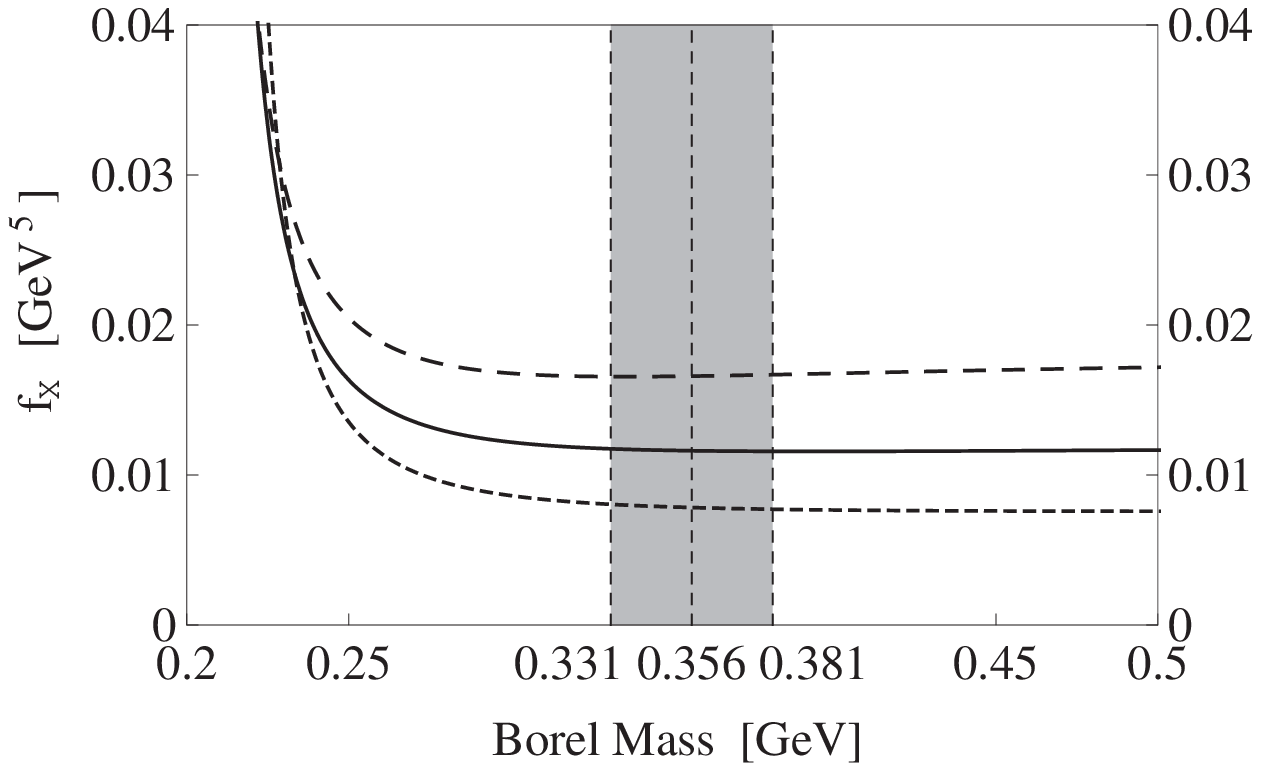}}
\caption{The variations of $\overline{\Lambda}_{\Lambda_c,2,0,\lambda\lambda}$ (left) and $f_{\Lambda_c,2,0,\lambda\lambda}$ (right) with respect to the
Borel mass $T$, when $J^{\alpha}_{3/2,+,\Lambda_c,2,0,\lambda\lambda}$ is used.
The short-dashed, solid, and long-dashed curves are obtained by fixing $\omega_c = 2.3$, 2.5, and 2.7 GeV, respectively.}
\label{fig:leading1}
\end{center}
\end{figure}

Still using the current $J^{\alpha}_{3/2,+,\Lambda_c,2,0,\lambda\lambda}$ as an example, we show the variations of CVG and PC, as defined in Eqs.~(\ref{eq_convergence}) and (\ref{eq_pole}), with respect to the Borel mass $T$ in Fig.~\ref{fig:pole}, and the variations of $\overline{\Lambda}_{\Lambda_c,2,0,\lambda\lambda}$ and $f_{\Lambda_c,2,0,\lambda\lambda}$ with respect to $T$ in Fig.~\ref{fig:leading1}, where $\omega_c$ is chosen to be 2.5 GeV. Now the Borel window is $0.331$ GeV $< T < 0.381$ GeV, and we obtain the following numerical results:
\begin{eqnarray}
\overline{\Lambda}_{\Lambda_c,2,0,\lambda\lambda} &=& 1.113 \mbox{ GeV} \, ,
\\ \nonumber f_{\Lambda_c,2,0,\lambda\lambda} &=& 0.012 \mbox{ GeV}^{5} \, ,
\end{eqnarray}
where the central values are obtained by choosing $T=0.356$ GeV and $\omega_c = 2.5$ GeV.

\begin{figure}[hbt]
\begin{center}
\scalebox{0.585}{\includegraphics{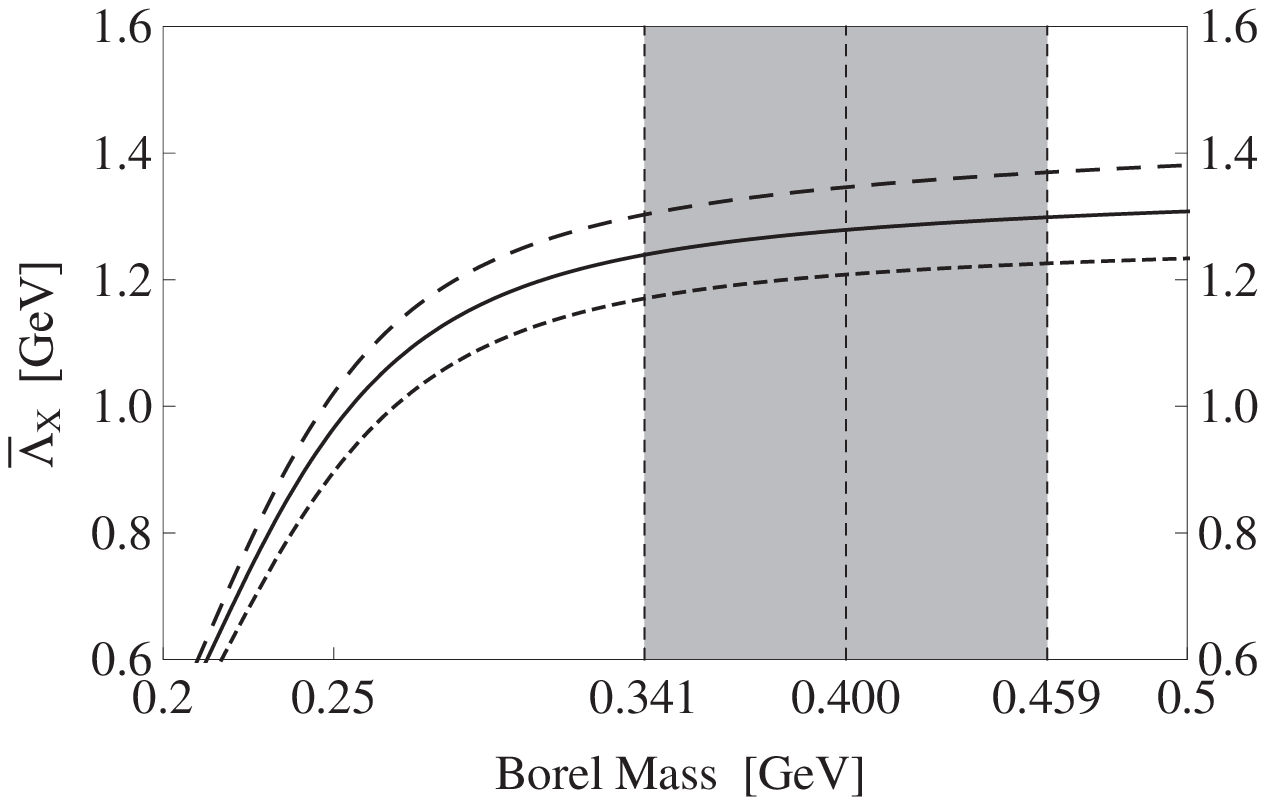}}
\scalebox{0.61}{\includegraphics{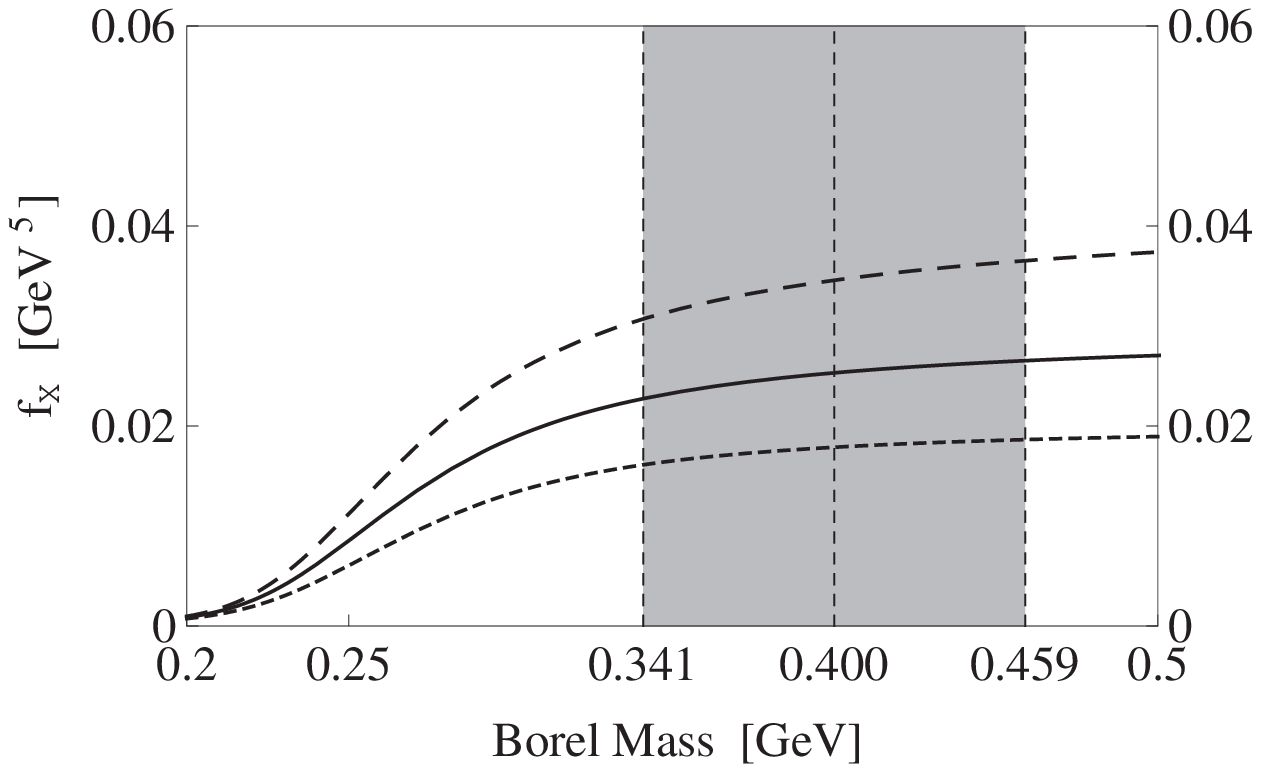}}
\caption{The variations of $\overline{\Lambda}_{\Xi_c,2,0,\lambda\lambda}$ (left) and $f_{\Xi_c,2,0,\lambda\lambda}$ (right) with respect to the
Borel mass $T$, when $J^{\alpha}_{3/2,+,\Xi_c,2,0,\lambda\lambda}$ is used.
The short-dashed, solid, and long-dashed curves are obtained by fixing $\omega_c = 2.8$, 3.0, and 3.2 GeV, respectively.}
\label{fig:leading2}
\end{center}
\end{figure}

We also show the variations of $\overline{\Lambda}_{\Xi_c,2,0,\lambda\lambda}$ and $f_{\Xi_c,2,0,\lambda\lambda}$ with respect to $T$ in Fig.~\ref{fig:leading2}, where $\omega_c$ is chosen to be 3.0 GeV. From these figures, we find the Borel window $0.341$ GeV $< T < 0.459$ GeV, and obtain the following numerical results:
\begin{eqnarray}
\overline{\Lambda}_{\Xi_c,2,0,\lambda\lambda} &=& 1.279 \mbox{ GeV} \, ,
\\ \nonumber f_{\Xi_c,2,0,\lambda\lambda} &=& 0.025 \mbox{ GeV}^{5} \, ,
\end{eqnarray}
where the central values are obtained by choosing $T=0.400$ GeV and $\omega_c = 3.0$ GeV.

\section{Sum Rules at the Order ${\mathcal O}(1/m_Q)$}
\label{sec:nexttoleading}

In this section we work up to the order ${\mathcal O}(1/m_Q)$ based on the HQET Lagrangian~\cite{Dai:1996qx,Dai:2003yg}:
\begin{eqnarray}
\mathcal{L}_{\rm eff} = \overline{h}_{v}iv\cdot Dh_{v} + \frac{1}{2m_{Q}}\mathcal{K} + \frac{1}{2m_{Q}}\mathcal{S} \, ,
\label{eq:next}
\end{eqnarray}
where $\mathcal{K}$ is the operator of nonrelativistic kinetic energy, and $\mathcal S$ is the Pauli term describing the chromomagnetic interaction:
\begin{eqnarray}
\mathcal{K} &=& \overline{h}_{v}(iD_{t})^{2}h_{v} \, ,
\\ \nonumber
\mathcal{S} &=& \frac{g}{2} C_{mag} (m_{Q}/\mu) \overline{h}_{v} \sigma_{\mu\nu} G^{\mu\nu} h_{v} \, .
\end{eqnarray}
Here $C_{mag} (m_{Q}/\mu) = [ \alpha_s(m_Q) / \alpha_s(\mu) ]^{3/\beta_0}$ with $\beta_0 = 11 - 2 n_f /3$.

The correlation function at the hadron level, Eq.~(\ref{eq:pole}), can be written up to the order ${\mathcal O}(1/m_Q)$ as
\begin{eqnarray}
\Pi(\omega)_{pole} &=& \frac{2(f+\delta f)^{2}}{2(\overline{\Lambda}+\delta m)-\omega}
\label{eq:correction}
\\ \nonumber &=& \frac{2f^{2}}{2\overline{\Lambda}-\omega}-\frac{4\delta mf^{2}}{(2\overline{\Lambda}-\omega)^{2}}+\frac{4f\delta f}{2\overline{\Lambda}-\omega} \, ,
\end{eqnarray}
where $\delta m_{j, P, F,j_l,s_l,\rho\rho/\lambda\lambda/\rho\lambda}$ is the correction to the mass $m_{j, P, F,j_l,s_l,\rho\rho/\lambda\lambda/\rho\lambda}$, and can be evaluated using the three-point correlation functions:
\begin{eqnarray}
\nonumber \delta_{O}\Pi_{j,P,F,j_l,s_l,\rho\rho/\lambda\lambda/\rho\lambda}^{\alpha_{1}\cdots\alpha_{j-1/2},\beta_{1}\cdots\beta_{j-1/2}}(\omega , \omega ')
&=& i^{2}\int d^{4}xd^{4}ye^{ik\cdot x-ik'\cdot y}\times\langle0|T[J_{j,P,F,j_l,s_l,\rho\rho/\lambda\lambda/\rho\lambda}^{\alpha_{1}\cdots \alpha_{j-1/2}}(x)O(0) \bar J_{j,P,F,j_l,s_l,\rho\rho/\lambda\lambda/\rho\lambda}^{\beta_{1}\cdots \beta_{j-1/2}}(y)]|0\rangle
\label{eq:nextpi}
\\ &=& \mathbb{S} [ g_t^{\alpha_1 \beta_1} \cdots g_t^{\alpha_{j-1/2} \beta_{j-1/2}} ] \delta_{O} \Pi_{j,P,F,j_l,s_l,\rho\rho/\lambda\lambda/\rho\lambda} (\omega) \, ,
\end{eqnarray}
where $O = \mathcal{K}$ or $\mathcal{S}$.
Based on the Lagrangian (\ref{eq:next}), these correlation functions can be written at the hadron level as
\begin{eqnarray}
\delta_{\mathcal{K}}\Pi(\omega,\omega')_{j,P,F,j_l,s_l,\rho\rho/\lambda\lambda/\rho\lambda} &=& \frac{2f^{2}K_{F,j_l,s_l,\rho\rho/\lambda\lambda/\rho\lambda}}{(2\overline{\Lambda}-\omega)(2\overline{\Lambda}-\omega')}
+\frac{2f^{2}G_{\mathcal{K}}(\omega')}{2\overline{\Lambda}-\omega}
+\frac{2f^{2}G_{\mathcal{K}}(\omega)}{2\overline{\Lambda}-\omega'} \, ,
\label{eq:K}
\\ \delta_{\mathcal{S}}\Pi(\omega,\omega')_{j,P,F,j_l,s_l,\rho\rho/\lambda\lambda/\rho\lambda} &=& \frac{2d_{M}f^{2}\Sigma_{F,j_l,s_l,\rho\rho/\lambda\lambda/\rho\lambda}}{(2\overline{\Lambda}-\omega)(2\overline{\Lambda}-\omega')}
+\frac{2d_{M}f^{2}G_{\mathcal{S}}(\omega')}{2\overline{\Lambda}-\omega} \,
+\frac{2d_{M}f^{2}G_{\mathcal{S}}(\omega)}{2\overline{\Lambda}-\omega'} \, ,
\label{eq:S}
\end{eqnarray}
where the following definitions have been used:
\begin{eqnarray}
\nonumber K_{F,j_l,s_l,\rho\rho/\lambda\lambda/\rho\lambda} &\equiv& \langle j,P,F,j_l,s_l,\rho\rho/\lambda\lambda/\rho\lambda|\overline{h}_{v}(iD_{\bot})^{2}h_{v}|j,P,F,j_l,s_l,\rho\rho/\lambda\lambda/\rho\lambda\rangle \, ,
\\ \nonumber d_{M}\Sigma_{F,j_l,s_l,\rho\rho/\lambda\lambda/\rho\lambda} &\equiv& \langle j,P,F,j_l,s_l,\rho\rho/\lambda\lambda/\rho\lambda| {g\over2} \overline{h}_{v}\sigma_{\mu\nu}G^{\mu\nu}h_{v}|j,P,F,j_l,s_l,\rho\rho/\lambda\lambda/\rho\lambda\rangle \, ,
\\ d_{M} &\equiv& d_{j,j_{l}} \, ,
\\ \nonumber d_{j_{l}-1/2,j_{l}} &=& 2j_{l}+2\, ,
\\ \nonumber d_{j_{l}+1/2,j_{l}} &=& -2j_{l} \, .
\end{eqnarray}
Then we fix $\omega = \omega^\prime$ and use Eqs.~(\ref{eq:correction}), (\ref{eq:K}), and (\ref{eq:S}) to obtain
\begin{eqnarray}
\delta m_{j, P, F,j_l,s_l,\rho\rho/\lambda\lambda/\rho\lambda} = -\frac{1}{4m_{Q}}(K_{F,j_l,s_l,\rho\rho/\lambda\lambda/\rho\lambda} + d_{M}C_{mag}\Sigma_{F,j_l,s_l,\rho\rho/\lambda\lambda/\rho\lambda} ) \, .
\end{eqnarray}
From this equation we find that only the term $\mathcal S$ ($\Sigma_{F,j_l,s_l,\rho\rho/\lambda\lambda/\rho\lambda}$) can cause a mass splitting within the same doublet.

The three-point correlation functions defined in Eq.~(\ref{eq:nextpi}) can also be evaluated at the quark and gluon level using the method of operator product expansion~\cite{Dai:1996qx,Dai:2003yg}.
Still using the currents $J^{\alpha}_{3/2,+,\Lambda_c,2,0,\lambda\lambda}$ and $J^{\alpha}_{3/2,+,\Xi_c,2,0,\lambda\lambda}$ as examples, we insert Eqs.~(\ref{eq:current19}) and (\ref{eq:current20}) into Eqs.~(\ref{eq:nextpi}),
make a double Borel transformation for both $\omega$ and $\omega^\prime$, take the two Borel parameters to be
equal, and then obtain:
\begin{eqnarray}
&& f_{\Lambda_c,2,0,\lambda\lambda}^2 K_{\Lambda_c,2,0,\lambda\lambda} e^{-2 \bar \Lambda_{\Lambda_c,2,0,\lambda\lambda} / T}
\label{eq:Kc1}
= \int_{0}^{\omega_c} [ - \frac{127}{10644480\pi^4}\omega^{11} - \frac{\langle g_s^2 GG \rangle}{17280 \pi^4}\omega^7 ]e^{-\omega/T}d\omega \, ,
\\ && f_{\Lambda_c,2,0,\lambda\lambda}^2 \Sigma_{\Lambda_c,2,0,\lambda\lambda} e^{-2 \bar \Lambda_{\Lambda_c,2,0,\lambda\lambda} / T}
\label{eq:Sc1}
= \int_{0}^{\omega_c} [ \frac{\langle g_s^2 GG \rangle}{24192 \pi^4}\omega^7 ]e^{-\omega/T}d\omega \, ,
\\ && f_{\Xi_c,2,0,\lambda\lambda}^2 K_{\Xi_c,2,0,\lambda\lambda} e^{-2 \bar \Lambda_{\Xi_c,2,0,\lambda\lambda} / T}
\label{eq:Kc2}
\\ \nonumber \quad &&
= \int_{2m_s}^{\omega_c} [ - \frac{127}{10644480\pi^4}\omega^{11} + \frac{307m_s^2}{483840\pi^4}\omega^9 + \frac{37m_s\langle \bar q q \rangle}{5040 \pi^2}\omega^7 - \frac{233m_s\langle \bar s s \rangle}{20160 \pi^2}\omega^7
\\ \nonumber \quad && ~~~~
- \frac{\langle g_s^2 GG \rangle}{17280 \pi^4}\omega^7 - \frac{1019m_s^2\langle g_s^2 GG \rangle}{184320 \pi^4}\omega^5 - \frac{13m_s\langle g_s^2 GG \rangle\langle \bar q q \rangle}{648 \pi^2}\omega^3 + \frac{67m_s\langle g_s^2 GG \rangle\langle \bar s s \rangle}{2304 \pi^2}\omega^3  ]e^{-\omega/T}d\omega  \, ,
\\ && f_{\Xi_c,2,0,\lambda\lambda}^2 \Sigma_{\Xi_c,2,0,\lambda\lambda} e^{-2 \bar \Lambda_{\Xi_c,2,0,\lambda\lambda} / T}
\label{eq:Sc2}
\\ \nonumber \quad && = \int_{2m_s}^{\omega_c} [ \frac{\langle g_s^2 GG \rangle}{24192 \pi^4}\omega^7 - \frac{m_s^2\langle g_s^2 GG \rangle}{1536 \pi^4}\omega^5 + \frac{5m_s\langle g_s^2 GG \rangle\langle \bar s s \rangle}{864 \pi^2}\omega^3  ]e^{-\omega/T}d\omega \, ,
\end{eqnarray}
Sum rules for other currents are shown in Appendix~\ref{sec:others}.

\begin{figure}[hbt]
\begin{center}
\scalebox{0.6}{\includegraphics{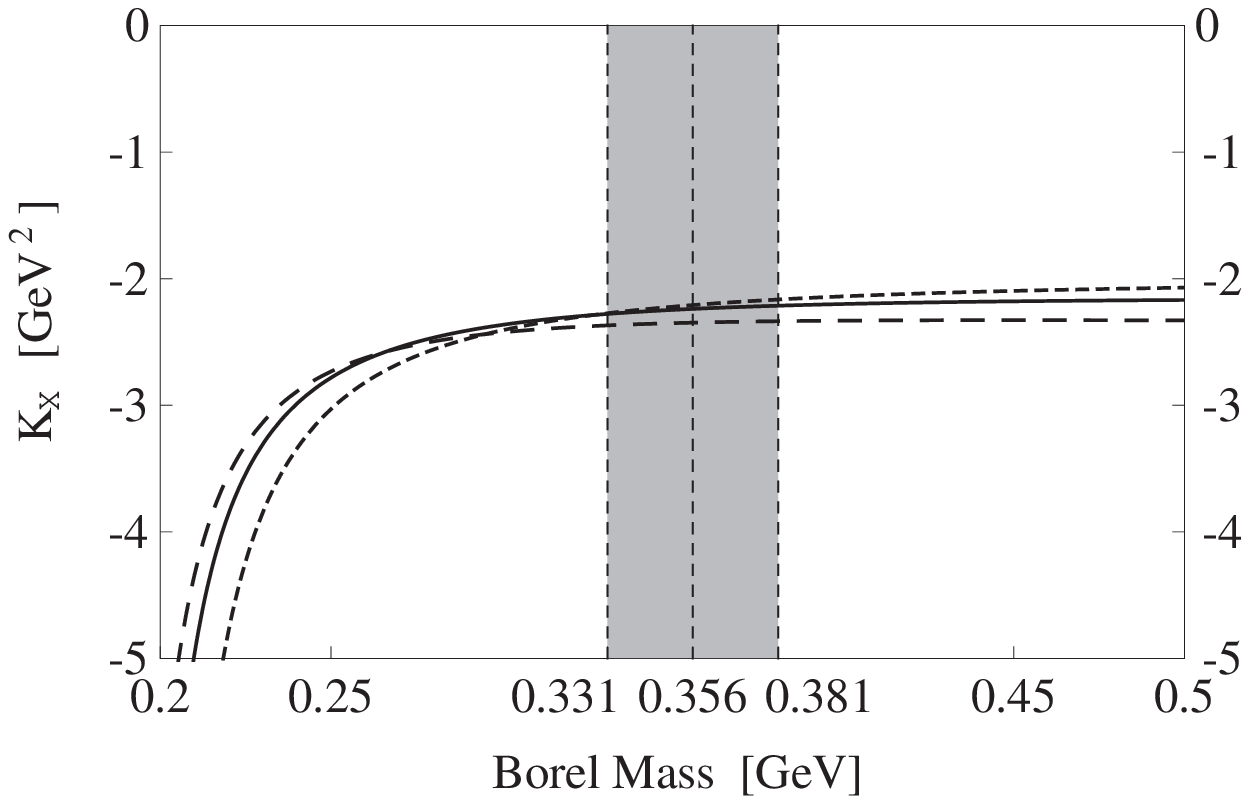}}
\scalebox{0.6}{\includegraphics{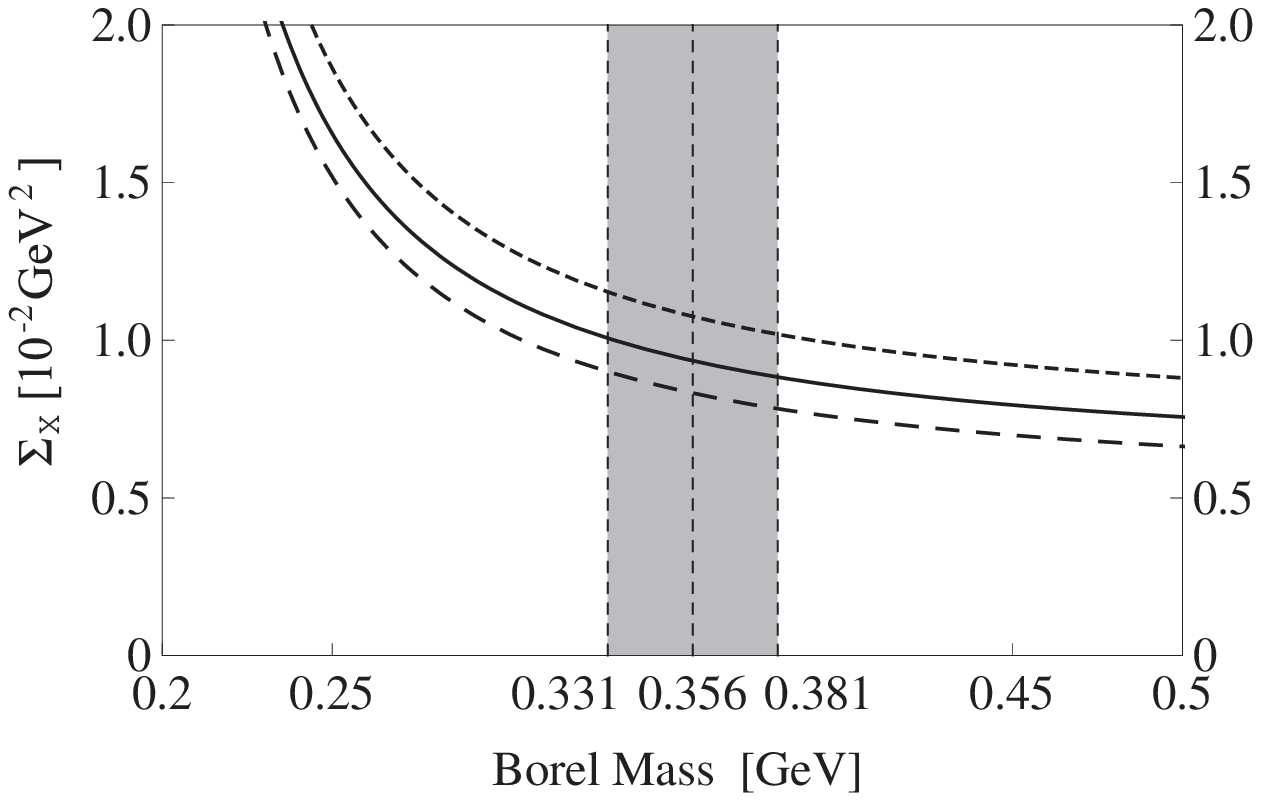}} \caption{The variations
of $K_{\Lambda_c,2,0,\lambda\lambda}$ (left) and $\Sigma_{\Lambda_c,2,0,\lambda\lambda}$ (right) with respect to the Borel mass
$T$, when $J^{\alpha}_{3/2,+,\Lambda_c,2,0,\lambda\lambda}$ is used. The short-dashed, solid and
long-dashed curves are obtained by fixing $\omega_c = 2.3$, 2.5 and 2.7 GeV, respectively.} \label{fig:next1}
\end{center}
\end{figure}

Finally, we obtain $K_{\Lambda_c,2,0,\lambda\lambda}$ and
$\Sigma_{\Lambda_c,2,0,\lambda\lambda}$ by simply dividing Eqs.~(\ref{eq:Kc1}) and (\ref{eq:Sc1}) by Eq.~(\ref{eq:ope1}). Their variations are shown in Fig.~\ref{fig:next1} with respect to the Borel mass $T$. We find their dependence on $T$ is weak in the Borel window $0.331$ GeV $< T < 0.381$ GeV, and obtain the following numerical results:
\begin{eqnarray}
K_{\Lambda_c,2,0,\lambda\lambda} &=& -2.239 \mbox{ GeV}^2 \, ,
\\ \nonumber \Sigma_{\Lambda_c,2,0,\lambda\lambda} &=& 0.014 \mbox{ GeV}^{2} \, ,
\end{eqnarray}
where the central values are obtained by choosing $T=0.356$ GeV and $\omega_c = 2.5$ GeV.

\begin{figure}[hbt]
\begin{center}
\scalebox{0.6}{\includegraphics{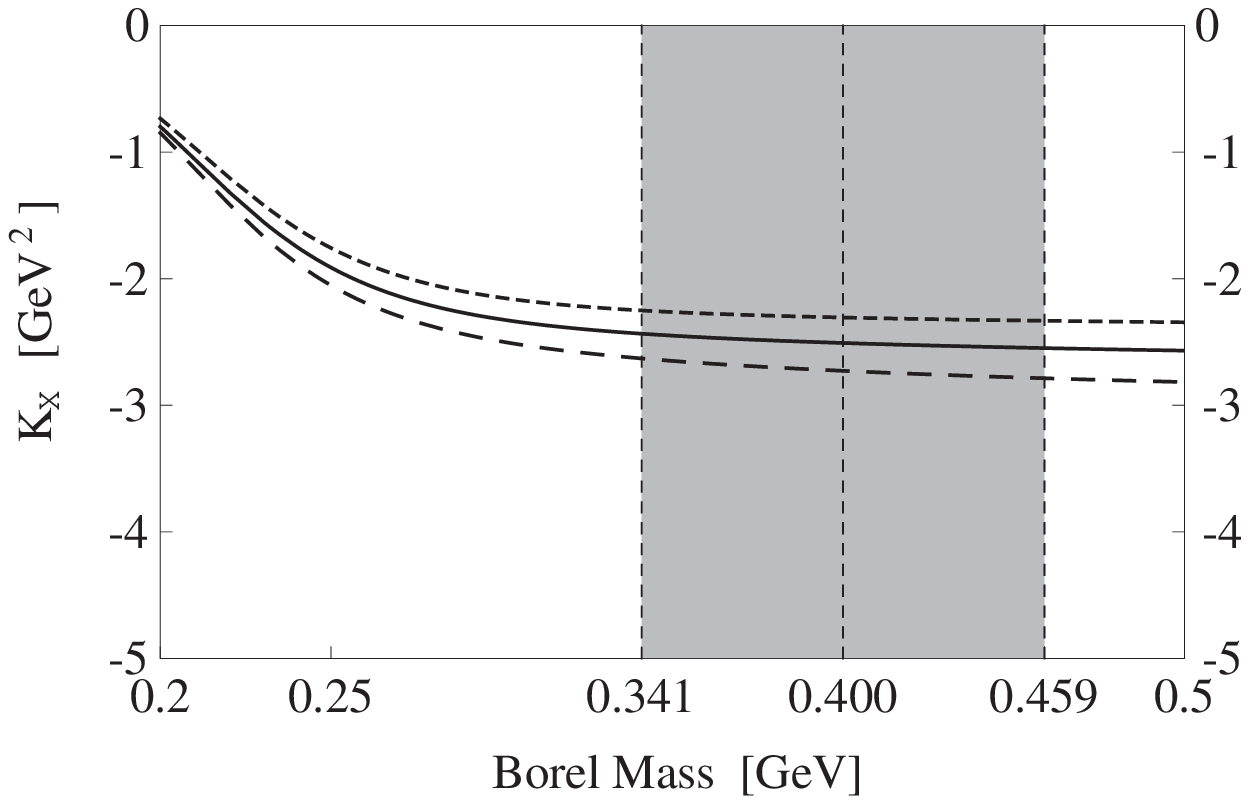}}
\scalebox{0.6}{\includegraphics{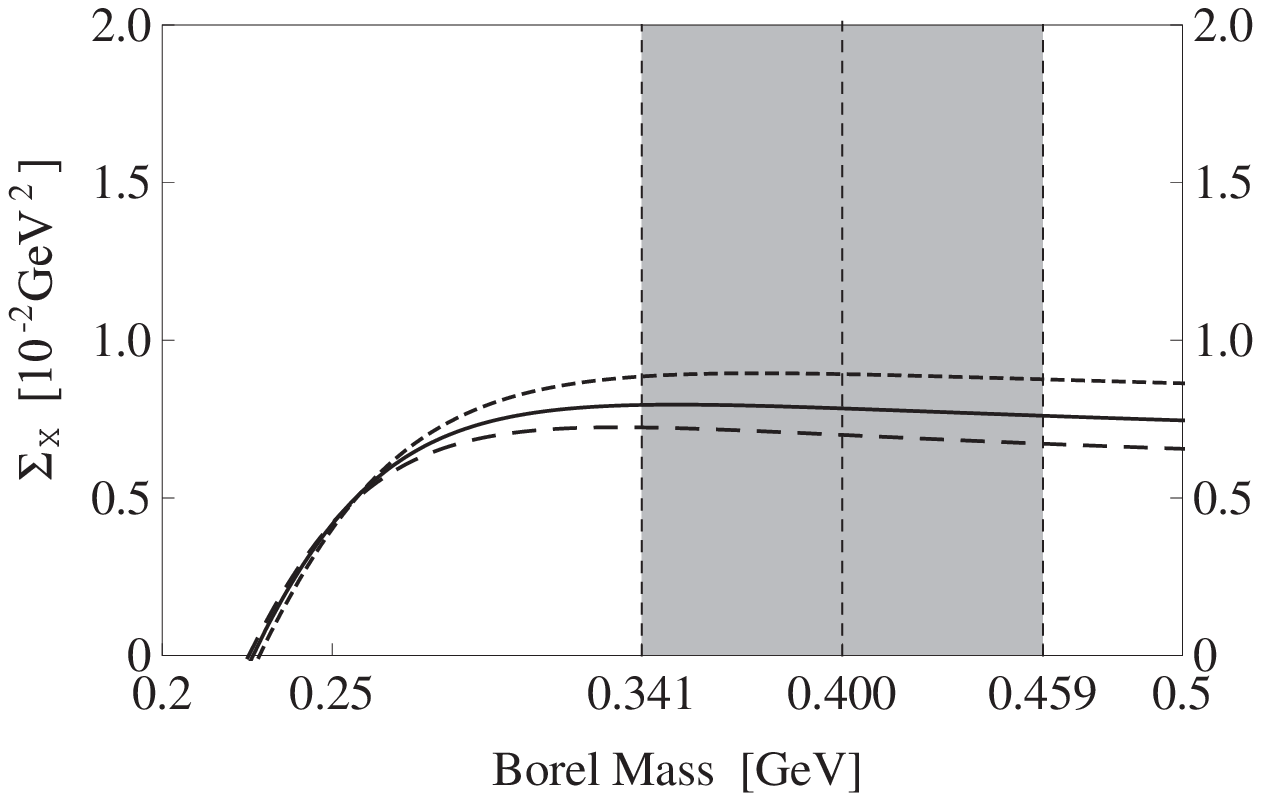}} \caption{The variations
of $K_{\Xi_c,2,0,\lambda\lambda}$ (left) and $\Sigma_{\Xi_c,2,0,\lambda\lambda}$ (right) with respect to the Borel mass
$T$, when $J^{\alpha}_{3/2,+,\Xi_c,2,0,\lambda\lambda}$ is used. The short-dashed, solid and
long-dashed curves are obtained by fixing $\omega_c = 2.8$, 3.0 and 3.2 GeV, respectively.} \label{fig:next2}
\end{center}
\end{figure}

We also obtain $K_{\Xi_c,2,0,\lambda\lambda}$ and
$\Sigma_{\Xi_c,2,0,\lambda\lambda}$ by simply dividing Eqs.~(\ref{eq:Kc2}) and (\ref{eq:Sc2}) by Eq.~(\ref{eq:ope2}), and show their variations in Fig.~\ref{fig:next2} with respect to the Borel mass $T$. We find their dependence on $T$ is weak in the Borel window $0.341$ GeV $< T < 0.459$ GeV, and obtain the following numerical results:
\begin{eqnarray}
K_{\Xi_c,2,0,\lambda\lambda} &=& -2.508 \mbox{ GeV}^2 \, ,
\\ \nonumber \Sigma_{\Xi_c,2,0,\lambda\lambda} &=& 0.008 \mbox{ GeV}^{2} \, ,
\end{eqnarray}
where the central values are obtained by choosing $T=0.400$ GeV and $\omega_c = 3.0$ GeV.

\section{Numerical Results and Discussions}
\label{sec:numerical}

\begin{figure}[hbt]
\begin{center}
\begin{tabular}{c}
\scalebox{0.6}{\includegraphics{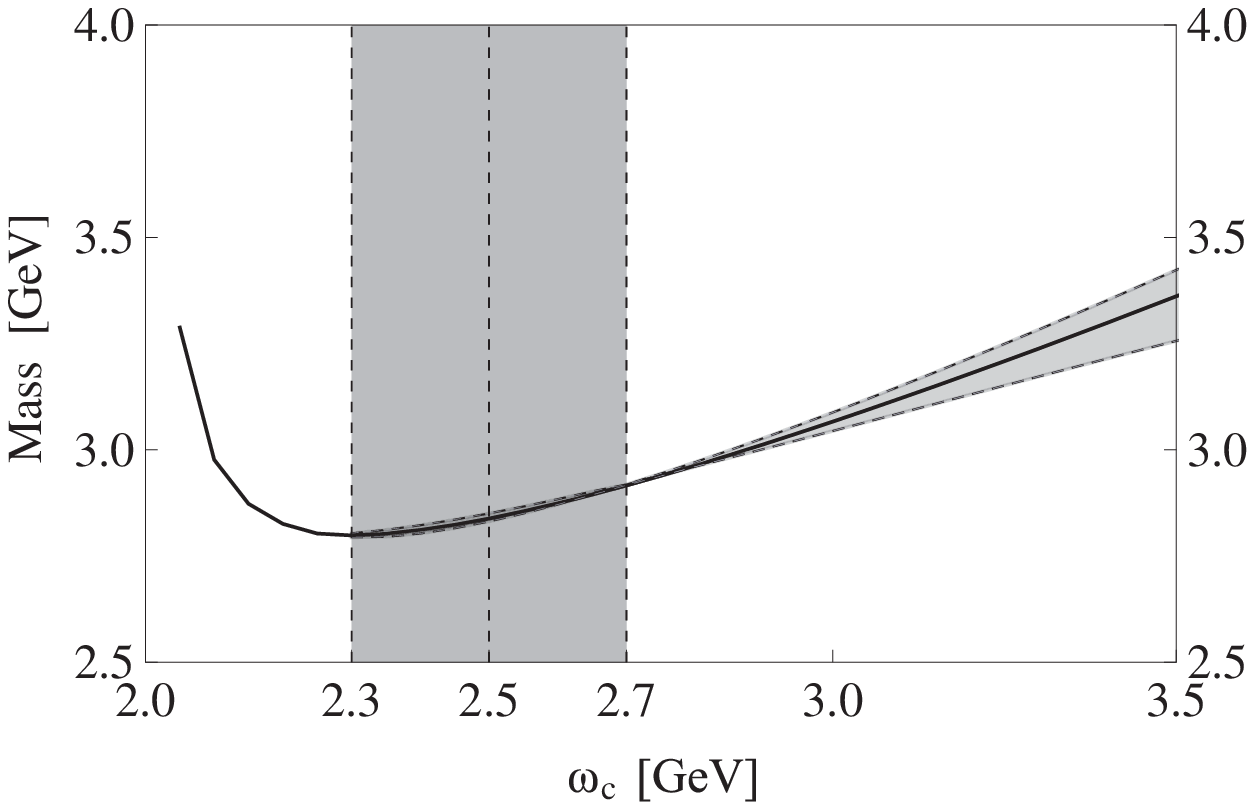}}
\scalebox{0.6}{\includegraphics{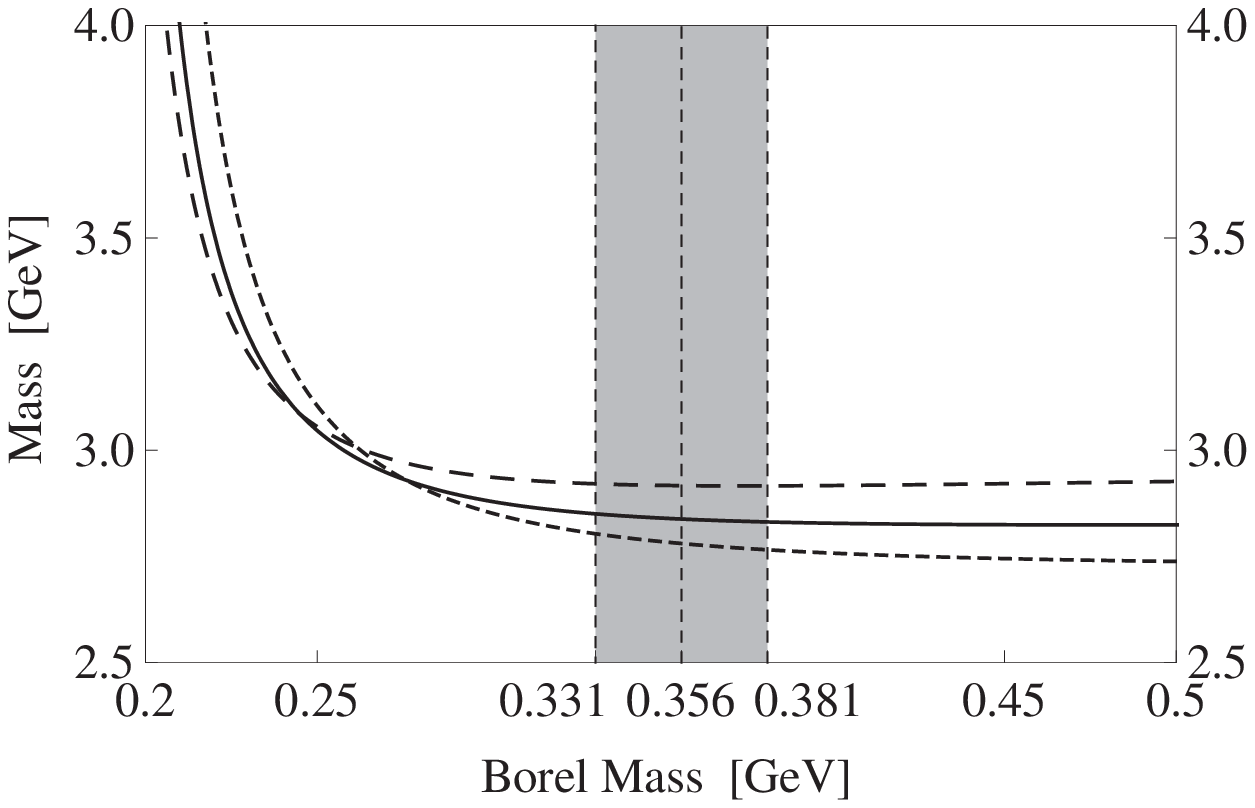}}
%\\
%\scalebox{0.6}{\includegraphics{lambda6F21wc.eps}}
\end{tabular}
\caption{Variations of $m_{\Lambda_c(5/2^+)}$ with respect to the threshold value $\omega_c$ (left) and the Borel mass $T$ (right), calculated using the charmed baryon doublet $[\Lambda_c,2,0,\lambda\lambda]$.
In the left panel, the shady band is obtained by changing $T$ inside Borel windows. The mass curves have minimum against $\omega_c$ around 2.2 GeV, where the $\omega_c$ dependence of the mass prediction is the weakest. However, at this point there does not exist any non-vanishing working region of the Borel mass $T$. We find that there exist non-vanishing working regions of $T$ as long as $\omega_c \geq 2.3$ GeV, and the $\omega_c$ dependence is still weak and acceptable in the region $2.3$ GeV$<\omega_c<2.7$ GeV. The results for $\omega_c < 2.3$ GeV are also shown, for which cases we choose the Borel mass $T$ when the PC, as defined in Eq.~(34), is around 10\%.
In the right figure, the short-dashed, solid and long-dashed curves are obtained by fixing $\omega_c = 2.3$, 2.5 and 2.7 GeV, respectively.
}
\label{fig:mass2002ud}
\end{center}
\end{figure}

Combining the results obtained in Sec.~\ref{sec:leading} and Sec.~\ref{sec:nexttoleading}, we obtain the masses of the heavy baryon doublet $[\Lambda_c,2,0,\lambda\lambda]$ satisfying:
\begin{eqnarray}
m_{\Lambda_c({3/2}^+)} &=& m_c + \overline{\Lambda}_{\Lambda_c,2,0,\lambda\lambda} - {1 \over 4m_c} [ K_{\Lambda_c,2,0,\lambda\lambda} + d_{3/2,2} \Sigma_{\Lambda_c,2,0,\lambda\lambda} ] \, ,
\\ \nonumber
m_{\Lambda_c({5/2}^+)} &=& m_c + \overline{\Lambda}_{\Lambda_c,2,0,\lambda\lambda} - {1 \over 4m_c} [ K_{\Lambda_c,2,0,\lambda\lambda} + d_{5/2,2} \Sigma_{\Lambda_c,2,0,\lambda\lambda} ] \, .
\end{eqnarray}
After inserting $d_{3/2,2} = 6$ and $d_{5/2,2} = -4$, we arrive at:
\begin{eqnarray}
{1\over10} \Big ( 4 m_{\Lambda_c({3/2}^+)} + 6 m_{\Lambda_c({5/2}^+)} \Big) &=& m_c + 1.113 \mbox{ GeV } - {1 \over 4m_c} [ - 2.239 \mbox{ GeV}^2 ] \, ,
\\ \nonumber
m_{\Lambda_c({5/2}^+)} - m_{\Lambda_c({3/2}^+)} &=& {1 \over 4m_c} \times {10} \times [ 0.014 \mbox{ GeV}^2 ] \, ,
\end{eqnarray}
where $\Lambda_c({3/2}^+)$ and $\Lambda_c({5/2}^+)$ are the two baryons contained in this doublet.
Clearly, the $\mathcal{O}(1/m_Q)$ corrections can not be neglected.
Then we use the PDG value $m_c = 1.275 \pm 0.025$ GeV~\cite{pdg} for the charm quark mass in the $\overline{\rm MS}$ scheme
to obtain numerical results:
\begin{eqnarray}
\nonumber m_{\Lambda_c({3/2}^+)} &=& 2.81 \mbox{ GeV} \, ,
\\ m_{\Lambda_c({5/2}^+)} &=& 2.84 \mbox{ GeV} \, ,
\\ \nonumber m_{\Lambda_c({5/2}^+)} - m_{\Lambda_c({3/2}^+)} &=& 28 \mbox{ MeV} \, .
\end{eqnarray}
These values are obtained for $\omega_c = 2.5$ GeV.
We change the threshold value $\omega_c$ and redo the same procedures.
We note that our third criterion is to require the dependence of $m_{j,P,F,j_l,s_l,\rho\rho/\lambda\lambda/\rho\lambda}$ (mass of the heavy baryon state) on this parameter $\omega_c$ to be weak.
Accordingly, we show the variation of $m_{\Lambda_c(5/2^+)}$ with respect to $\omega_c$ in the left panel of Fig.~\ref{fig:mass2002ud} in a large region 2.0 GeV$ < \omega_c < 3.5$ GeV.
The mass curves have minimum against $\omega_c$ around 2.2 GeV, where the $\omega_c$ dependence of the mass prediction is the weakest. However, at this point we apply the two criteria on the Borel mass $T$ (see discussions in Sec.~\ref{sec:leading}) but can not obtain any non-vanishing working region of $T$. We find that there exist non-vanishing working regions of $T$ as long as $\omega_c \geq 2.3$ GeV, and the $\omega_c$ dependence is still weak and acceptable in the region $2.3$ GeV$<\omega_c<2.7$ GeV.
Hence, we choose $2.3$ GeV$<\omega_c<2.7$ GeV and $0.331$ GeV $< T < 0.381$ GeV as our working regions, and obtain the following numerical results for the baryon doublet $[\Lambda_c,2,0,\lambda\lambda]$:
\begin{eqnarray}
\nonumber m_{\Lambda_c({3/2}^+)} &=& 2.81^{+0.33}_{-0.18} \mbox{ GeV} \, ,
\\ m_{\Lambda_c({5/2}^+)} &=& 2.84^{+0.37}_{-0.20} \mbox{ GeV} \, ,
\\ \nonumber m_{\Lambda_c({5/2}^+)} - m_{\Lambda_c({3/2}^+)} &=& 28^{+45}_{-24} \mbox{ MeV} \, ,
\end{eqnarray}
whose central values correspond to $T=0.356$ GeV and $\omega_c = 2.5$ GeV, and the
uncertainties are due to the Borel mass $T$, the threshold value $\omega_c$, the charm quark mass $m_c$ and the quark and gluon condensates.
We also show the variation of $m_{\Lambda_c(5/2^+)}$ with respect to the Borel mass $T$ in the right panel of Fig.~\ref{fig:mass2002ud}, in a broad region $0.2$ GeV$< T < 0.5$ GeV, where these curves are more stable inside the Borel window $0.331$ GeV $< T < 0.381$ GeV.
The mass of the $\Lambda_c({5/2}^+)$ in the doublet $[\Lambda_c,2,0,\lambda\lambda]$ is consistent with the mass of the $\Lambda_c(2880)$~\cite{pdg}:
\begin{eqnarray}
m^{\rm exp}_{\Lambda_c(2880), 5/2^+} = 2881.53 \pm 0.35 \mbox{ MeV} \, ,
\end{eqnarray}
and supports it to be a $D$-wave charmed baryon of $J^P=5/2^+$.
Our result further suggests that the $\Lambda_c(2880)$ of $J^P=5/2^+$ has a partner state, the $\Lambda_c(3/2^+)$ of $J^P=3/2^+$. Its mass is $2.81^{+0.33}_{-0.18}$ GeV, and the mass difference between it and the $\Lambda_c(2880)$ is $28^{+45}_{-24}$ MeV.
We note that there are large theoretical uncertainties in our results for the masses of the heavy baryons, but their differences within the same doublet are produced with much less theoretical uncertainty because they do not depend much on the charm quark mass and the threshold value~\cite{Chen:2015kpa,Mao:2015gya}.

\begin{figure}[hbt]
\begin{center}
\begin{tabular}{c}
\scalebox{0.6}{\includegraphics{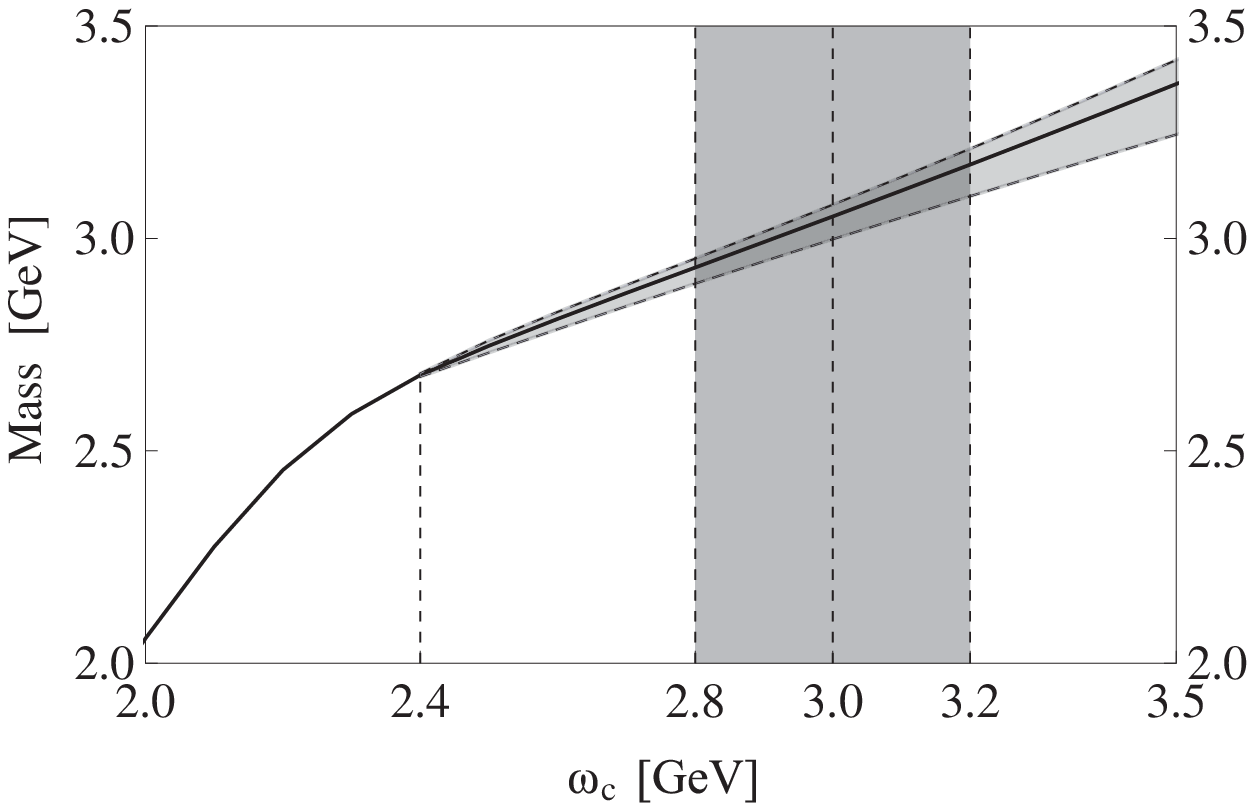}}
\scalebox{0.6}{\includegraphics{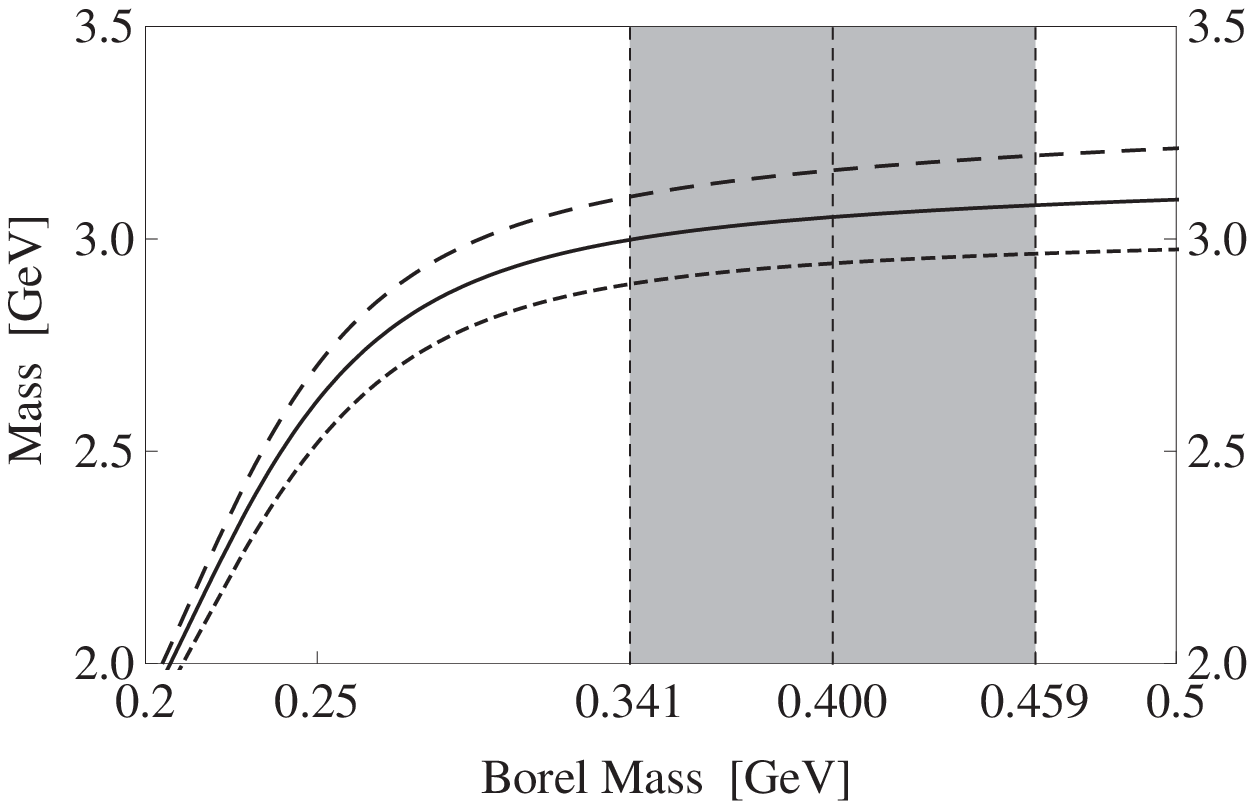}}
%\\
%\scalebox{0.6}{\includegraphics{lambda6F21wc.eps}}
\end{tabular}
\caption{Variations of $m_{\Xi_c(5/2^+)}$ with respect to the threshold value $\omega_c$ (left) and the Borel mass $T$ (right), calculated using the charmed baryon doublet $[\Xi_c,2,0,\lambda\lambda]$.
In the left panel, the shady band is obtained by changing $T$ inside Borel windows, which exist as long as $\omega_c \geq 2.4$ GeV. We properly fine-tune the threshold value $\omega_c$ to be around 3.0 GeV so that $\omega_c(\Xi_c(5/2^+)) - \omega_c(\Lambda_c(5/2^+)) = 0.5$ GeV, which value is the same as those used in our previous studies on $P$-wave heavy baryons~\cite{Chen:2015kpa,Mao:2015gya}.
In the right figure, the short-dashed, solid and long-dashed curves are obtained by fixing $\omega_c = 2.8$, 3.0 and 3.2 GeV, respectively.
}
\label{fig:mass2002us}
\end{center}
\end{figure}

We follow the same procedures to study the baryon doublet $[\Xi_c,2,0,\lambda\lambda]$, and show the variation of $m_{\Xi_c(5/2^+)}$ with respect to the threshold value $\omega_c$ in the left panel of Fig.~\ref{fig:mass2002us}. Different from the case of $m_{\Lambda_c(5/2^+)}$, the mass curves do not have minimum against $\omega_c$, but the $\omega_c$ dependence of the mass prediction is still not strong when $\omega_c > 2.5$ GeV where there exist Borel windows.
We properly fine-tune the threshold value $\omega_c$ to be around 3.0 GeV so that $\omega_c(\Xi_c(5/2^+)) - \omega_c(\Lambda_c(5/2^+)) = 0.5$ GeV, which value is the same as those used in our previous studies on $P$-wave heavy baryons~\cite{Chen:2015kpa,Mao:2015gya}.
Together we choose $2.8$ GeV$<\omega_c<3.2$ GeV and $0.341$ GeV $< T < 0.459$ GeV as our working regions, and obtain the following numerical results for the baryon doublet $[\Xi_c,2,0,\lambda\lambda]$:
\begin{eqnarray}
\nonumber m_{\Xi_c({3/2}^+)} &=& 3.04^{+0.15}_{-0.15} \mbox{ GeV} \, ,
\\ m_{\Xi_c({5/2}^+)} &=& 3.05^{+0.15}_{-0.16} \mbox{ GeV} \, ,
\\ \nonumber m_{\Xi_c({5/2}^+)} - m_{\Xi_c({3/2}^+)} &=& 15^{+16}_{-13} \mbox{ MeV} \, ,
\end{eqnarray}
whose central values correspond to $T=0.400$ GeV and $\omega_c = 3.0$ GeV.
We also show the variation of $m_{\Xi_c(5/2^+)}$ with respect to the Borel mass $T$ in the right panel of Fig.~\ref{fig:mass2002us}, where these curves are stable inside the Borel window $0.341$ GeV $< T < 0.459$ GeV.
The masses of the $\Xi_c({3/2}^+)$ and $\Xi_c({5/2}^+)$ in the doublet $[\Xi_c,2,0,\lambda\lambda]$ are consistent with the masses of the $\Xi_c(3055)$ and $\Xi_c(3080)$~\cite{pdg}
as well as their difference:
\begin{eqnarray}
\nonumber && m^{\rm exp}_{\Xi_c(3055)^+} = 3055.1 \pm 1.7 \mbox{ MeV} \, ,
\\ && m^{\rm exp}_{\Xi_c(3080)^+} = 3076.94 \pm 0.28 \mbox{ MeV} \, , m^{\rm exp}_{\Xi_c(3080)^0} = 3079.9 \pm 1.4 \mbox{ MeV} \, ,
\\ \nonumber && m^{\rm exp}_{\Xi_c(3080)^+} - m^{\rm exp}_{\Xi_c(3055)^+} = 21.8 \pm 1.7 \mbox{ MeV} \, .
\end{eqnarray}
This suggests that the $\Xi_c(3055)$ and $\Xi_c(3080)$ have quantum number $J^P = 3/2^+$ and $5/2^+$, respectively, which assignments have been proposed or discussed in detail in Refs.~\cite{Ebert:2011kk,Chen:2014nyo,Cheng:2015rra}.

\begin{table}[hbt]
\begin{center}
\renewcommand{\arraystretch}{1.6}
\caption{Masses of the $D$-wave charmed baryons obtained using the baryon doublets $[\mathbf{\bar 3}_F,2,0,\rho\rho]$, $[\mathbf{\bar 3}_F,2,0,\lambda\lambda]$, $[\mathbf{\bar 3}_F,1,1,\rho\lambda]$, $[\mathbf{\bar 3}_F,2,1,\rho\lambda]$ and $[\mathbf{\bar 3}_F,3,1,\rho\lambda]$.
As discussed at the end of Sec.~\ref{sec:current}, a) for the baryon doublet $[\mathbf{\bar 3}_F,3,1,\rho\lambda]$ containing $\Lambda_c(5/2^+,7/2^+)$ and $\Xi_c(5/2^+,7/2^+)$, we only evaluate their average masses ${1\over14} ( 6 m_{\Lambda_c({5/2}^+)} + 8 m_{\Lambda_c({7/2}^+)} )$ and ${1\over14} ( 6 m_{\Xi_c({5/2}^+)} + 8 m_{\Xi_c({7/2}^+)} )$; b) for the baryon doublet $[\mathbf{\bar 3}_F, 2, 1, \rho\lambda]$ ($s_l = 1$ and $j_l = 2$), we estimate their masses by simply averaging between $[\mathbf{\bar 3}_F,1,1,\rho\lambda]$ ($s_l = 1$ and $j_l = 1$) and $[\mathbf{\bar 3}_F,3,1,\rho\lambda]$ ($s_l = 1$ and $j_l = 3$).
We assume that free parameters $\omega_c$ in the same multiplet satisfy the relation $\omega_c(\Xi_c)- \omega_c(\Lambda_c) = 0.5$ GeV, except for the $[\mathbf{\bar 3}_F,1,1,\rho\lambda]$.
}
\begin{tabular}{c | c | c | c | c c | c c | c | c | c}
\hline\hline
\multirow{2}{*}{Multiplets} & \multirow{2}{*}{B} & $\omega_c$ & Working region & $\overline{\Lambda}$ & $f$ & $K$ & $\Sigma$ & Baryons & Mass & Difference
\\ & & (GeV) & (GeV) & (GeV) & (GeV$^{5}$) & (GeV$^2$) & (GeV$^2$) & ($j^P$) & (GeV) & (MeV)
\\ \hline\hline
\multirow{4}{*}{$[\mathbf{\bar 3}_F,2,0,\lambda\lambda]$} & \multirow{2}{*}{$\Lambda_c$} & \multirow{2}{*}{2.5} & \multirow{2}{*}{$0.331< T < 0.381$} & \multirow{2}{*}{$1.113$} & \multirow{2}{*}{$0.012$} & \multirow{2}{*}{$-2.239$} & \multirow{2}{*}{$0.014$} & $\Lambda_c(3/2^+)$ & $2.81^{+0.33}_{-0.18}$ & \multirow{2}{*}{$28^{+45}_{-24}$}
\\ \cline{9-10}
& & & & & & & & $\Lambda_c(5/2^+)$ & $2.84^{+0.37}_{-0.20}$ &
\\ \cline{2-11}
& \multirow{2}{*}{$\Xi_c$} & \multirow{2}{*}{3.0} & \multirow{2}{*}{$0.341< T < 0.459$} & \multirow{2}{*}{$1.279$} & \multirow{2}{*}{$0.025$} & \multirow{2}{*}{$-2.508$} & \multirow{2}{*}{$0.008$}
& $\Xi_c(3/2^+)$ & $3.04^{+0.15}_{-0.15}$ & \multirow{2}{*}{$15^{+16}_{-13}$}
\\ \cline{9-10}
& & & & & & & & $\Xi_c(5/2^+)$ & $3.05^{+0.15}_{-0.16}$ &
\\ \hline
\multirow{4}{*}{$[\mathbf{\bar 3}_F,2,0,\rho\rho]$} & \multirow{2}{*}{$\Lambda_c$} & \multirow{2}{*}{3.4} & \multirow{2}{*}{$0.358< T < 0.499$} & \multirow{2}{*}{$1.650$} & \multirow{2}{*}{$0.065$} & \multirow{2}{*}{$-1.742$} & \multirow{2}{*}{$0.011$} & $\Lambda_c(3/2^+)$ & $3.25^{+1.72}_{-0.28}$ & \multirow{2}{*}{$22^{+120}_{-20}$}
\\ \cline{9-10}
& & & & & & & & $\Lambda_c(5/2^+)$ & $3.28^{+1.83}_{-0.30}$ &
\\ \cline{2-11}
& \multirow{2}{*}{$\Xi_c$} & \multirow{2}{*}{3.9} & \multirow{2}{*}{$0.502< T < 0.591$} & \multirow{2}{*}{$1.723$} & \multirow{2}{*}{$0.10$} & \multirow{2}{*}{$-1.308$} & \multirow{2}{*}{$0.006$}
& $\Xi_c(3/2^+)$ & $3.25^{+0.16}_{-0.14}$ & \multirow{2}{*}{$11^{+16}_{-9}$}
\\ \cline{9-10}
& & & & & & & & $\Xi_c(5/2^+)$ & $3.26^{+0.17}_{-0.15}$ &
\\ \hline
\multirow{4}{*}{$[\mathbf{\bar 3}_F,1,1,\rho\lambda]$} & \multirow{2}{*}{$\Lambda_c$} & \multirow{2}{*}{3.0} & \multirow{2}{*}{$0.397< T < 0.457$} & \multirow{2}{*}{$1.335$} & \multirow{2}{*}{$0.044$} & \multirow{2}{*}{$-2.116$} & \multirow{2}{*}{$0.006$} & $\Lambda_c(1/2^+)$ & $3.02^{+0.19}_{-0.14}$ & \multirow{2}{*}{$7^{+11}_{-6}$}
\\ \cline{9-10}
& & & & & & & & $\Lambda_c(3/2^+)$ & $3.03^{+0.20}_{-0.14}$ &
\\ \cline{2-11}
& \multirow{2}{*}{$\Xi_c$} & \multirow{2}{*}{4.0} & \multirow{2}{*}{$0.526< T < 0.599$} & \multirow{2}{*}{$1.887$} & \multirow{2}{*}{$0.21$} & \multirow{2}{*}{$-2.954$} & \multirow{2}{*}{$0.003$}
& $\Xi_c(1/2^+)$ & $3.74^{+0.14}_{-0.13}$ & \multirow{2}{*}{$3^{+3}_{-3}$}
\\ \cline{9-10}
& & & & & & & & $\Xi_c(3/2^+)$ & $3.74^{+0.14}_{-0.13}$ &
\\ \hline
\multirow{4}{*}{$\begin{array}{c}[\mathbf{\bar 3}_F,2,1,\rho\lambda] \\ {\rm (estimated)}\end{array}$} & \multirow{2}{*}{$\Lambda_c$} & \multirow{2}{*}{--} & \multirow{2}{*}{--} & \multirow{2}{*}{--} & \multirow{2}{*}{--} & \multirow{2}{*}{--} & \multirow{2}{*}{--} & $\Lambda_c(3/2^+)$ & \multirow{2}{*}{$\sim3.20$} & \multirow{2}{*}{--}
\\ \cline{9-9}
& & & & & & & & $\Lambda_c(5/2^+)$ & &
\\ \cline{2-11}
& \multirow{2}{*}{$\Xi_c$} & \multirow{2}{*}{--} & \multirow{2}{*}{--} & \multirow{2}{*}{--} & \multirow{2}{*}{--} & \multirow{2}{*}{--} & \multirow{2}{*}{--}
& $\Xi_c(3/2^+)$ & \multirow{2}{*}{$\sim3.76$} & \multirow{2}{*}{--}
\\ \cline{9-9}
& & & & & & & & $\Xi_c(5/2^+)$ & &
\\ \hline
\multirow{4}{*}{$\begin{array}{c}[\mathbf{\bar 3}_F,3,1,\rho\lambda] \\ {\rm (simplified)}\end{array}$} & \multirow{2}{*}{$\Lambda_c$} & \multirow{2}{*}{3.6} & \multirow{2}{*}{$0.496< T < 0.542$} & \multirow{2}{*}{$1.628$} & \multirow{2}{*}{$0.022$} & \multirow{2}{*}{$-2.939$} & \multirow{2}{*}{--} & $\Lambda_c(5/2^+)$ & \multirow{2}{*}{$3.48^{+0.33}_{-0.18}$} & \multirow{2}{*}{--}
\\ \cline{9-9}
& & & & & & & & $\Lambda_c(7/2^+)$ & &
\\ \cline{2-11}
& \multirow{2}{*}{$\Xi_c$} & \multirow{2}{*}{4.1} & \multirow{2}{*}{$0.556< T < 0.609$} & \multirow{2}{*}{$1.920$} & \multirow{2}{*}{$0.045$} & \multirow{2}{*}{$-3.105$} & \multirow{2}{*}{--}
& $\Xi_c(5/2^+)$ & \multirow{2}{*}{$3.80^{+0.20}_{-0.16}$} & \multirow{2}{*}{--}
\\ \cline{9-9}
& & & & & & & & $\Xi_c(7/2^+)$ & &
\\ \hline \hline
\end{tabular}
\label{tab:results}
\end{center}
\end{table}

We also study the other four baryon doublets, $[\mathbf{\bar 3}_F,2,0,\rho\rho]$, $[\mathbf{\bar 3}_F,1,1,\rho\lambda]$, $[\mathbf{\bar 3}_F,2,1,\rho\lambda]$ and $[\mathbf{\bar 3}_F,3,1,\rho\lambda]$.
Two important notes are:
\begin{enumerate}

\item It is too complicated to directly use $J^{\alpha_1\alpha_2\alpha_3}_{7/2,+,\mathbf{\bar 3}_F,3,1,\rho\lambda}$, defined in Eq.~(\ref{eq:current18}), to
perform QCD sum rule analyses, so we shall use its simplified version without the projection operator $\Gamma^{\alpha_1\alpha_2\alpha_3,\mu_1\mu_2\mu_3}$:
\begin{eqnarray}
J^{\prime\alpha_1\alpha_2\alpha_3}_{7/2,+,\mathbf{\bar 3}_F,3,1,\rho\lambda}(x)
&=& \mathbb{S}^\prime[ i^2 \epsilon_{abc} \Big ( [\mathcal{D}^{\alpha_1} \mathcal{D}^{\alpha_2} q^{aT}(x)] \mathbb{C} \gamma_{\alpha_3} q^b(x) - q^{aT}(x) \mathbb{C} \gamma_{\alpha_3} [\mathcal{D}^{\alpha_1} \mathcal{D}^{\alpha_2} q^b(x)] \Big ) \times h_v^c(x) ] \, ,
\label{eq:current21}
\end{eqnarray}
where $\mathbb{S}^\prime[\cdots]$ is used to denote symmetrization and subtracting the trace
terms in the sets $(\alpha_1 \cdots \alpha_3)$.
Using this current, we can well calculate sum rules at the leading order as well as the $\mathcal{K}$ correction ($K_{\mathbf{\bar 3}_F,3,1,\rho\lambda}$) at the order ${\mathcal O}(1/m_Q)$, but the $\mathcal S$ correction ($\Sigma_{\mathbf{\bar 3}_F,3,1,\rho\lambda}$) at the order ${\mathcal O}(1/m_Q)$ can not be evaluated.

\item Because we failed to construct the two currents belonging to the baryon doublet $[\mathbf{\bar 3}_F, 2, 1, \rho\lambda]$ with $s_l = 1$ and $j_l = 2$, we shall estimate their masses by averaging between $[\mathbf{\bar 3}_F,1,1,\rho\lambda]$ ($s_l = 1$ and $j_l = 1$) and $[\mathbf{\bar 3}_F,3,1,\rho\lambda]$ ($s_l = 1$ and $j_l = 3$), weighted by the spin-orbital splittings:
\begin{eqnarray}
l_\rho \otimes s_l = {1\over2} \big( j_l(j_l+1) - l_\rho(l_\rho+1) - s_l(s_l+1) \big) = {1\over2} \big( j_l(j_l+1) - 4 \big) \, .
\end{eqnarray}
Hence, we obtain
\begin{eqnarray}
{\rm Mass}(j_l = 2) = {3\over5} \times {\rm Mass}(j_l = 1) + {2\over5} \times {\rm Mass}(j_l = 3) \, .
\end{eqnarray}

\end{enumerate}
However, their obtained results are difficult to explain the $\Lambda_c(2880)$, $\Xi_c(3055)$ and $\Xi_c(3080)$ at the same time:
\begin{enumerate}

\item The baryon doublet $[\mathbf{\bar 3}_F,2,0,\rho\rho]$ contains $\Lambda_c(3/2^+,5/2^+)$ and $\Xi_c(3/2^+,5/2^+)$. We use them to perform QCD sum rule analyses, and show variations of $m_{\Lambda_c(5/2^+)}$ and $m_{\Xi_c(5/2^+)}$ with respect to the threshold value $\omega_c$ in Fig.~\ref{fig:mass2020}. The obtained masses are listed in Table~\ref{tab:results}:
\begin{eqnarray}
m_{\Lambda_c({3/2}^+)} &=& 3.25^{+1.72}_{-0.28} \mbox{ GeV} \, , m_{\Lambda_c({5/2}^+)} = 3.28^{+1.83}_{-0.30} \mbox{ GeV} \, , \Delta m = 22^{+120}_{-20} \mbox{ MeV} \, ,
\\ \nonumber m_{\Xi_c({3/2}^+)} &=& 3.25^{+0.16}_{-0.14} \mbox{ GeV} \, , m_{\Xi_c({5/2}^+)} = 3.26^{+0.17}_{-0.15} \mbox{ GeV} \, , \Delta m = 11^{+16}_{-9} \mbox{ MeV} \, ,
\end{eqnarray}
    whose values are significantly larger than the masses of the $\Lambda_c(2880)$, $\Xi_c(3055)$ and $\Xi_c(3080)$.

\begin{figure}[hbt]
\begin{center}
\begin{tabular}{c}
\scalebox{0.6}{\includegraphics{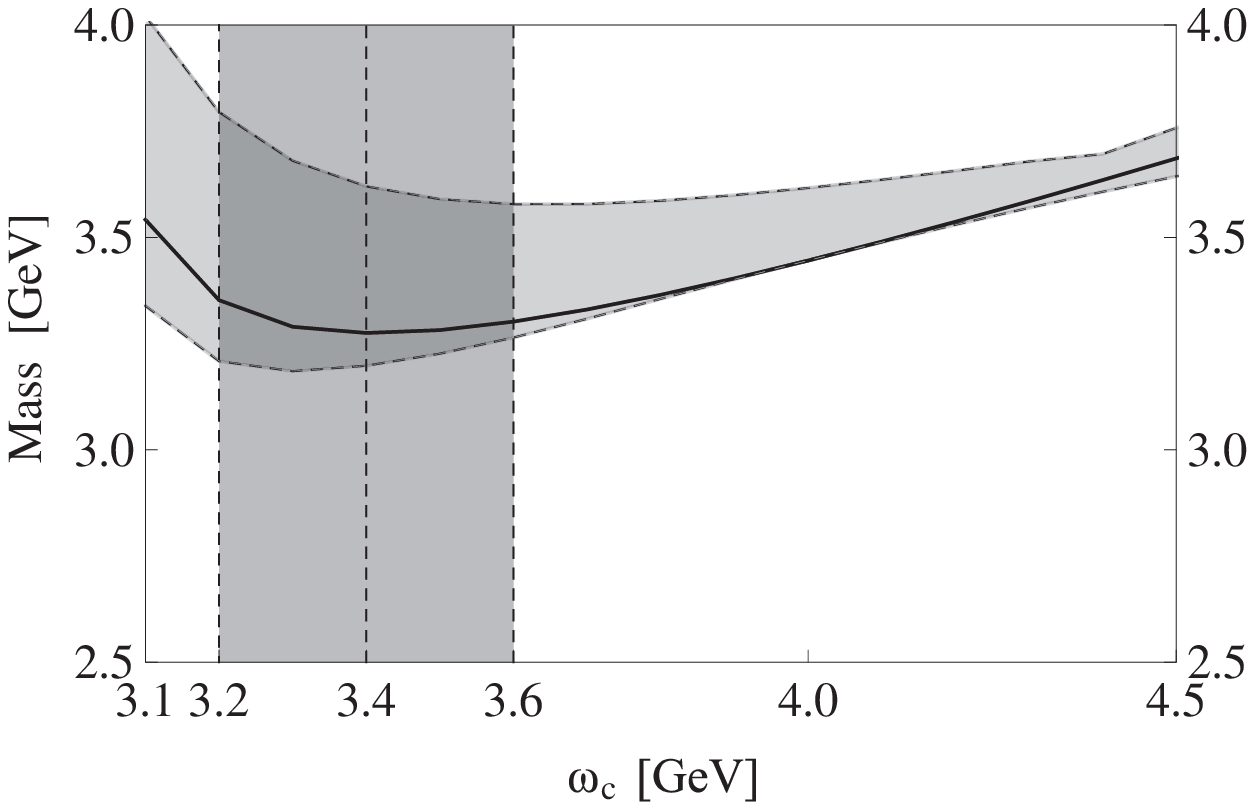}}
\scalebox{0.6}{\includegraphics{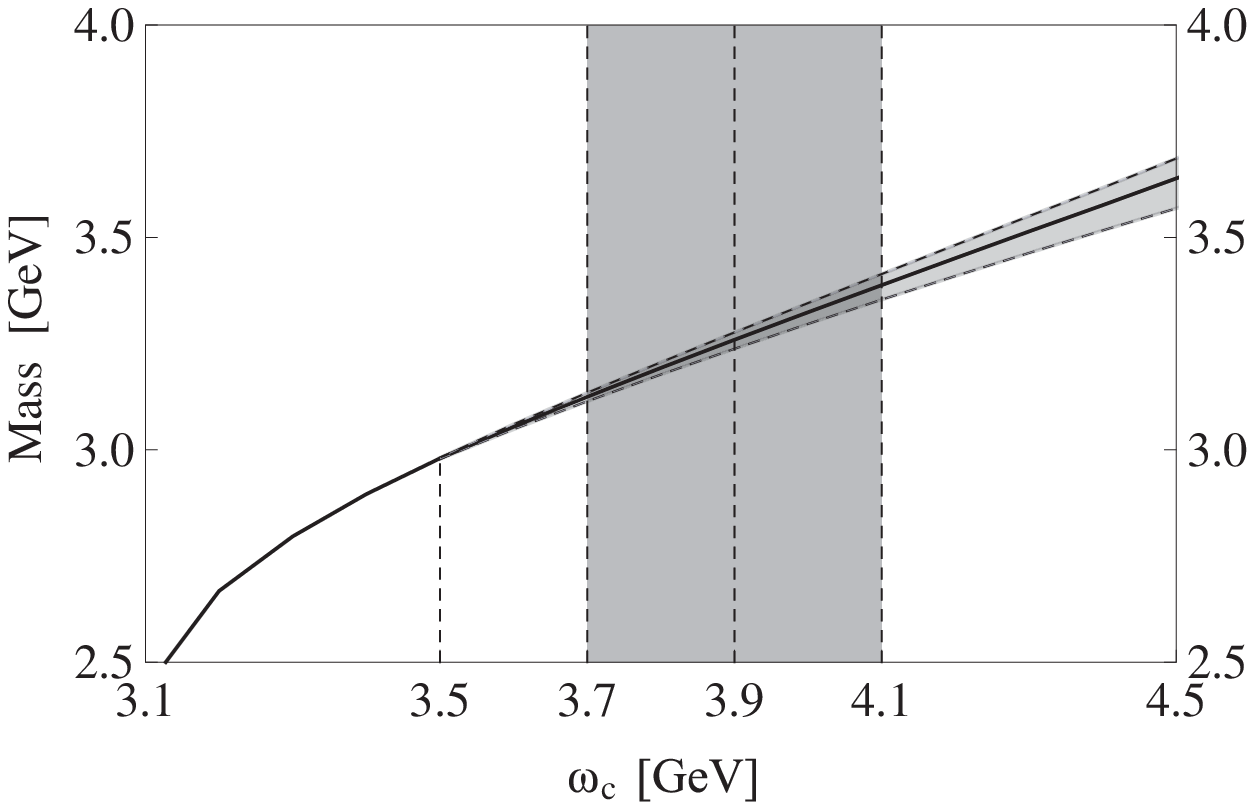}}
%\\
%\scalebox{0.6}{\includegraphics{lambda6F21wc.eps}}
\end{tabular}
\caption{Variations of $m_{\Lambda_c(5/2^+)}$ (left) and $m_{\Xi_c(5/2^+)}$ (right) with respect to the threshold value $\omega_c$, calculated using the charmed baryon doublet $[\mathbf{\bar 3}_F,2,0,\rho\rho]$.
The shady band is obtained by changing $T$ inside Borel windows, which exist as long as $\omega_c \geq 3.1$ GeV (left) and $\omega_c \geq 3.5$ GeV (right).
In the left panel we choose $\omega_c$ to be around 3.4 GeV, where the mass curves have minimum against it.
In the right panel we properly fine-tune $\omega_c$ to be around 3.9 GeV so that $\omega_c(\Xi_c(5/2^+)) - \omega_c(\Lambda_c(5/2^+)) = 0.5$ GeV.
}
\label{fig:mass2020}
\end{center}
\end{figure}

\item The baryon doublet $[\mathbf{\bar 3}_F,1,1,\rho\lambda]$ contains $\Lambda_c(1/2^+,3/2^+)$ and $\Xi_c(1/2^+,3/2^+)$. This doublet does not contain any baryon of $J^P = 5/2^+$. We use them to perform QCD sum rule analyses, and show variations of $m_{\Lambda_c(3/2^+)}$ and $m_{\Xi_c(3/2^+)}$ with respect to the threshold value $\omega_c$ in Fig.~\ref{fig:mass1111}. The obtained masses are listed in Table~\ref{tab:results}:
\begin{eqnarray}
m_{\Lambda_c({1/2}^+)} &=& 3.02^{+0.19}_{-0.14} \mbox{ GeV} \, , m_{\Lambda_c({3/2}^+)} = 3.03^{+0.20}_{-0.14} \mbox{ GeV} \, , \Delta m = 7^{+11}_{-6} \mbox{ MeV} \, ,
\\ \nonumber m_{\Xi_c({1/2}^+)} &=& 3.74^{+0.14}_{-0.13} \mbox{ GeV} \, , m_{\Xi_c({3/2}^+)} = 3.74^{+0.14}_{-0.13} \mbox{ GeV} \, , \Delta m = 3^{+3}_{-3} \mbox{ MeV} \, .
\end{eqnarray}

\begin{figure}[hbt]
\begin{center}
\begin{tabular}{c}
\scalebox{0.6}{\includegraphics{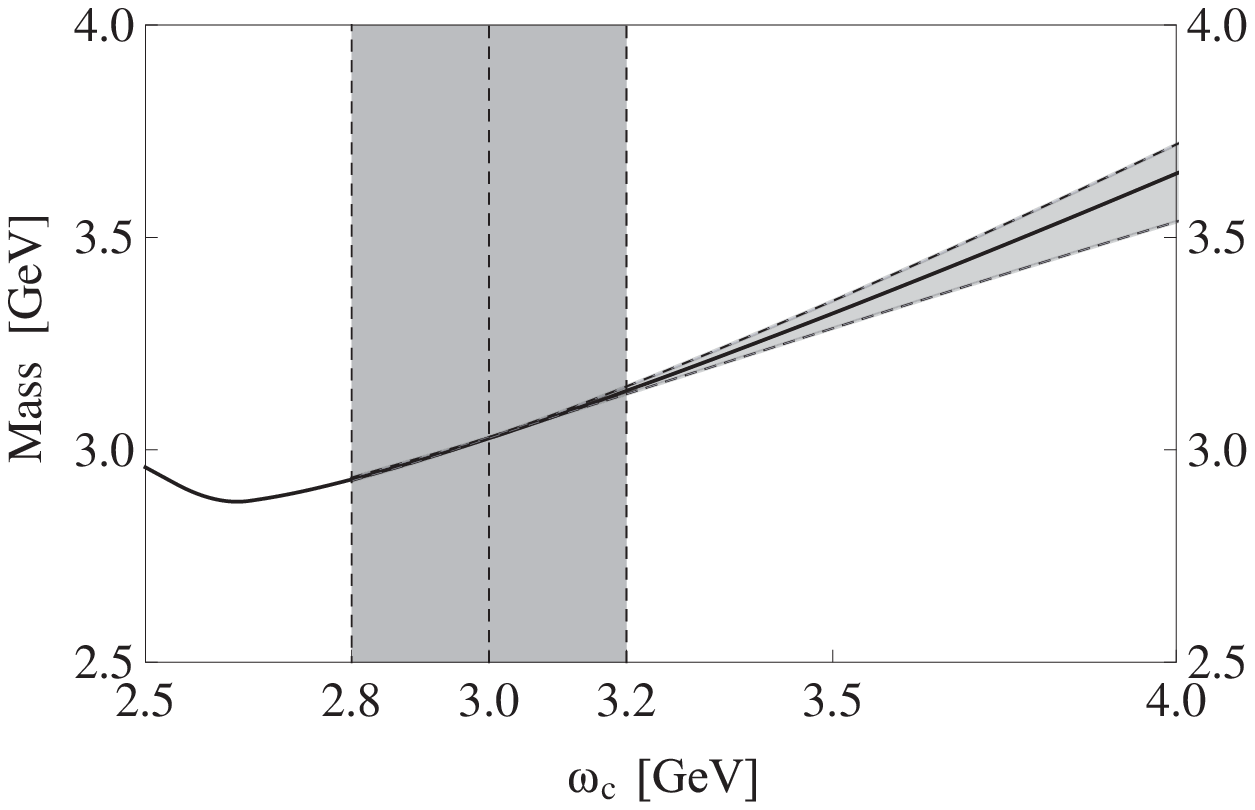}}
\scalebox{0.6}{\includegraphics{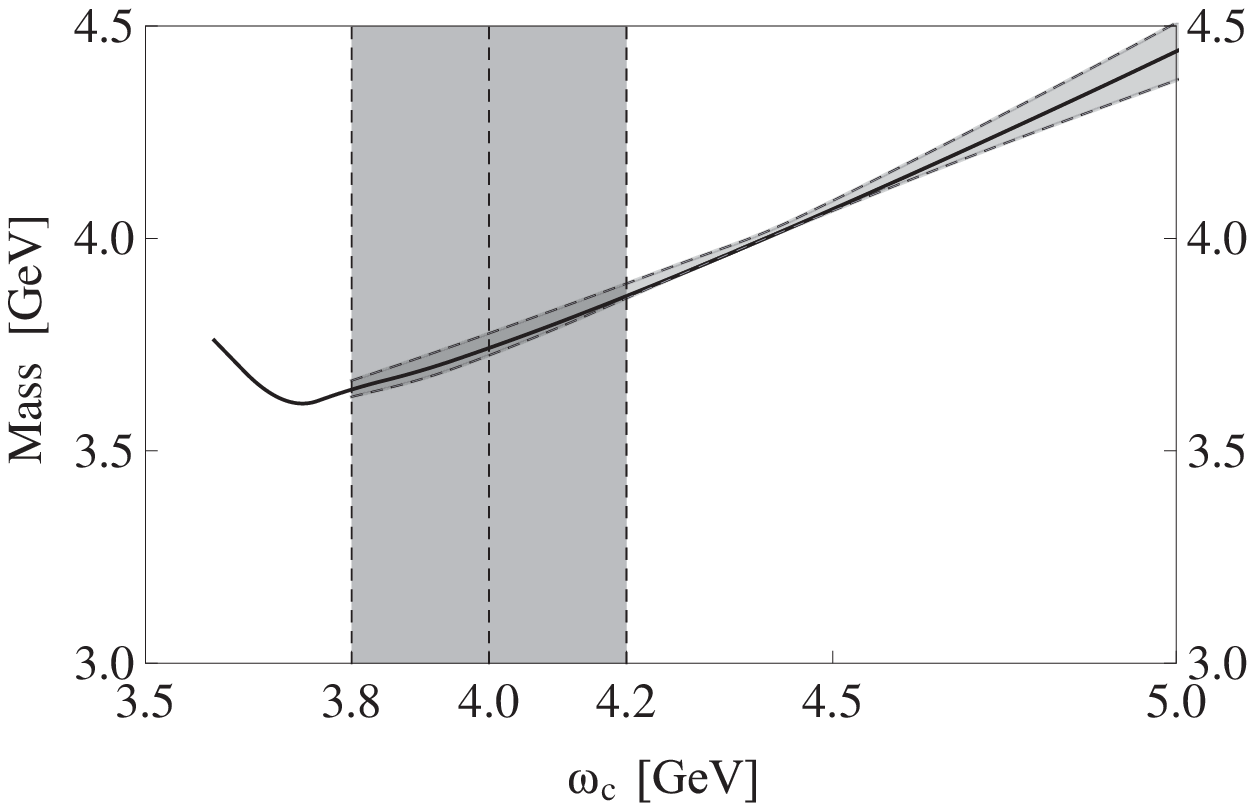}}
%\\
%\scalebox{0.6}{\includegraphics{lambda6F21wc.eps}}
\end{tabular}
\caption{Variations of $m_{\Lambda_c(3/2^+)}$ (left) and $m_{\Xi_c(3/2^+)}$ (right) with respect to the threshold value $\omega_c$, calculated using the charmed baryon doublet $[\mathbf{\bar 3}_F,1,1,\rho\lambda]$.
The shady band is obtained by changing $T$ inside Borel windows.
Although the mass curves have minimum against $\omega_c$ around 2.6 GeV (left) and 3.7 GeV (right), there exist Borel windows as long as $\omega_c \geq 2.8$ GeV (left) and $\omega_c \geq 3.8$ GeV (right), and the $\omega_c$ dependence is still weak and acceptable in the region 2.8 GeV $<\omega_c<$ 3.2 GeV (left) and 3.8 GeV $<\omega_c<$ 4.2 GeV (right).
}
\label{fig:mass1111}
\end{center}
\end{figure}

\item The baryon doublet $[\mathbf{\bar 3}_F,3,1,\rho\lambda]$ contains $\Lambda_c(5/2^+,7/2^+)$ and $\Xi_c(5/2^+,7/2^+)$. We use them to perform QCD sum rule analyses. As discussed at the end of Sec.~\ref{sec:current}, we can only calculate their average masses. Using the following formulae
\begin{eqnarray}
m_{\Lambda_c({5/2}^+)} &=& m_c + \overline{\Lambda}_{\Lambda_c,3,1,\rho\lambda} - {1 \over 4m_c} [ K_{\Lambda_c,3,1,\rho\lambda} + d_{5/2,3} \Sigma_{\Lambda_c,3,1,\rho\lambda} ]  \, ,
\\ \nonumber
m_{\Lambda_c({7/2}^+)} &=& m_c + \overline{\Lambda}_{\Lambda_c,3,1,\rho\lambda} - {1 \over 4m_c} [ K_{\Lambda_c,3,1,\rho\lambda} + d_{7/2,3} \Sigma_{\Lambda_c,3,1,\rho\lambda} ]  \, ,
\end{eqnarray}
and similar formulae for the $\Xi_c({5/2}^+)$ and $\Xi_c({7/2}^+)$, we can obtain    
\begin{eqnarray}
{1\over14} \Big ( 6 m_{\Lambda_c({5/2}^+)} + 8 m_{\Lambda_c({7/2}^+)} \Big) &=& 3.48^{+0.33}_{-0.18} \mbox{ GeV }  \, ,
\\ \nonumber
{1\over14} \Big ( 6 m_{\Xi_c({5/2}^+)} + 8 m_{\Xi_c({7/2}^+)} \Big) &=& 3.80^{+0.20}_{-0.16} \mbox{ GeV } \, .
\end{eqnarray}
These values are listed in Table~\ref{tab:results}, which are significantly larger than the masses of the $\Lambda_c(2880)$, $\Xi_c(3055)$ and $\Xi_c(3080)$. We also show their variations with respect to the threshold value $\omega_c$ in Fig.~\ref{fig:mass3111}.

\begin{figure}[hbt]
\begin{center}
\begin{tabular}{c}
\scalebox{0.6}{\includegraphics{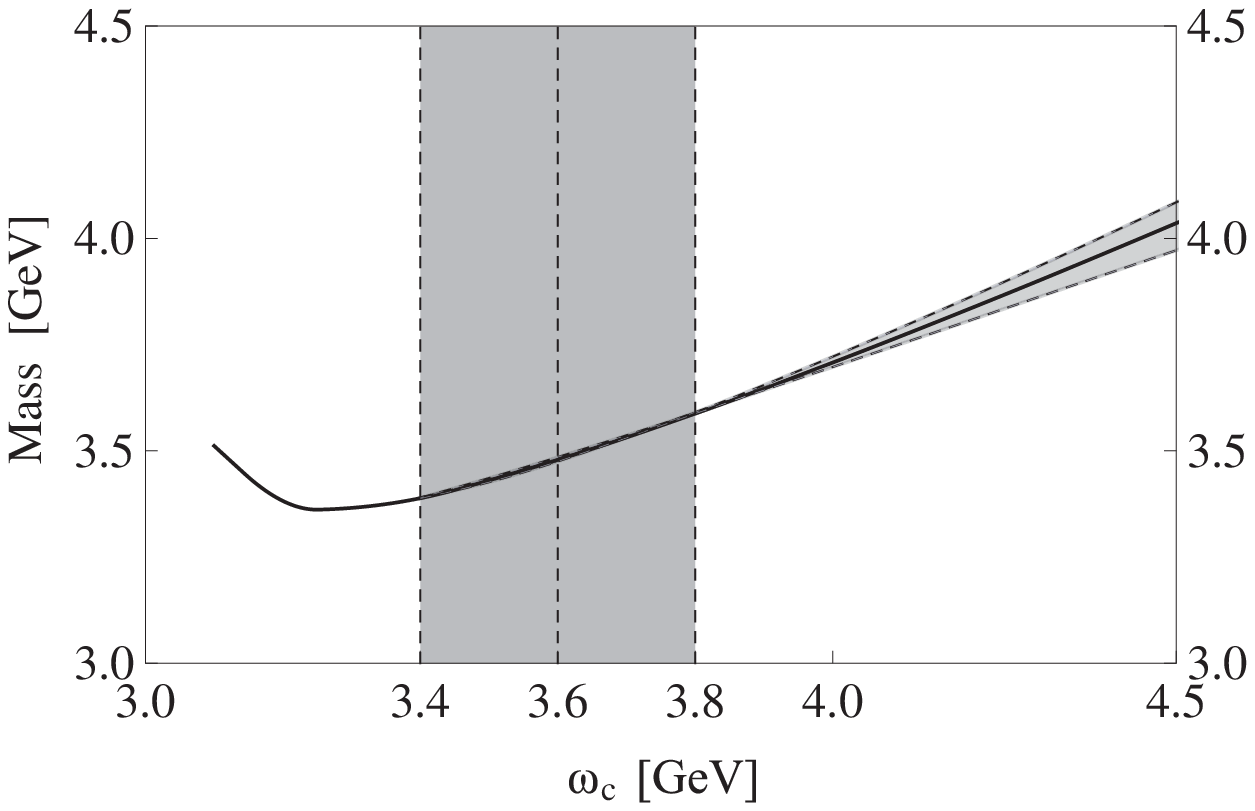}}
\scalebox{0.6}{\includegraphics{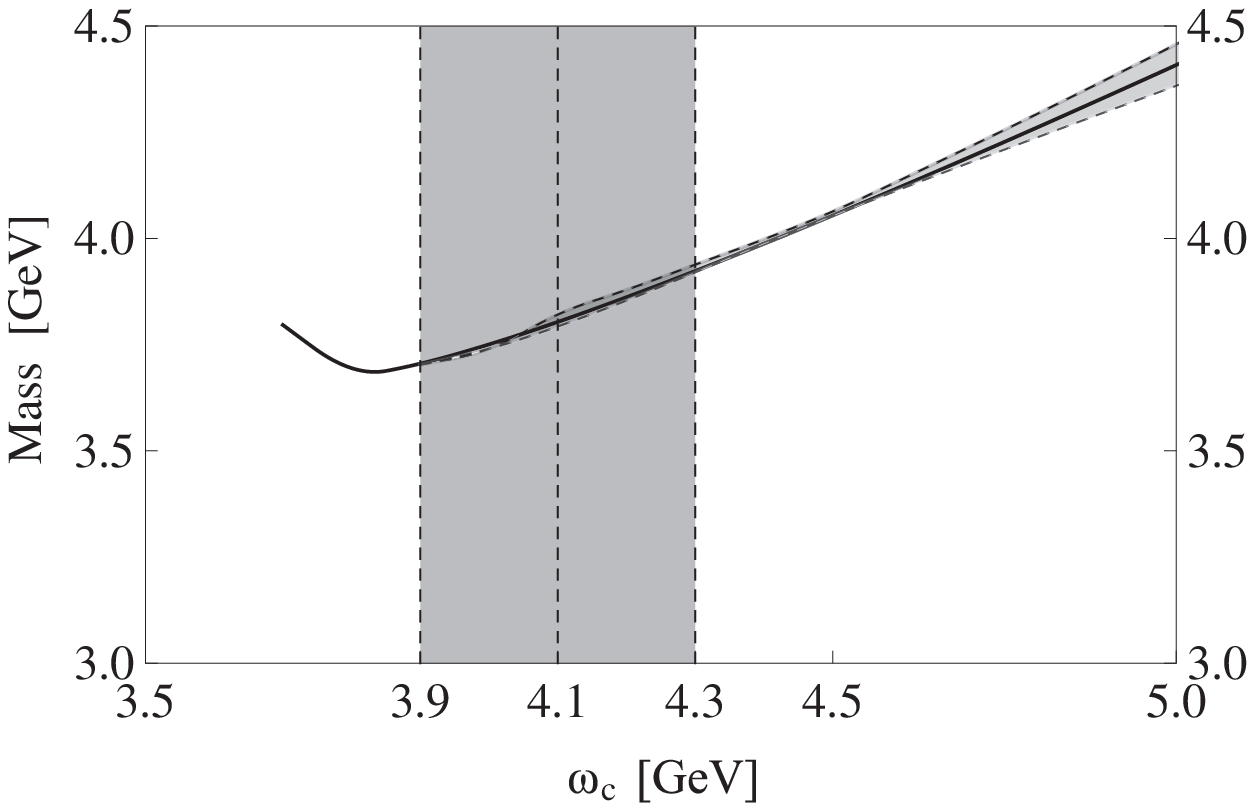}}
%\\
%\scalebox{0.6}{\includegraphics{lambda6F21wc.eps}}
\end{tabular}
\caption{Variations of $m_{[\Lambda_c,3,1,\rho\lambda]}$ (left) and $m_{[\Xi_c,3,1,\rho\lambda]}$ (right) with respect to the threshold value $\omega_c$, calculated using the charmed baryon doublet $[\mathbf{\bar 3}_F,3,1,\rho\lambda]$.
The shady band is obtained by changing $T$ inside Borel windows.
Although the mass curves have minimum against $\omega_c$ around 3.2 GeV (left) and 3.8 GeV (right), there exist Borel windows as long as $\omega_c \geq 3.4$ GeV (left) and $\omega_c \geq 3.9$ GeV (right), and the $\omega_c$ dependence is still weak and acceptable in the region 3.4 GeV $<\omega_c<$ 3.8 GeV (left) and 3.9 GeV $<\omega_c<$ 4.3 GeV (right). Moreover, these two threshold values satisfy $\omega_c(\Xi_c) - \omega_c(\Lambda_c) = 0.5$ GeV.
}
\label{fig:mass3111}
\end{center}
\end{figure}

\item The baryon doublet $[\mathbf{\bar 3}_F, 2, 1, \rho\lambda]$ contains $\Lambda_c(3/2^+,5/2^+)$ and $\Xi_c(3/2^+,5/2^+)$. We estimate their masses by averaging between $[\mathbf{\bar 3}_F,1,1,\rho\lambda]$ ($s_l = 1$ and $j_l = 1$) and $[\mathbf{\bar 3}_F,3,1,\rho\lambda]$ ($s_l = 1$ and $j_l = 3$), weighted by the spin-orbital splittings, to be:
\begin{eqnarray}
m_{[\Lambda_c, 2, 1, \rho\lambda]} &\sim& 3.20 \mbox{ GeV }  \, ,
\\ \nonumber m_{[\Xi_c, 2, 1, \rho\lambda]} &\sim& 3.76 \mbox{ GeV } \, ,
\end{eqnarray}
    These values are listed in Table~\ref{tab:results}, which are significantly larger than the masses of the $\Lambda_c(2880)$, $\Xi_c(3055)$ and $\Xi_c(3080)$.

\end{enumerate}

\section{Summary}
\label{sec:summary}

Summarizing all these results, we have studied the $D$-wave charmed baryons of $SU(3)$ flavor $\mathbf{\bar3}_F$ using the method of QCD sum rules within HQET.
We have calculated their masses up to the order $\mathcal{O}(1/m_Q)$ with large theoretical uncertainty, and we have also calculated their mass splittings within the same doublet with much less theoretical uncertainty.
Our results suggest that the $\Lambda_c(2880)$, $\Xi_c(3055)$ and $\Xi_c(3080)$ can be well described by the baryon doublet $[\mathbf{\bar 3}_F,2,0,\lambda\lambda]$ with $l_\rho = 0$, $l_\lambda = 2$ and $s_l=0$: a) the $\Lambda_c(2880)$ has $J^P=5/2^+$, it has a partner state the $\Lambda_c(3/2^+)$ of $J^P=3/2^+$ with a mass around $2.81 ^{+0.33}_{-0.18}$ GeV, and their mass difference is $28^{+45}_{-24}$ MeV;
b) the $\Xi_c(3055)$ and $\Xi_c(3080)$ have quantum number $J^P = 3/2^+$ and $5/2^+$, respectively.
The first conclusion (a) is consistent with the recent reference~\cite{Lu:2016ctt} by L\"{u} {\it et al}.

\begin{table}[hbt]
\begin{center}
\renewcommand{\arraystretch}{1.6}
\caption{Masses of the $D$-wave bottom baryons obtained using the baryon doublets $[\mathbf{\bar 3}_F,2,0,\rho\rho]$, $[\mathbf{\bar 3}_F,2,0,\lambda\lambda]$, $[\mathbf{\bar 3}_F,1,1,\rho\lambda]$, $[\mathbf{\bar 3}_F,2,1,\rho\lambda]$ and $[\mathbf{\bar 3}_F,3,1,\rho\lambda]$.
}
\begin{tabular}{c | c | c | c | c c | c c | c | c | c}
\hline\hline
\multirow{2}{*}{Multiplets} & \multirow{2}{*}{B} & $\omega_c$ & Working region & $\overline{\Lambda}$ & $f$ & $K$ & $\Sigma$ & Baryons & Mass & Difference
\\ & & (GeV) & (GeV) & (GeV) & (GeV$^{5}$) & (GeV$^2$) & (GeV$^2$) & ($j^P$) & (GeV) & (MeV)
\\ \hline\hline
\multirow{4}{*}{$[\mathbf{\bar 3}_F,2,0,\lambda\lambda]$} & \multirow{2}{*}{$\Lambda_b$} & \multirow{2}{*}{2.5} & \multirow{2}{*}{$0.331< T < 0.381$} & \multirow{2}{*}{$1.113$} & \multirow{2}{*}{$0.012$} & \multirow{2}{*}{$-2.239$} & \multirow{2}{*}{$0.014$} & $\Lambda_b(3/2^+)$ & $6.01^{+0.20}_{-0.12}$ & \multirow{2}{*}{$6^{+10}_{-5}$}
\\ \cline{9-10}
& & & & & & & & $\Lambda_b(5/2^+)$ & $6.01^{+0.20}_{-0.13}$ &
\\ \cline{2-11}
& \multirow{2}{*}{$\Xi_b$} & \multirow{2}{*}{3.0} & \multirow{2}{*}{$0.341< T < 0.459$} & \multirow{2}{*}{$1.279$} & \multirow{2}{*}{$0.025$} & \multirow{2}{*}{$-2.508$} & \multirow{2}{*}{$0.008$}
& $\Xi_b(3/2^+)$ & $6.19^{+0.10}_{-0.12}$ & \multirow{2}{*}{$3^{+3}_{-3}$}
\\ \cline{9-10}
& & & & & & & & $\Xi_b(5/2^+)$ & $6.19^{+0.10}_{-0.12}$ &
\\ \hline
\multirow{4}{*}{$[\mathbf{\bar 3}_F,2,0,\rho\rho]$} & \multirow{2}{*}{$\Lambda_b$} & \multirow{2}{*}{3.4} & \multirow{2}{*}{$0.358< T < 0.499$} & \multirow{2}{*}{$1.650$} & \multirow{2}{*}{$0.065$} & \multirow{2}{*}{$-1.742$} & \multirow{2}{*}{$0.011$} & $\Lambda_b(3/2^+)$ & $6.52^{+1.55}_{-0.26}$ & \multirow{2}{*}{$5^{+26}_{-4}$}
\\ \cline{9-10}
& & & & & & & & $\Lambda_b(5/2^+)$ & $6.52^{+1.58}_{-0.27}$ &
\\ \cline{2-11}
& \multirow{2}{*}{$\Xi_b$} & \multirow{2}{*}{3.9} & \multirow{2}{*}{$0.502< T < 0.591$} & \multirow{2}{*}{$1.723$} & \multirow{2}{*}{$0.10$} & \multirow{2}{*}{$-1.308$} & \multirow{2}{*}{$0.006$}
& $\Xi_b(3/2^+)$ & $6.57^{+0.16}_{-0.12}$ & \multirow{2}{*}{$2^{+3}_{-2}$}
\\ \cline{9-10}
& & & & & & & & $\Xi_b(5/2^+)$ & $6.57^{+0.16}_{-0.12}$ &
\\ \hline
\multirow{4}{*}{$[\mathbf{\bar 3}_F,1,1,\rho\lambda]$} & \multirow{2}{*}{$\Lambda_b$} & \multirow{2}{*}{3.0} & \multirow{2}{*}{$0.397< T < 0.457$} & \multirow{2}{*}{$1.335$} & \multirow{2}{*}{$0.044$} & \multirow{2}{*}{$-2.116$} & \multirow{2}{*}{$0.006$} & $\Lambda_b(1/2^+)$ & $6.22^{+0.18}_{-0.12}$ & \multirow{2}{*}{$1^{+2}_{-1}$}
\\ \cline{9-10}
& & & & & & & & $\Lambda_b(3/2^+)$ & $6.23^{+0.18}_{-0.12}$ &
\\ \cline{2-11}
& \multirow{2}{*}{$\Xi_b$} & \multirow{2}{*}{4.0} & \multirow{2}{*}{$0.526< T < 0.599$} & \multirow{2}{*}{$1.887$} & \multirow{2}{*}{$0.21$} & \multirow{2}{*}{$-2.954$} & \multirow{2}{*}{$0.003$}
& $\Xi_b(1/2^+)$ & $6.82^{+0.11}_{-0.09}$ & \multirow{2}{*}{$1^{+1}_{-1}$}
\\ \cline{9-10}
& & & & & & & & $\Xi_b(3/2^+)$ & $6.82^{+0.11}_{-0.09}$ &
\\ \hline
\multirow{4}{*}{$\begin{array}{c}[\mathbf{\bar 3}_F,2,1,\rho\lambda] \\ {\rm (estimated)}\end{array}$} & \multirow{2}{*}{$\Lambda_b$} & \multirow{2}{*}{--} & \multirow{2}{*}{--} & \multirow{2}{*}{--} & \multirow{2}{*}{--} & \multirow{2}{*}{--} & \multirow{2}{*}{--} & $\Lambda_b(3/2^+)$ & \multirow{2}{*}{$\sim6.36$} & \multirow{2}{*}{--}
\\ \cline{9-9}
& & & & & & & & $\Lambda_b(5/2^+)$ & &
\\ \cline{2-11}
& \multirow{2}{*}{$\Xi_b$} & \multirow{2}{*}{--} & \multirow{2}{*}{--} & \multirow{2}{*}{--} & \multirow{2}{*}{--} & \multirow{2}{*}{--} & \multirow{2}{*}{--}
& $\Xi_b(3/2^+)$ & \multirow{2}{*}{$\sim6.84$} & \multirow{2}{*}{--}
\\ \cline{9-9}
& & & & & & & & $\Xi_b(5/2^+)$ & &
\\ \hline
\multirow{4}{*}{$\begin{array}{c}[\mathbf{\bar 3}_F,3,1,\rho\lambda] \\ {\rm (simplified)}\end{array}$} & \multirow{2}{*}{$\Lambda_b$} & \multirow{2}{*}{3.6} & \multirow{2}{*}{$0.496< T < 0.542$} & \multirow{2}{*}{$1.628$} & \multirow{2}{*}{$0.022$} & \multirow{2}{*}{$-2.939$} & \multirow{2}{*}{--} & $\Lambda_b(5/2^+)$ & \multirow{2}{*}{$6.56^{+0.28}_{-0.15}$} & \multirow{2}{*}{--}
\\ \cline{9-9}
& & & & & & & & $\Lambda_b(7/2^+)$ & &
\\ \cline{2-11}
& \multirow{2}{*}{$\Xi_b$} & \multirow{2}{*}{4.1} & \multirow{2}{*}{$0.556< T < 0.609$} & \multirow{2}{*}{$1.920$} & \multirow{2}{*}{$0.045$} & \multirow{2}{*}{$-3.105$} & \multirow{2}{*}{--}
& $\Xi_b(5/2^+)$ & \multirow{2}{*}{$6.86^{+0.18}_{-0.13}$} & \multirow{2}{*}{--}
\\ \cline{9-9}
& & & & & & & & $\Xi_b(7/2^+)$ & &
\\ \hline \hline
\end{tabular}
\label{tab:bottomresults}
\end{center}
\end{table}

We have also evaluated the masses of the $D$ bottom baryons of $SU(3)$ flavor $\mathbf{\bar3}_F$. The results are listed in Table~\ref{tab:bottomresults}, where we have used the pole mass of the bottom quark, i.e., $m_b = 4.78 \pm 0.06$ GeV~\cite{pdg}.
We note again that the obtained bottom baryon masses significantly depend on the bottom quark mass, so have large theoretical uncertainty, but their splittings within the same doublet have much less theoretical uncertainty. Especially, the results obtained by using the baryon doublet $[\mathbf{\bar 3}_F,2,0,\lambda\lambda]$ are
\begin{eqnarray}
\nonumber m_{\Lambda_b({3/2}^+)} &=& 6.01^{+0.20}_{-0.12} \mbox{ GeV} \, ,
\\ \nonumber m_{\Lambda_b({5/2}^+)} &=& 6.01^{+0.20}_{-0.13} \mbox{ GeV} \, ,
\\ m_{\Lambda_b({5/2}^+)} - m_{\Lambda_b({3/2}^+)} &=& 6^{+10}_{-5} \mbox{ MeV} \, ,
\\ \nonumber m_{\Xi_b({3/2}^+)} &=& 6.19^{+0.10}_{-0.12} \mbox{ GeV} \, ,
\\ \nonumber m_{\Xi_b({5/2}^+)} &=& 6.19^{+0.10}_{-0.12} \mbox{ GeV} \, ,
\\ \nonumber m_{\Xi_b({5/2}^+)} - m_{\Xi_b({3/2}^+)} &=& 3^{+3}_{-3} \mbox{ MeV} \, .
\end{eqnarray}
We suggest to search for them in further experiments.

To end our paper, we would like to note that not only masses but also decay and production
properties are useful to clarify the nature of the heavy baryons, and an experimental project of such studies is planned at J-PARC~\cite{e50}.
Accordingly, in the following studies we plan to study the $D$-wave charmed baryons of $SU(3)$ flavor $\mathbf{6}_F$ and the $D$-wave bottom baryons.
We also plan to study decay properties of the excited heavy baryons, which can probably provide more useful information.

\section*{ACKNOWLEDGMENTS}

We thank Cheng-Ping Shen for useful discussions.
This project is supported by
the National Natural Science Foundation of China under Grants No. 11205011, No. 11475015, No. 11375024, No. 11222547, No. 11175073, and No. 11261130311,
the Ministry of Education of China (SRFDP under Grant No. 20120211110002 and the Fundamental Research Funds for the Central Universities),
the Fok Ying-Tong Education Foundation (Grant No. 131006), and
the National Program for Support of Top-notch Young Professionals.
Q.M. is supported by the Key Natural Science Research Program of Anhui Educational Committee (Grant No. KJ2016A774).
A.H. is supported in part by Grants-in-Aid for Scientific Research of JSPS, No. JP26400273(C).

\appendix

\section{Several Projection Operators}
\label{sec:project}

For the interpolating field, the projection operator projecting into pure spin 1 is:
\begin{eqnarray}
\Gamma_t^{\mu,\nu} &=& g_t^{\mu\nu} - {1 \over 3} \gamma_t^{\mu}\gamma_t^{\nu} \, .
\end{eqnarray}
The projection operator projecting into pure spin 2 is:
\begin{eqnarray}
\Gamma_t^{\mu_1\mu_2,\nu_1\nu_2} &=& g_t^{\mu_1\nu_1} g_t^{\mu_2\nu_2} + g_t^{\mu_1\nu_2} g_t^{\mu_2\nu_1} - {2 \over 15} g_t^{\mu_1\mu_2} g_t^{\nu_1\nu_2}
\\ \nonumber && - {1 \over 3} g_t^{\mu_1\nu_1} \gamma_t^{\mu_2}\gamma_t^{\nu_2} - {1 \over 3} g_t^{\mu_1\nu_2} \gamma_t^{\mu_2}\gamma_t^{\nu_1} - {1 \over 3} g_t^{\mu_2\nu_1} \gamma_t^{\mu_1}\gamma_t^{\nu_2} - {1 \over 3} g_t^{\mu_2\nu_2} \gamma_t^{\mu_1}\gamma_t^{\nu_1}
\\ \nonumber && + {1 \over 15} \gamma_t^{\mu_1} \gamma_t^{\nu_1} \gamma_t^{\mu_2} \gamma_t^{\nu_2} + {1 \over 15} \gamma_t^{\mu_1} \gamma_t^{\nu_2} \gamma_t^{\mu_2} \gamma_t^{\nu_1} + {1 \over 15} \gamma_t^{\mu_2} \gamma_t^{\nu_1} \gamma_t^{\mu_1} \gamma_t^{\nu_2} + {1 \over 15} \gamma_t^{\mu_2} \gamma_t^{\nu_2} \gamma_t^{\mu_1} \gamma_t^{\nu_1} \, .
\end{eqnarray}
The projection operator projecting into pure spin 3 is:
\begin{eqnarray}
\nonumber \Gamma_t^{\mu_1\mu_2\mu_3,\nu_1\nu_2\nu_3} &=& \mathbb{S}^{\prime\prime} \Big[ g_t^{\mu_1\nu_1} g_t^{\mu_2\nu_2} g_t^{\mu_3\nu_3}
+ c_1 \times g_t^{\mu_1\nu_1} g_t^{\mu_2\mu_3} g_t^{\nu_2\nu_3}
+ c_2 \times g_t^{\mu_1\nu_1} g_t^{\mu_2\nu_2} \gamma_t^{\mu_3}\gamma_t^{\nu_3}
+ c_3 \times g_t^{\mu_1\mu_2} g_t^{\nu_1\nu_2} \gamma_t^{\mu_3}\gamma_t^{\nu_3}
\\ &&
+ c_4 \times g_t^{\mu_1\nu_1} \gamma_t^{\mu_2} \gamma_t^{\nu_2} \gamma_t^{\mu_3} \gamma_t^{\nu_3}
+ c_5 \gamma_t^{\mu_1} \gamma_t^{\nu_1} \gamma_t^{\mu_2} \gamma_t^{\nu_2} \gamma_t^{\mu_3} \gamma_t^{\nu_3} \Big] \, ,
\end{eqnarray}
where $\mathbb{S}^{\prime\prime} [\cdots]$ denotes symmetrization and subtracting the trace terms in the sets $(\mu_1 \mu_2 \mu_3)$ and $(\nu_1\nu_2\nu_3)$. The five coefficients $c_{1,2,3,4,5}$ can be obtained by solving $\gamma^t_{\mu_1} \Gamma_t^{\mu_1\mu_2\mu_3,\nu_1\nu_2\nu_3} = 0$, which is not an easy task so we do not solve it here.

In the present study we do not need to always use these projection operators. For example, $J^{\alpha}_{3/2,+,\mathbf{\bar 3}_F,2,0,\rho\rho}$ defined in Eq.~(\ref{eq:current1}) naturally satisfies
\begin{eqnarray}
\gamma^{t}_{\alpha} J^{\alpha}_{3/2,+,\mathbf{\bar 3}_F,2,0,\rho\rho}(x) = 0 \, ,
\end{eqnarray}
so it has pure spin $3/2$.

The situation is much simpler for the two-point correlation function
\begin{eqnarray}
\Pi^{\alpha_1\cdots\alpha_{j-1/2},\beta_1\cdots\beta_{j-1/2}}_{F,j_l,s_l,\rho\rho/\lambda\lambda/\rho\lambda} (\omega)
&=& i \int d^4 x e^{i k x} \langle 0 |
T[J^{\alpha_1\cdots\alpha_{j-1/2}}_{j,P,F,j_l,s_l,\rho\rho/\lambda\lambda/\rho\lambda}(x)
\bar J^{\beta_1\cdots\beta_{j-1/2}}_{j,P,F,j_l,s_l,\rho\rho/\lambda\lambda/\rho\lambda}(0)] | 0 \rangle
\\ \nonumber &=& \mathbb{S} [ g_t^{\alpha_1 \beta_1} \cdots g_t^{\alpha_{j-1/2} \beta_{j-1/2}} ] {1 + v\!\!\!\slash \over 2} \Pi_{F,j_l,s_l,\rho\rho/\lambda\lambda/\rho\lambda} (\omega) + \cdots \, ,
\end{eqnarray}
that its leading term, $\Pi_{F,j_l,s_l,\rho\rho/\lambda\lambda/\rho\lambda} (\omega)$, only contains the highest spin $j$ component, while $\cdots$ contains other spin components.
$\mathbb{S} [\cdots]$ has been defined to denote symmetrization and subtracting the trace terms in the sets $(\alpha_1 \cdots \alpha_{j-1/2})$ and $(\beta_1 \cdots \beta_{j-1/2})$.

\section{Other Sum Rules}
\label{sec:others}

In this appendix we show the sum rules for other currents with different quark contents:

\begin{eqnarray}
&& \Pi_{\Lambda_c,2,0,\rho\rho} = f_{\Lambda_c,2,0,\rho\rho}^{2} e^{-2 \bar \Lambda_{\Lambda_c,2,0,\rho\rho} / T}
=\int_{0}^{\omega_c} [\frac{5}{145152\pi^4}\omega^9 - \frac{5\langle g_s^2 GG \rangle}{1728 \pi^4} \omega^5]e^{-\omega/T}d\omega \, ,
\\ && f_{\Lambda_c,2,0,\rho\rho}^{2} K_{\Lambda_c,2,0,\rho\rho} e^{-2 \bar \Lambda_{\Lambda_c,2,0,\rho\rho} / T}
=\int_{0}^{\omega_c} [ - \frac{41}{6386688\pi^4}\omega^{11} + \frac{59\langle g_s^2 GG \rangle}{90720\pi^4} \omega^7]e^{-\omega/T}d\omega \, ,
\\ && f_{\Lambda_c,2,0,\rho\rho}^{2} \Sigma_{\Lambda_c,2,0,\rho\rho} e^{-2 \bar \Lambda_{\Lambda_c,2,0,\rho\rho} / T}
=\int_{0}^{\omega_c} [\frac{\langle g_s^2 GG \rangle}{24192\pi^4} \omega^7]e^{-\omega/T}d\omega \, .
\end{eqnarray}

\begin{eqnarray}
&& \Pi_{\Xi_c,2,0,\rho\rho} = f_{\Xi_c,2,0,\rho\rho}^{2} e^{-2 \bar \Lambda_{\Xi_c,2,0,\rho\rho} / T}
\\ \nonumber && =\int_{2m_s}^{\omega_c} [ \frac{5}{145152\pi^4} \omega^9 - \frac{m_s^2}{672\pi^4} \omega^7 - \frac{m_s\langle \bar q q \rangle}{72 \pi^2} \omega^5 + \frac{m_s\langle \bar s s \rangle}{48 \pi^2} \omega^5
\\ \nonumber && ~~~~ -\frac{5\langle g_s^2 GG \rangle}{1728 \pi^4} \omega^5 + \frac{5m_s^2\langle g_s^2 GG \rangle}{192 \pi^4} \omega^3 - \frac{5m_s\langle g_s^2 GG \rangle\langle \bar s s \rangle}{72 \pi^2} \omega ]e^{-\omega/T}d\omega \, ,
\\ && f_{\Xi_c,2,0,\rho\rho}^{2} K_{\Xi_c,2,0,\rho\rho} e^{-2 \bar \Lambda_{\Xi_c,2,0,\rho\rho} / T}
\\ \nonumber && = \int_{2m_s}^{\omega_c} [ - \frac{41}{6386688\pi^4} \omega^{11} + \frac{197m_s^2}{483840\pi^4} \omega^9 + \frac{37m_s\langle \bar q q \rangle}{5040\pi^2} \omega^7 - \frac{277m_s\langle \bar s s \rangle}{20160\pi^2} \omega^7 + \frac{11m_s\langle g_s \bar q \sigma Gq \rangle}{180\pi^2} \omega^5
\\ \nonumber && ~~~~ + \frac{1921\langle g_s^2 GG \rangle}{2903040\pi^4} \omega^7 - \frac{7169m_s^2\langle g_s^2 GG \rangle}{552960\pi^4} \omega^5 - \frac{13m_s\langle g_s^2 GG \rangle\langle \bar q q \rangle}{216\pi^2} \omega^3 + \frac{2381m_s\langle g_s^2 GG \rangle\langle \bar s s \rangle}{20736\pi^2} \omega^3
\\ \nonumber && ~~~~ - \frac{121m_s\langle g_s^2 GG \rangle\langle g_s \bar q \sigma Gq \rangle}{1728\pi^2} \omega]e^{-\omega/T}d\omega\, ,
\\ && f_{\Xi_c,2,0,\rho\rho}^{2} \Sigma_{\Xi_c,2,0,\rho\rho} e^{-2 \bar \Lambda_{\Xi_c,2,0,\rho\rho} / T}
\\ \nonumber && = \int_{2m_s}^{\omega_c} [ \frac{\langle g_s^2 GG \rangle}{24192 \pi^4} \omega^7 - \frac{m_s^2\langle g_s^2 GG \rangle}{1536 \pi^4} \omega^5 + \frac{5m_s\langle g_s^2 GG \rangle\langle \bar s s \rangle}{864 \pi^2} \omega^3]e^{-\omega/T}d\omega\, .
\end{eqnarray}

\begin{eqnarray}
&&\Pi_{\Lambda_c,1,1,\rho\lambda}= f_{\Lambda_c,1,1,\rho\lambda}^2 e^{-2 \bar \Lambda_{\Lambda_c,1,1,\rho\lambda} / T}
= \int_{0}^{\omega_c} [ \frac{13}{161280\pi^4} \omega^9 - \frac{43\langle g_s^2 GG \rangle}{15360 \pi^4} \omega^5 ]e^{-\omega/T}d\omega  \, ,
\\ && f_{\Lambda_c,1,1,\rho\lambda}^2 K_{\Lambda_c,1,1,\rho\lambda} e^{-2 \bar \Lambda_{\Lambda_c,1,1,\rho\lambda} / T}
=\int_{0}^{\omega_c} [ - \frac{461}{17740800\pi^4} \omega^{11} + \frac{383\langle g_s^2 GG \rangle}{322560 \pi^4} \omega^7 ]e^{-\omega/T}d\omega \, ,
\\ && f_{\Lambda_c,1,1,\rho\lambda}^2 \Sigma_{\Lambda_c,1,1,\rho\lambda} e^{-2 \bar \Lambda_{\Lambda_c,1,1,\rho\lambda} / T}
=\int_{0}^{\omega_c} [ \frac{\langle g_s^2 GG \rangle}{26880 \pi^4} \omega^7 ]e^{-\omega/T}d\omega \, .
\end{eqnarray}

\begin{eqnarray}
&& \Pi_{\Xi_c,1,1,\rho\lambda}= f_{\Xi_c,1,1,\rho\lambda}^2 e^{-2 \bar \Lambda_{\Xi_c,1,1,\rho\lambda} / T}
\\ \nonumber && = \int_{2m_s}^{\omega_c} [ \frac{13}{161280\pi^4} \omega^9 - \frac{9m_s^2}{2240 \pi^4} \omega^7 - \frac{m_s\langle \bar q q \rangle}{16 \pi^2} \omega^5 + \frac{3m_s\langle \bar s s \rangle}{32 \pi^2} \omega^5 - \frac{3m_s\langle g_s \bar q \sigma Gq \rangle}{8 \pi^2} \omega^3
\\ \nonumber && ~~~~ - \frac{43\langle g_s^2 GG \rangle}{15360 \pi^4} \omega^5 + \frac{9m_s^2\langle g_s^2 GG \rangle}{256 \pi^4} \omega^3 - \frac{9m_s\langle g_s^2 GG \rangle\langle \bar s s \rangle}{64 \pi^2} \omega]e^{-\omega/T}d\omega \, ,
\\ && f_{\Xi_c,1,1,\rho\lambda}^2 K_{\Xi_c,1,1,\rho\lambda} e^{-2 \bar \Lambda_{\Xi_c,1,1,\rho\lambda} / T}
\\ \nonumber && = \int_{2m_s}^{\omega_c} [ - \frac{461}{17740800\pi^4} \omega^{11} + \frac{283m_s^2}{161280\pi^4} \omega^9 + \frac{m_s\langle \bar q q \rangle}{32 \pi^2} \omega^7 - \frac{29m_s\langle \bar s s \rangle}{448\pi^2} \omega^7 + \frac{5m_s\langle g_s \bar q \sigma Gq \rangle}{ 16\pi^2} \omega^5
\\ \nonumber && ~~~~ + \frac{383\langle g_s^2 GG \rangle}{ 322560\pi^4} \omega^7 - \frac{191m_s^2\langle g_s^2 GG \rangle}{ 7680\pi^4} \omega^5 - \frac{5m_s\langle g_s^2 GG \rangle\langle \bar q q \rangle}{ 72\pi^2} \omega^3 + \frac{133m_s\langle \bar s s \rangle\langle g_s^2 GG \rangle}{ 768\pi^2} \omega^3
\\ \nonumber && ~~~~ - \frac{m_s\langle g_s \bar q \sigma Gq \rangle\langle g_s^2 GG \rangle}{ 48\pi^2} \omega ]e^{-\omega/T}d\omega \, ,
\\ && f_{\Xi_c,1,1,\rho\lambda}^2 \Sigma_{\Xi_c,1,1,\rho\lambda} e^{-2 \bar \Lambda_{\Xi_c,1,1,\rho\lambda} / T}
\\ \nonumber && = \int_{2m_s}^{\omega_c} [ \frac{\langle g_s^2 GG \rangle}{26880 \pi^4} \omega^7 - \frac{m_s^2\langle g_s^2 GG \rangle}{960 \pi^4} \omega^5 + \frac{5m_s\langle g_s^2 GG \rangle\langle \bar s s \rangle}{288 \pi^2} \omega^3]e^{-\omega/T}d\omega \, .
\end{eqnarray}

\begin{eqnarray}
&& \Pi_{\Lambda_c,3,1,\rho\lambda}= f_{\Lambda_c,3,1,\rho\lambda}^2 e^{-2 \bar \Lambda_{\Lambda,3,1,\rho\lambda} / T} = \int_{0}^{\omega_c} [ \frac{1}{322560\pi^4} \omega^9 - \frac{\langle g_s^2 GG \rangle}{3840 \pi^4} \omega^5]e^{-\omega/T}d\omega \, ,
\\ && f_{\Lambda_c,3,1,\rho\lambda}^2 K_{\Lambda_c,3,1,\rho\lambda} e^{-2 \bar \Lambda_{\Lambda,3,1,\rho\lambda} / T} = \int_{0}^{\omega_c} [ - \frac{1}{1075200\pi^4} \omega^{11} + \frac{53\langle g_s^2 GG \rangle}{552960 \pi^4} \omega^7]e^{-\omega/T}d\omega \, .
\end{eqnarray}

\begin{eqnarray}
&& \Pi_{\Xi_c,3,1,\rho\lambda}= f_{\Xi_c,3,1,\rho\lambda}^2 e^{-2 \bar \Lambda_{\Xi_c,3,1,\rho\lambda} / T}
\\ \nonumber && = \int_{2m_s}^{\omega_c} [ \frac{1}{322560\pi^4} \omega^9 - \frac{m_s^2}{6720\pi^4} \omega^7 - \frac{m_s\langle \bar q q \rangle}{480 \pi^2} \omega^5 + \frac{m_s\langle \bar s s \rangle}{320 \pi^2} \omega^5 - \frac{m_s\langle g_s \bar q \sigma Gq \rangle}{96 \pi^2} \omega^3
\\ \nonumber && ~~~~ - \frac{\langle g_s^2 GG \rangle}{3840 \pi^4} \omega^5 + \frac{m_s^2\langle g_s^2 GG \rangle}{512 \pi^4} \omega^3 - \frac{m_s\langle \bar s s \rangle\langle g_s^2 GG \rangle}{128 \pi^2} \omega]e^{-\omega/T}d\omega \, ,
\\ && f_{\Xi_c,3,1,\rho\lambda}^2 K_{\Xi_c,3,1,\rho\lambda} e^{-2 \bar \Lambda_{\Xi_c,3,1,\rho\lambda} / T}
\\ \nonumber && = \int_{2m_s}^{\omega_c} [ - \frac{1}{1075200\pi^4} \omega^{11} + \frac{29m_s^2}{483840\pi^4} \omega^9 + \frac{m_s\langle \bar q q \rangle}{960 \pi^2} \omega^7 - \frac{9m_s\langle \bar s s \rangle}{4480 \pi^2} \omega^7 + \frac{3m_s\langle g_s \bar q \sigma G q\rangle}{320 \pi^2} \omega^5
\\ \nonumber && ~~~~ + \frac{53\langle g_s^2 GG \rangle}{552960 \pi^4} \omega^7 - \frac{217m_s^2\langle g_s^2 GG \rangle}{184320 \pi^4} \omega^5 - \frac{5m_s\langle \bar q q \rangle\langle g_s^2 GG \rangle}{1728 \pi^2} \omega^3 + \frac{65m_s\langle \bar s s \rangle\langle g_s^2 GG \rangle}{6912 \pi^2} \omega^3
\\ \nonumber && ~~~~ - \frac{m_s\langle g_s^2 GG \rangle\langle g_s \bar q \sigma G q\rangle}{1152 \pi^2} \omega]e^{-\omega/T}d\omega \, .
\end{eqnarray}

\end{document}